\documentclass[10pt, twocolumn,prb,amsmath,amssymb]{revtex4}
\usepackage{graphicx}
\usepackage{dcolumn}
\usepackage{bm}

\begin{document}
\title{Differential voltage amplification from ferroelectric negative capacitance}

\author{ Asif I. Khan$^{1,\dashv,\ast}$, Michael Hoffmann$^{2,\dashv}$, Korok Chatterjee$^{2}$,\\  Zhongyuan Lu$^{2}$, Ruijuan Xu$^{3}$, Claudy Serrao$^{4}$, Samuel Smith$^{2}$, Lane W. Martin$^{3,5}$,\\Chenming C. Hu$^{2}$, Ramamoorthy Ramesh$^{3,4,5}$, Sayeef Salahuddin$^{2,5\ast}$ \\
\normalsize{$^1$ School of Electrical and Computer Engineering, Georgia Institute of Technology, Atlanta, GA 30326}\\
\normalsize{$^{2}$Department of Electrical Engineering and Computer Sciences, University of California, Berkeley, CA 94720, USA}\\
\normalsize{$^{3}$Department of Material Science and Engineering, University of California, Berkeley, CA 94720, USA}\\
\normalsize{$^{4}$Department of Physics, University of California, Berkeley, CA 94720, USA}\\
\normalsize{$^{5}$Material Science Division, Lawrence Berkeley National Laboratory, Berkeley, CA 94720, USA}\\
\normalsize{$^\dashv$These authors contributed equally.}\\
\normalsize{$^\ast$To whom correspondence should be addressed; }\\\normalsize{E-mail: asif.khan@ece.gatech.edu, sayeef@berkeley.edu.}
}


\date{\today}

\begin{abstract}
\noindent It is well known that one needs an external source of energy to provide voltage amplification. Because of this, conventional circuit elements such as resistors, inductors or capacitors cannot provide amplification all by themselves. Here, we demonstrate that a ferroelectric can cause a differential amplification without needing such an external energy source. As the ferroelectric switches from one polarization state to the other, a transfer of energy takes place from the ferroelectric to the dielectric, determined by the ratio of their capacitances, which, in turn, leads to the differential amplification. {This amplification is very different in nature from conventional inductor-capacitor based circuits where an oscillatory amplification can be observed. The demonstration of differential voltage amplification from completely passive capacitor elements only, has fundamental ramifications for next generation electronics.}  \end{abstract}

\maketitle

Amplification forms the cornerstone of modern electronics. Amplification boosts an otherwise weak change of an electrical signal into a measurable quantity.  Conventional circuit blocks are made of three different elements: resistors, capacitors and inductors. Working alone, none of these elements can provide amplification. Due to this reason, these are often called the `passive' elements. Amplification in today's electronics is provided by transistors (often called the `active' elements), that draw additional energy from an external voltage source for amplifying signals. An exception to this rule came in 1953 with the invention of tunnel diodes by L. Esaki \cite{esaki1958new}. The tunnel diode can provide a negative differential resistance in the region where quantum mechanical tunneling is blocked by mis-alignment of the bands. It was shown \cite{esaki1963fundamentals} that in this region of negative differential resistance (NDR), a differential voltage amplification is possible. The tunnel diode amplifiers have found myriad of applications in electronics. Nonetheless, they remain to be the only example of an element that can provide a differential amplification by itself. In this work, we show that the recently discovered negative differential capacitance (NDC) \cite{salahuddin2008use, khan2015negative} in a ferroelectric material can provide a similar differential amplification of a voltage signal. Unlike, the tunnel diodes, however, the amplification comes from the imaginary part of the impedance (ND{\it C} vs ND{\it R}). Also, NDC is obtained in a simple capacitance configuration, an insulator sandwiched between two metallic plates, rather than needing a p-n junction.

\begin{figure}[!b]
\begin{center}
 \includegraphics[width=3.in]{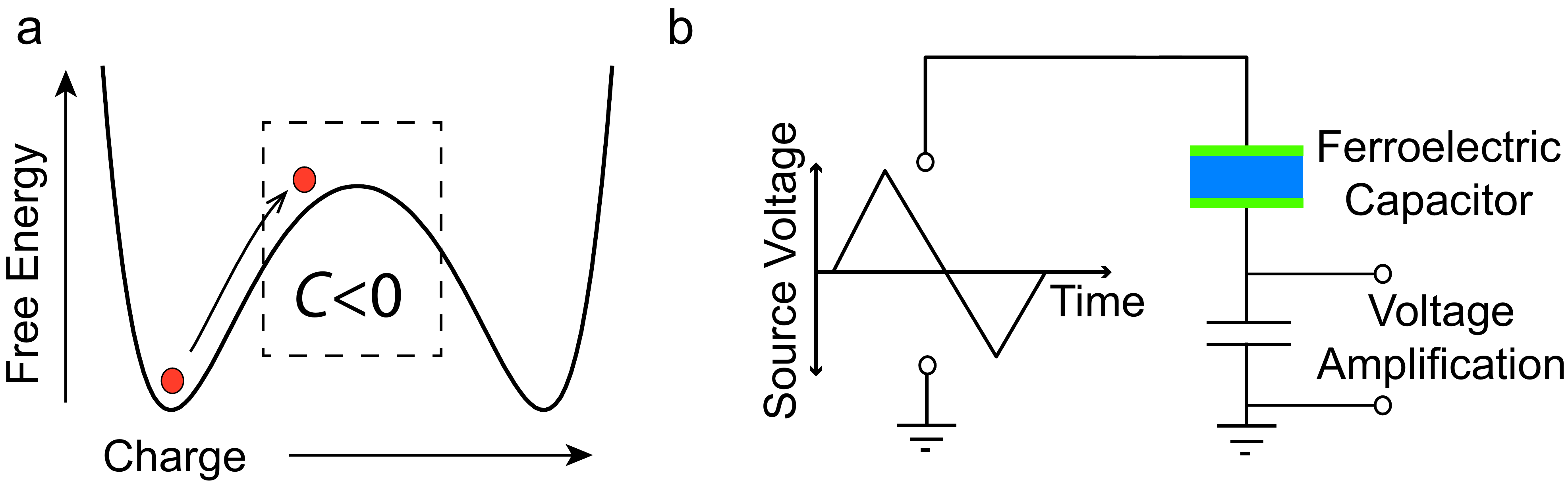}\vspace{-.2in}
 \end{center}
\caption{ {\bf  Voltage amplification due to ferroelectric negative capacitance:} (a) Energy landscape of a ferroelectric capacitor. The capacitance, $C$, is negative in the region enclosed by the dashed box. (b)  The experimental setup. }
 \label{figure1}
\end{figure} 

A ferroelectric material is characterized by a double well energy ($U$) landscape (Fig. \ref{figure1}a) \cite{lines1977principles}. Typically, the material is stabilized in one of the minima of the landscape. But when it is switched, it goes through the maximum where the susceptibility $[\partial^2 U/\partial Q^2]^{-1}$ is negative. The fact that the ferroelectric material can exhibit such a state of negative susceptibility or capacitance, has already been demonstrated \cite{salahuddin2008use, khan2015negative,khan2015extending,khan2011experimental,gao2014room,rusu2010metal,hoffmann2016direct,appleby2014experimental,dasguptasub,khan2016negative,jo2015negative,kasamatsu2016emergence,zubko2016negative,lee2015prospects, zhou2016ferroelectric, xiao2012simulation, nourbakhsh2017subthreshold}. Therefore, if another dielectric capacitor is placed in series with the ferroelectric (as shown schematically in Fig. \ref{figure1}b), a differential voltage amplification is expected across the dielectric capacitor. To see it, we note that the rate of change in the voltage across the dielectric capacitor ($V_D$) can be written in terms of the applied voltage ($V_S$) and the capacitance of the ferroelectric ($C_{FE}$) in the following way: 

\begin{equation}
\frac{dV_D}{dt}=\frac{dV_S}{dt}-\frac{1}{C_{FE}}\frac{dQ}{dt}=\frac{dV_S}{dt}+\frac{1}{|C_{FE}|}\frac{dQ}{dt}
\end{equation}

Here $C_{FE}$ is a function of time ($t$). Note also that the amplification is differential ($dV_D/dt>dV_S/dt$) and not absolute (\emph{i.e.}, $V_D$ may not be larger than $V_S$).

\begin{figure*}[!t]
\begin{center}
 \includegraphics[width=4.5in]{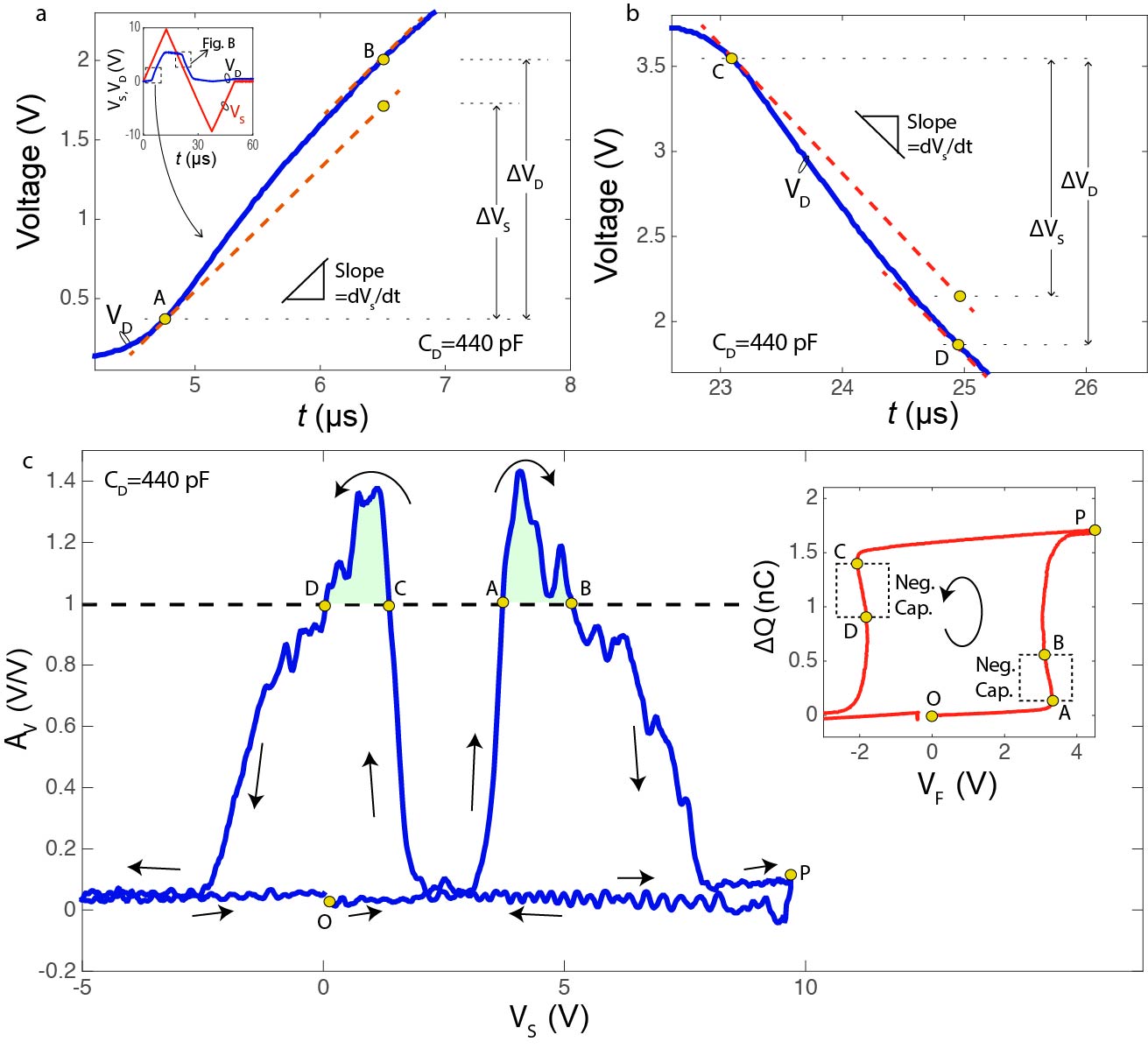}\vspace{-.3in}
 \end{center}
 \caption{ {\bf Voltage amplification in a ferroelectric-dielectric series circuit.} (a,b) The waveforms corresponding the voltage across the positive capacitor $V_D$ to a bipolar triangular voltage pulse $V_S$: 0 V $\rightarrow$ +10 V $\rightarrow$ -10 V $\rightarrow$ 0 V with period $T$=50 $\mu$s during 4.2 $\mu$s$<t<$6.7 $\mu$s (a)  and 22.6 $\mu$s$<t<$25.3 $\mu$s  (b). $C_D$=440 pF. The dashed red line has the same slew rate as that of $V_S(t)$ in these time frames.  The inset in Fig. a shows the waveforms corresponding to the source voltage $V_S$ and $V_D$ during the entire cycle. Differential amplification is observed in the regions corresponding to the green shades, segment $AB$ and $CD$ in Fig. a and b, respectively.   (c) Amplification $A_V$ (=$dV_D/dV_S$) as a function $V_S$.  The inset in Fig. c shows the extracted ferroelectric charge ($\Delta Q$)-voltage ($V_F$) characteristics extracted from the waveforms.} \label{figure2}\vspace{-.2in}
\end{figure*}

To test this hypothesis, we have built model systems where thin film capacitors of Pb(Zr$_{0.2}$Ti$_{0.8}$)O$_3$ (PZT) are placed in series with parallel plate capacitors with variable capacitance ($C_D$ = 77-440 pF). Epitaxial PZT  films (thickness=70-100 nm) are grown using the pulsed laser deposition technique on metallic SrRuO$_3$ buffered SrTiO$_3$ (001) substrates, and Ti/Au top metal electrodes were  sputtered and patterned. The series circuit is connected to a pulse function generator, and  the voltage across the dielectric ($V_D$) and the source voltage ($V_S$) are monitored with an oscilloscope. In Fig. \ref{figure2}a,  we  show the response of the series combination of a 100 nm thick PZT capacitor with area $A_F$ = (35 $\mu$m)$^2$ and a dielectric capacitor  $C_D$=440 pF to a bipolar triangular voltage pulse $V_S$: 0 V $\rightarrow$ +10 V $\rightarrow$ -10 V $\rightarrow$ 0 V with period $T$= 50 $\mu$s. The inset shows the $V_D$ and $V_S$ waveforms for the entire cycle and the main panel shows a blown-up version of the $V_D$ waveform during 4.2 $\mu$s $\le$ time ($t$) $\le$ 6.7 $\mu$s. The red dashed lines in the Fig. \ref{figure2}a main panel indicate the slope of $V_S$ in this time duration. We clearly observe that, during the segment $AB$, $V_S$ is increasing, and  $V_D$ increases faster than  $V_S$ ($i.e.$, $dV_D/dt>dV_S/dt$) indicating amplification of the source voltage at the node between the ferroelectric and the dielectric capacitor. In the segment $AB$, the change in $V_D$ ($\Delta V_D$) and $V_S$ ($\Delta V_S$) are  $\sim$1.7  V and $\sim$1.4 V, respectively, leading to an average amplification $\Delta V_D/\Delta V_S$ $\cong$1.21. Similarly, in the latter part of the transient response when $V_S$ is decreasing, an amplification in the $V_D$ waveform is observed in the segment $CD$, which is shown in Fig. \ref{figure2}b. In $CD$, $\Delta V_D \cong$ 1.66 V and $\Delta V_S \cong$ 1.4 V leading to  an average amplification of $\sim$1.19. Fig. \ref{figure2}c shows the amplification factor $A_V=dV_D/dV_S$ as a function of $V_S$ (see the caption of Fig. \ref{figure2}c and supplementary section 1 for $A_V$ calculation method). The $A_V-V_S$ curve has a butterfly shape, in which, $A_V>1$ in the segments, $AB$ and $CD$.  When the ferroelectric-dielectric combined system is in  an amplification state, the ferroelectric capacitor is essentially in a negative capacitance state. To demonstrate that, the ferroelectric charge ($\Delta Q$)-voltage ($V_F$) characteristics extracted from the waveforms is plotted in the inset of Fig. \ref{figure2}c (see supplementary section 1 for the extraction method). The extracted charge-voltage  curve of the ferroelectric capacitor is hysteretic and has distinctive negative slopes at the knees of the hysteresis loop  (segments $AB$ and $CD$) indicating negative capacitance in these regions.   Starting at the point $O$ at $t$ = 0 ($V_S$ = 0),  with the increase of $V_S$ ($V_S$: 0$\rightarrow$+10 V), the state of the ferroelectric capacitor ($(V_F$,$\Delta Q)$ pair)  traces the path $OABP$ in the hysteresis loop (Fig. \ref{figure2}c inset); when it is in the segment $AB$, the system responds with an amplification, which corresponds to  the $AB$ segment in the $V_D$ waveform in Fig. \ref{figure2}a and in the $A_V-V_S$ curve in the Fig. \ref{figure2}c main panel. During the latter part of the voltage pulse ($V_S$: +10 V$\rightarrow$ -10 V$\rightarrow$ 0 V), the ferroelectric goes through the rest of the hysteresis curve, and, when it goes through the other negative capacitance segment $CD$ (in Fig. \ref{figure2}c inset), the system again exhibits an amplification corresponding to the segment $CD$ in the $V_D$ waveform (Fig. \ref{figure2}b) and in the $A_V-V_S$ curve (Fig.  \ref{figure2}c main panel). We note that exactly similar shapes of ferroelectric hysteresis loops with negative capacitance was reported  in Ref. \cite{khan2015negative}  by using a completely different experimental setup with a series resistor. The observations made in Fig. \ref{figure2} confirm that the same  phenomenon in ferroelectric materials that results in negative capacitance transients in a ferroelectric capacitor-resistor series circuit leads to the amplification in a ferroelectric-dielectric series system.

\begin{figure}[!t]
\begin{center}
 \includegraphics[width=3in]{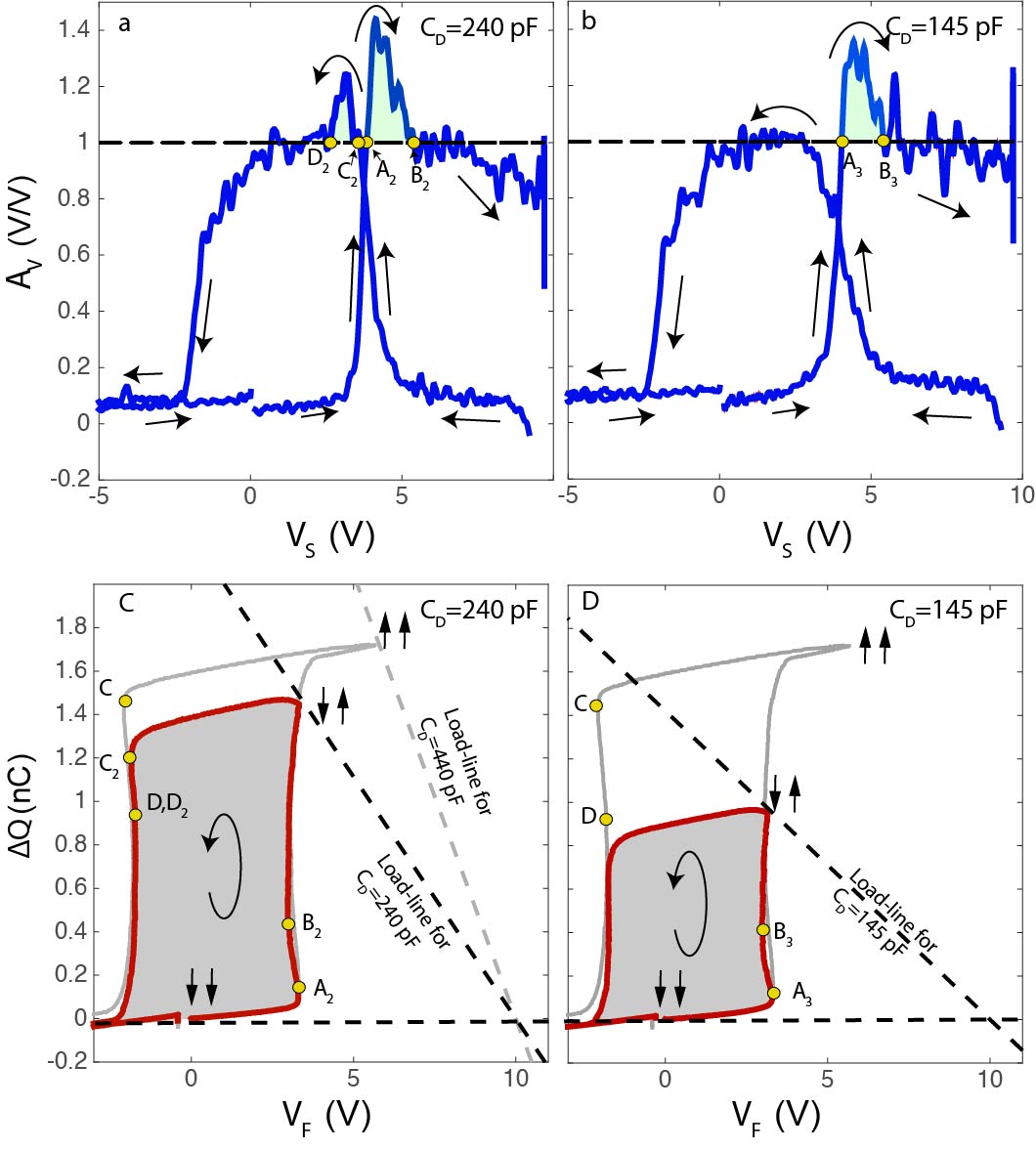}\vspace{-.3in}
 \end{center}
\caption{ {\bf Effect of dielectric capacitance on amplification.}   (a,b)  Amplification $A_V$ as a function of $V_S$ for  $C_D$=240 pF (a) and 145 pF (b). (c,d) The extracted charge-voltage characteristics of the ferroelectric capacitor for   $C_D$=240 pF (c) and 145 pF (d). These loops are overlaid on the ferroelectric charge-voltage characteristics extracted   $C_D$=440 pF, which is also shown in Fig. \ref{figure2}c inset. The load lines for the dielectric capacitor: $\Delta Q=C_D\times (V_{S}-V_F)$ corresponding to $V_S$=+10 V are plotted for $C_D$=240 pF  and 145 pF are plotted in Fig. c and d, respectively. The load-line ($V_S$=+10 V) for $C_D$=440 pF is plotted in both of them.}  \label{figure3}\vspace{-.2in}
\end{figure}

\begin{figure}[!t]
\begin{center}
 \includegraphics[width=3.in]{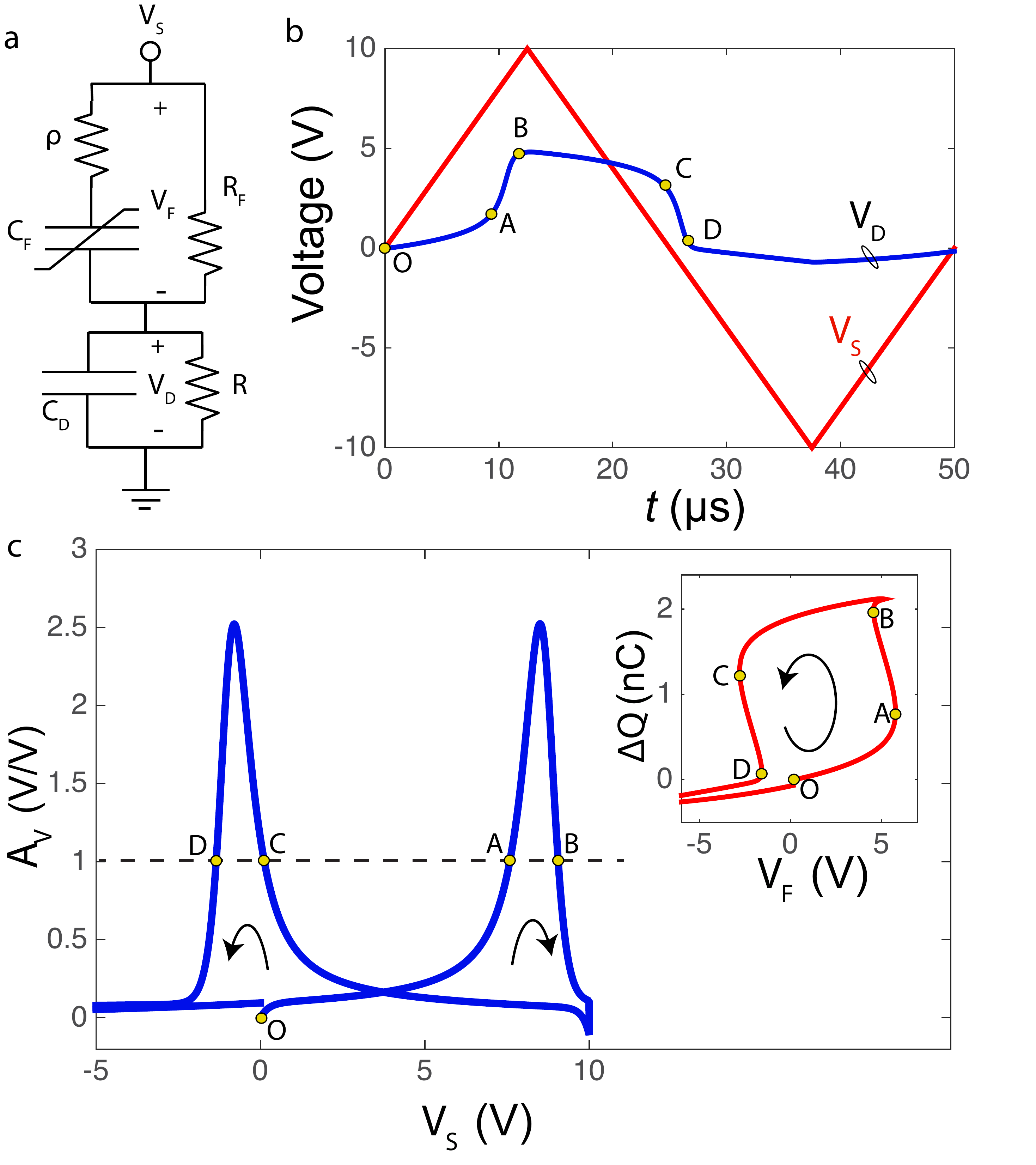}\vspace{-.3in}
 \end{center}
\caption{ {\bf Simulation results.}   (a) Circuit diagram of the simulation. $C_F$, $\rho$ and $R_F$ represent the  capacitance, the internal resistance and the leakage resistance of the ferroelectric capacitor. $C_D$ and $R$ represent the  capacitance and the leakage resistance of the dielectric capacitance.  (b) Simulated waveforms corresponding to $V_S$ and $V_D$ of the circuit shown in Fig. a in response to a bipolar triangular pulse $V_S$: 0 V $\rightarrow$ +10 V $\rightarrow$ -10 V $\rightarrow$ 0 V with period $T$=50 $\mu$s. Amplification  is observed in the segments, $AB$ and $CD$. (c) Simulated amplification $A_V$ as a function of $V_S$. $A_V\ge$1 in the segments, $AB$ and $CD$. The inset shows the ferroelectric charge-voltage characteristics extracted from the waveforms in Fig. b. }  \label{figure4} \vspace{-.2in}
\end{figure} 

We next study  the nature of the amplification response of the ferroelectric-dielectric system   by varying the dielectric capacitance $C_D$. Fig. \ref{figure3}a and \ref{figure3}b show the $A_V$-$V_S$ characteristics of the system for $C_D$=240 pF and 145 pF, respectively, and Fig. \ref{figure3}c and \ref{figure3}d plot the corresponding ferroelectric charge-voltage characteristics overlaid on the ferroelectric hysteresis loop calculated for $C_D$=440 pF.  The corresponding transient responses are  shown in supplementary Fig. S4 and S5. Comparing  Fig. \ref{figure2}c inset and \ref{figure3}a, we observe that the amplification in the ramp-down segment decreases when $C_D$ is changed from 440 pF to 240 pF. For $C_D$=145 pF, amplification ceases in  the ramp-down segment  as shown in Fig. \ref{figure3}b. In Fig. \ref{figure3}c and \ref{figure3}d, it is clear that, unlike the case for $C_D$=440 pF, the ferroelectric hysteresis loops do not saturate at the positive $V_F$ side for $C_D$=240 pF and 145 pF. As such, the ferroelectric capacitor traverses through minor loops  for these two smaller  $C_D$s. We note for the ferroelectric hysteresis loop for $C_D$=440 pF that the negative capacitance states occur only in a certain range of $(V_F, \Delta Q)$. For $C_D$=240 pF, the shape of the minor loop is such that it contains a smaller range of the negative capacitance $(V_F, \Delta Q)$ states in the reverse sweep path (compare the segments $CD$ and $C_2D_2$ on the ferroelectric loops for $C_D$=440 pF and 240 pF in Fig. \ref{figure3}c), thereby, resulting in a reduced amplification in the ramp-down compared to that for $C_D$=440 pF. For $C_D$=145 pF, the minor loop is much smaller, and does not contain any of the negative capacitance states, leading to no amplification during the ramp-down. Note that, in the forward sweep, the negative capacitance states remains intact on the ferroelectric loops for $C_D$=240 pF and 145 pF (see segments $A_2B_2$ and $A_3B_3$ in Fig. \ref{figure3}c and \ref{figure3}d). This is why, during the ramp-up, the system shows an amplification in the ramp-up for both 240 pF and 145 pF. 

To understand why the value of $C_D$ determines the shape of the ferroelectric hysteresis loop, we employ the load-line technique--a widely used method for analyzing  operating points in non-linear electronic circuits \cite{sedra1998microelectronic}. In Fig. \ref{figure3}c and \ref{figure3}d, the load lines for the dielectric capacitor: $\Delta Q=C_D\times (V_{S}-V_F)$ corresponding to $V_S$=+10 V are plotted for $C_D$=240 pF and 145 pF, respectively, along with that for $C_D$=440 pF in both of them. The intersection of the load-line and the ferroelectric  charge-voltage characteristics is the state of the ferroelectric when $V_S$=+10 V--the state with highest value of $\Delta Q$. With the decrease of $C_D$, the load-line becomes more slanted and intersects the hysteresis loop at smaller $\Delta Q$ values. Note that, at $t$=0, the ferroelectric capacitor is in a uniformly polarized state with the polarization pointing towards the SRO-PZT interface (indicated by $\downarrow\downarrow$ in Fig. \ref{figure3}c and \ref{figure3}d). For $C_D$=440 pF, the ferroelectric polarization completely switches at $V_S$=+10 V, which is indicated by $\uparrow\uparrow$ in Fig. \ref{figure3}c and \ref{figure3}d. On the other hand,   the ferroelectric polarization does not switch completely (indicated by $\downarrow\uparrow$ in Fig. \ref{figure3}c and \ref{figure3}d) at $V_S$=+10 V for $C_D$=240 pF and 145 pF, for which the ferroelectric charge-voltage characteristics trace a minor loop in the return path. This analysis shows that a smaller value of $C_D$ results in  a larger voltage drop across the dielectric capacitor ($\Delta Q/C_D$) and a smaller current that can flow in the circuit ($C_DdV_D/dt$), essentially, limiting the amount of charge that can transfer onto the ferroelectric capacitor from the source, which, for the particular cases of $C_D$=240 pF and 145 pF leads to incomplete switching of the polarization in response to the bipolar triangular pulse $V_S$: 0 V $\rightarrow$ +10 V $\rightarrow$ -10 V $\rightarrow$ 0 V. 
 
Similar functional behavior of the nature of the differential voltage amplification was also observed in the same system for much larger time periods ($T$= 500 $\mu$s and 5 ms), which  is detailed in supplementary section 3.  As a self-consistent check, we performed independent time-dependent measurements on ferroelectric-resistor series circuits using the same PZT capacitor and obtained ferroelectric charge-voltage hysteresis loops showing negative slope of the $P-V$ characteristics similar in shape to that in Fig. \ref{figure3}b (see supplementary section 8 and Fig. S32 for details). The robust behavior of amplification in the ferroelectric-dielectric series network over a span of time periods varying by 2 orders of magnitude as well negative capacitance transience in the ferroelectric-resistor series circuit provides a self-consistent proof of the underlying physics of the differential amplification occurring through negative capacitance.

{We also performed  transient simulations that consider both homogeneous and inhomogeneous switching of the ferroelectric polarization. In the presence of the intermediate metal electrode between the ferroelectric and the dielectric capacitor, the dielectric capacitor does not result in a depolarizing field in the ferroelectric at $V_S$=0--this is unlike the cases in ferroelectric-dielectric heterostructures without intermediate metallic layers [7,8,12,13,17]. As such,  the ferroelectric can be polarized even at $V_S$=0   [23].  For the homogeneous switching simulations (details in supplementary section 6), the Landau-Khalatnikov equation [22] and Kirchhoff's current and voltage laws are self-consistently solved; the simulated $V_D$ waveform, amplification $A_V$-$V_S$ curve and the ferroelectric charge-voltage characteristics extracted from the simulated waveforms shown in Fig. 4b and 4c are in qualitative agreement with the results shown in Fig. 2. The effect of inhomogeneous switching  on the amplification in our ferroelectric-dielectric system was simulated using a time-dependent Ginzburg-Landau (TDGL) framework which is described in the supplementary section 7. The TDGL based simulations can capture the quantitative features of our results as shown in supplementary Fig. S28 and S30.}

To summarize, we have directly measured differential voltage amplification in a combination of purely passive elements: a ferroelectric capacitor connected in series with an ordinary dielectric capacitor. As the ferroelectric switches from one state to the other, it imparts some of its stored energy onto the dielectric, leading to the amplification. We note that, in this process, there is no amplification of energy;  the dielectric eventually gives back that energy to the ferroelectric during the time when the amplification falls below 1. Since such energy transfer is dependent upon the combined potential energy landscape of the ferroelectric-dielectric combination, the amplification depends upon how well the capacitances are matched (note that the potential energy is linear with capacitance). Note that, while the amplification in Esaki diodes are based on the negative differential behavior of the {\it Real} or resistive part of the impedance, our work is based on the {\it imaginary} or reactive part of the impedance. Thus, together with the Esaki diodes, our work provides a complete picture: negative differential behavior in either part of the impedance will lead to a differential voltage amplification. The amplification demonstrated here can overcome the limits of voltage requirement in conventional transistors, often termed as the Boltzmann Tyranny \cite{meindl2001limits,zhirnov2003limits,theis2010s}, and therefore has a direct consequence for energy efficient electronics. Such amplification could also find applications in  high frequency transistors (by boosting the transconductance) and also for improving the sensitivity of sensor circuits beyond conventional limits. 

This work was supported in part by the Office of Naval Research (ONR), the Center for  Low Energy Systems Technology (LEAST), one of the six SRC STARnet Centers, sponsored by MARCO and DARPA and Entegris and Applied Materials under the I-Rice Center at the University of California, Berkeley. R. X. acknowledges support from the National Science Foundation under grant DMR-1608938. L.W.M. acknowledges support from the Army Research Office under grant W911NF-14-1-0104.

\clearpage

\begin{center}
\bf \Large Supplementary Online Information:\\ Differential voltage amplification from ferroelectric negative capacitance
\end{center}
\section{Amplification for different $C_D$}
\begin{figure}[!b]
\begin{center}
 \includegraphics[width=3.in]{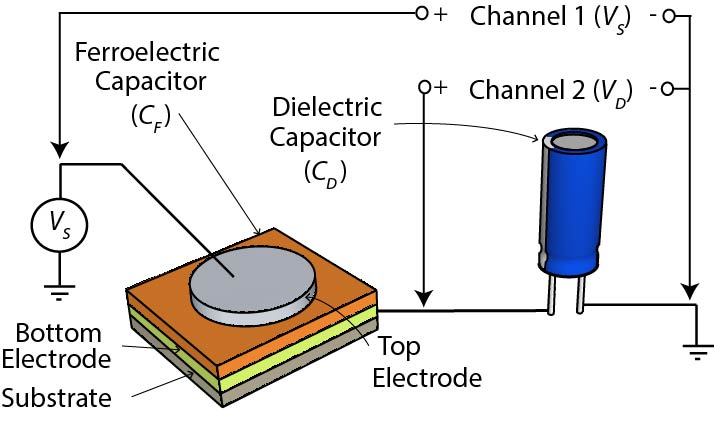}
 \end{center}
\caption{ Schematic diagram of the experimental set-up. }
 \label{setup}
\end{figure}

\begin{figure*}[!h]
\begin{center}
 \includegraphics[width=6.5in]{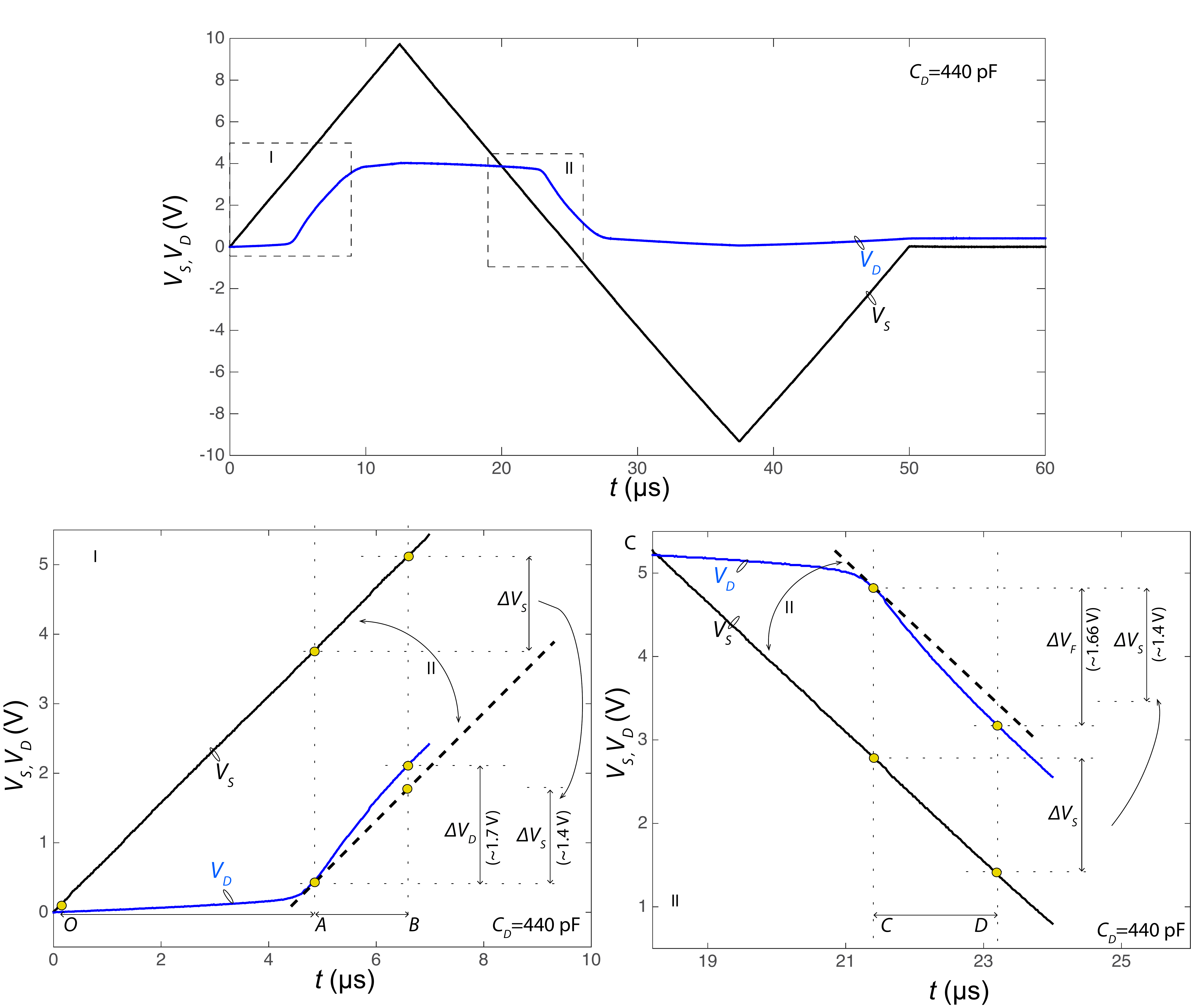}
 \end{center}
\caption{ $V_D$ and $V_S$ of the ferroelectric-dielectric series circuit in response to a bipolar triangular pulse $V_S$: 0$\rightarrow$+10 V$\rightarrow$ -10 V$\rightarrow$0 V with a period $T$=50 $\mu$s for $C_D$={440 pF}. Voltage amplification is observed in the main panel in the regions enclosed by boxes I and II. Magnified versions of the $V_D$ waveforms corresponding to boxes I and II are shown in the bottom panels separately.   In the bottom panels   the dashed  lines have the same slew rates as those of $V_S(t)$ in these durations. }
 \label{440pF}
\end{figure*}

\begin{figure*}[!h]
\begin{center}
 \includegraphics[width=6.5in]{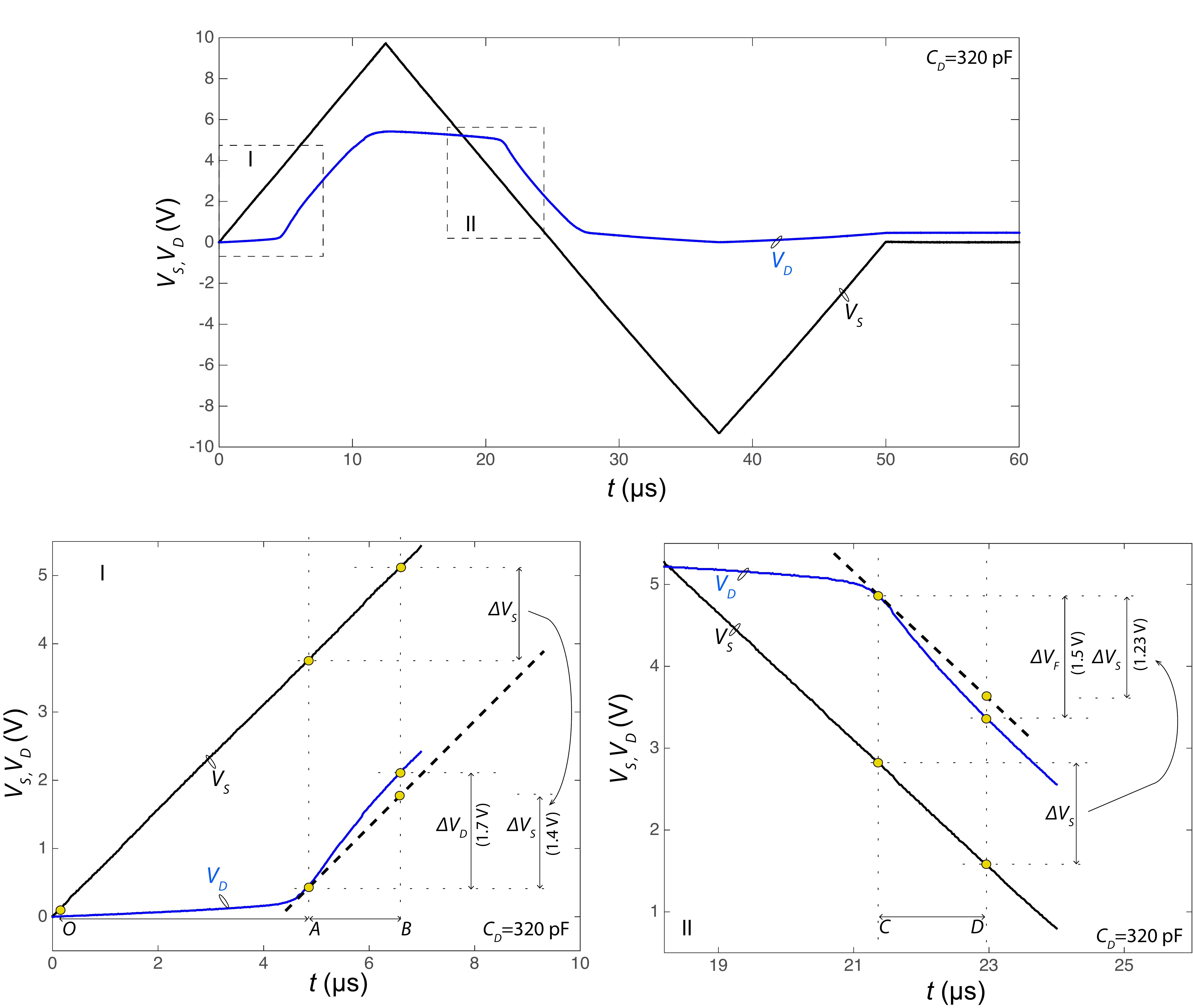}
 \end{center}
\caption{  $V_D$ and $V_S$ of the ferroelectric-dielectric series circuit in response to a bipolar triangular pulse $V_S$: 0$\rightarrow$+10 V$\rightarrow$ -10 V$\rightarrow$0 V with a period $T$=50 $\mu$s for $C_D$={320 pF}. Voltage amplification is observed in the main panel in the regions enclosed by boxes I and II. Magnified versions of the $V_D$ waveforms corresponding to boxes I and II are shown in the bottom panels separately.   In the bottom panels   the dashed  lines have the same slew rates as those of $V_S(t)$ in these durations.}
 \label{320pF}
\end{figure*}

\begin{figure*}[!h]
\begin{center}
 \includegraphics[width=6.5in]{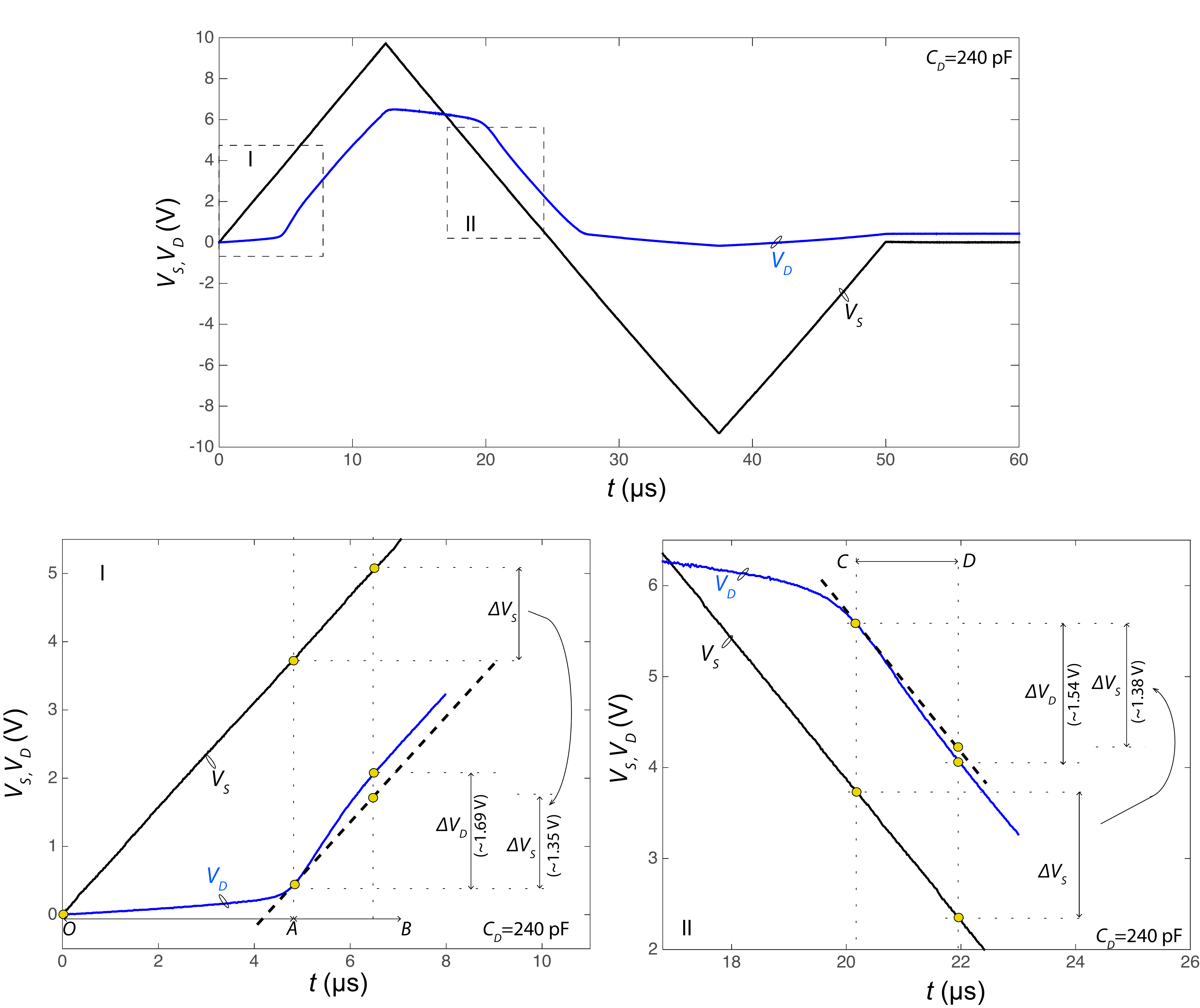}
 \end{center}
\caption{ $V_D$ and $V_S$ of the ferroelectric-dielectric series circuit in response to a bipolar triangular pulse $V_S$: 0$\rightarrow$+10 V$\rightarrow$ -10 V$\rightarrow$0 V with a period $T$=50 $\mu$s for $C_D$={240 pF}. Voltage amplification is observed in the main panel in the regions enclosed by boxes I and II. Magnified versions of the $V_D$ waveforms corresponding to boxes I and II are shown in the bottom panels separately.   In the bottom panels   the dashed  lines have the same slew rates as those of $V_S(t)$ in these durations. }
 \label{240pF}
\end{figure*}

\begin{figure*}[!h]
\begin{center}
 \includegraphics[width=6.5in]{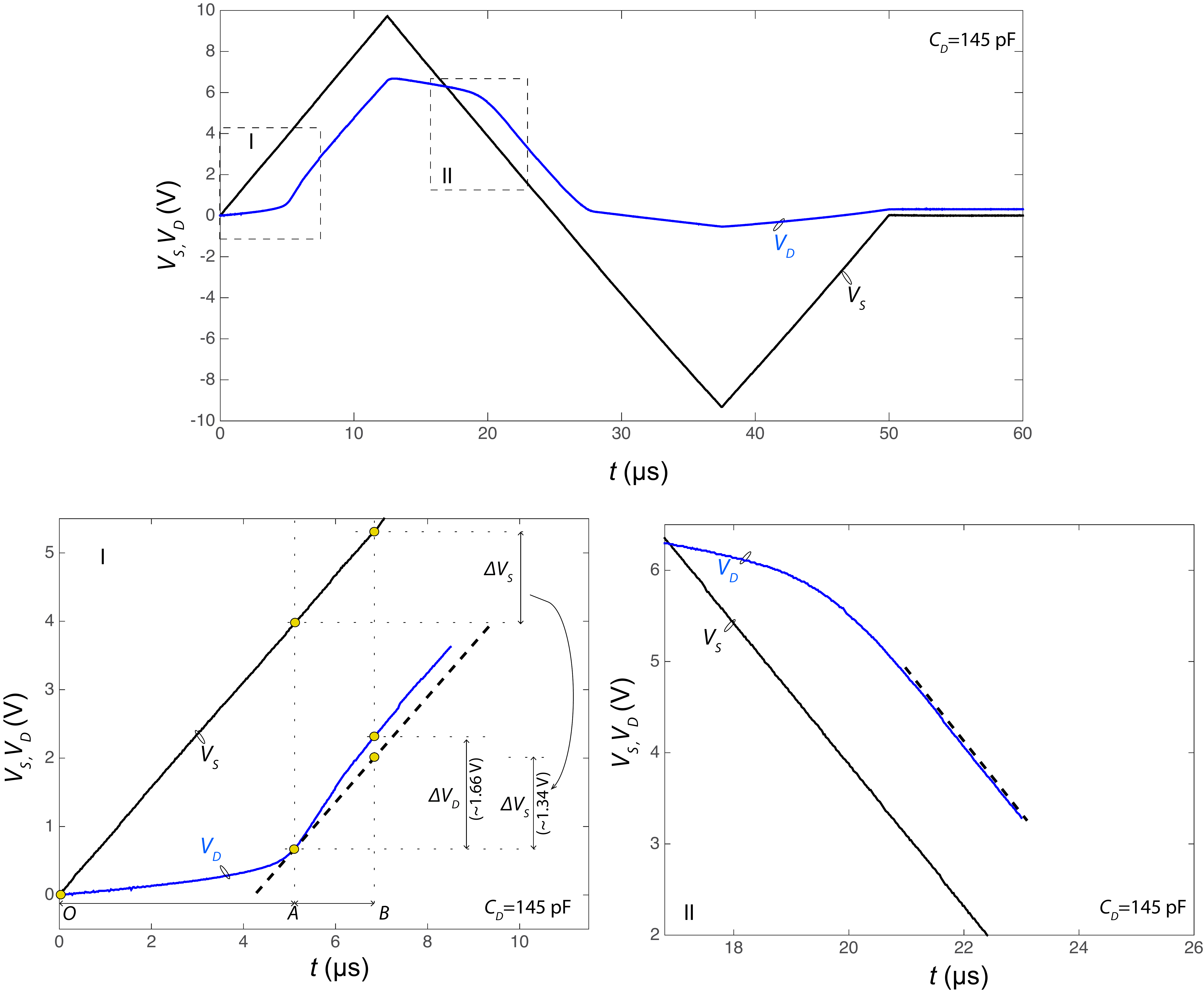}
 \end{center}
\caption{ $V_D$ and $V_S$ of the ferroelectric-dielectric series circuit in response to a bipolar triangular pulse $V_S$: 0$\rightarrow$+10 V$\rightarrow$ -10 V$\rightarrow$0 V with a period $T$=50 $\mu$s for $C_D$={145 pF}. Voltage amplification is observed in the main panel in the regions enclosed by boxes I and II. Magnified versions of the $V_D$ waveforms corresponding to boxes I and II are shown in the bottom panels separately.   In the bottom panels   the dashed  lines have the same slew rates as those of $V_S(t)$ in these durations.}
 \label{145pF}
\end{figure*}

\begin{figure*}[!h]
\begin{center}
 \includegraphics[width=6.5in]{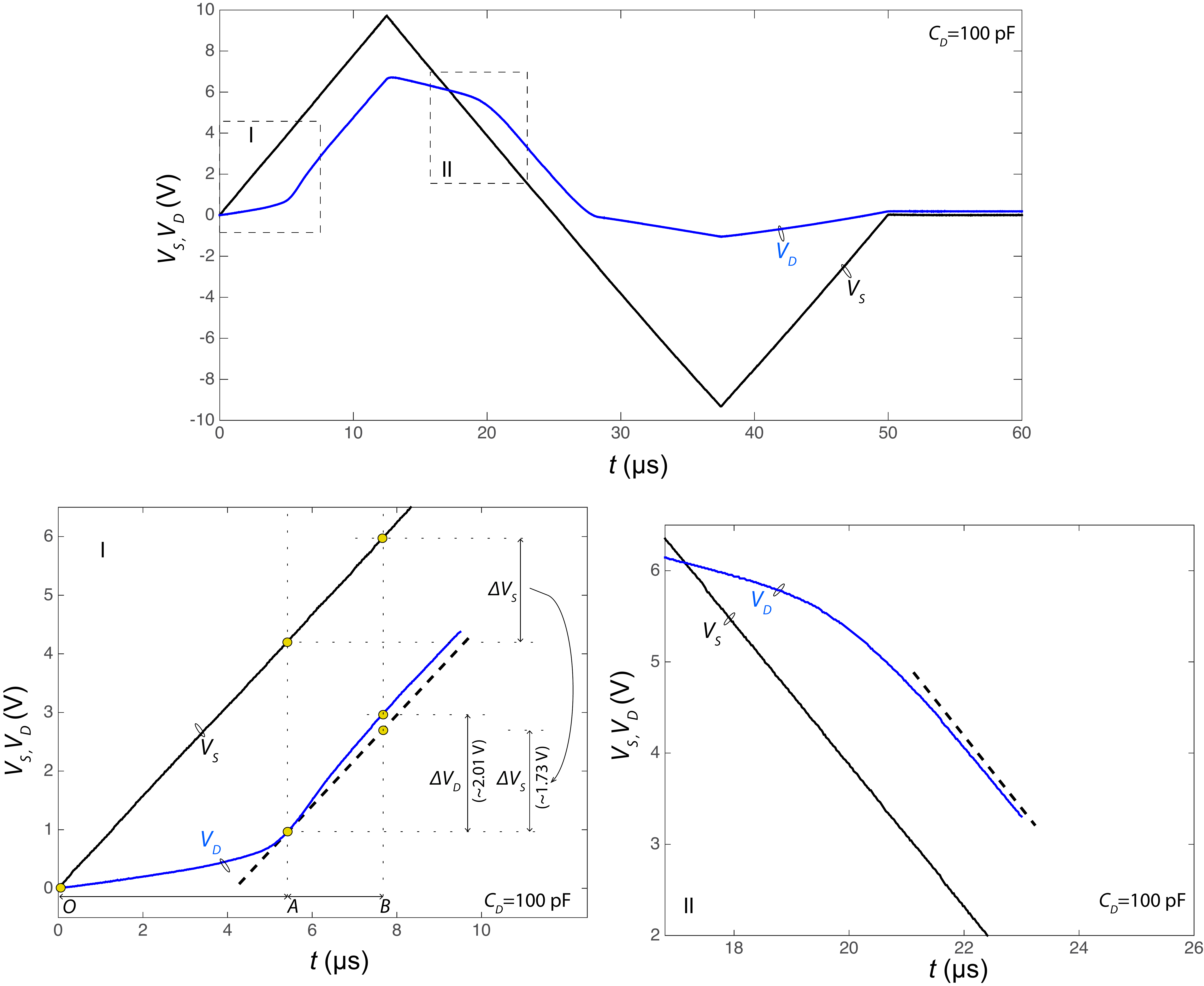}
 \end{center}
\caption{ $V_D$ and $V_S$ of the ferroelectric-dielectric series circuit in response to a bipolar triangular pulse $V_S$: 0$\rightarrow$+10 V$\rightarrow$ -10 V$\rightarrow$0 V with a period $T$=50 $\mu$s for $C_D$={100 pF}. Voltage amplification is observed in the main panel in the regions enclosed by boxes I and II. Magnified versions of the $V_D$ waveforms corresponding to boxes I and II are shown in the bottom panels separately.   In the bottom panels   the dashed  lines have the same slew rates as those of $V_S(t)$ in these durations. }
 \label{100pF}
\end{figure*}

\begin{figure*}[!h]
\begin{center}
 \includegraphics[width=6.5in]{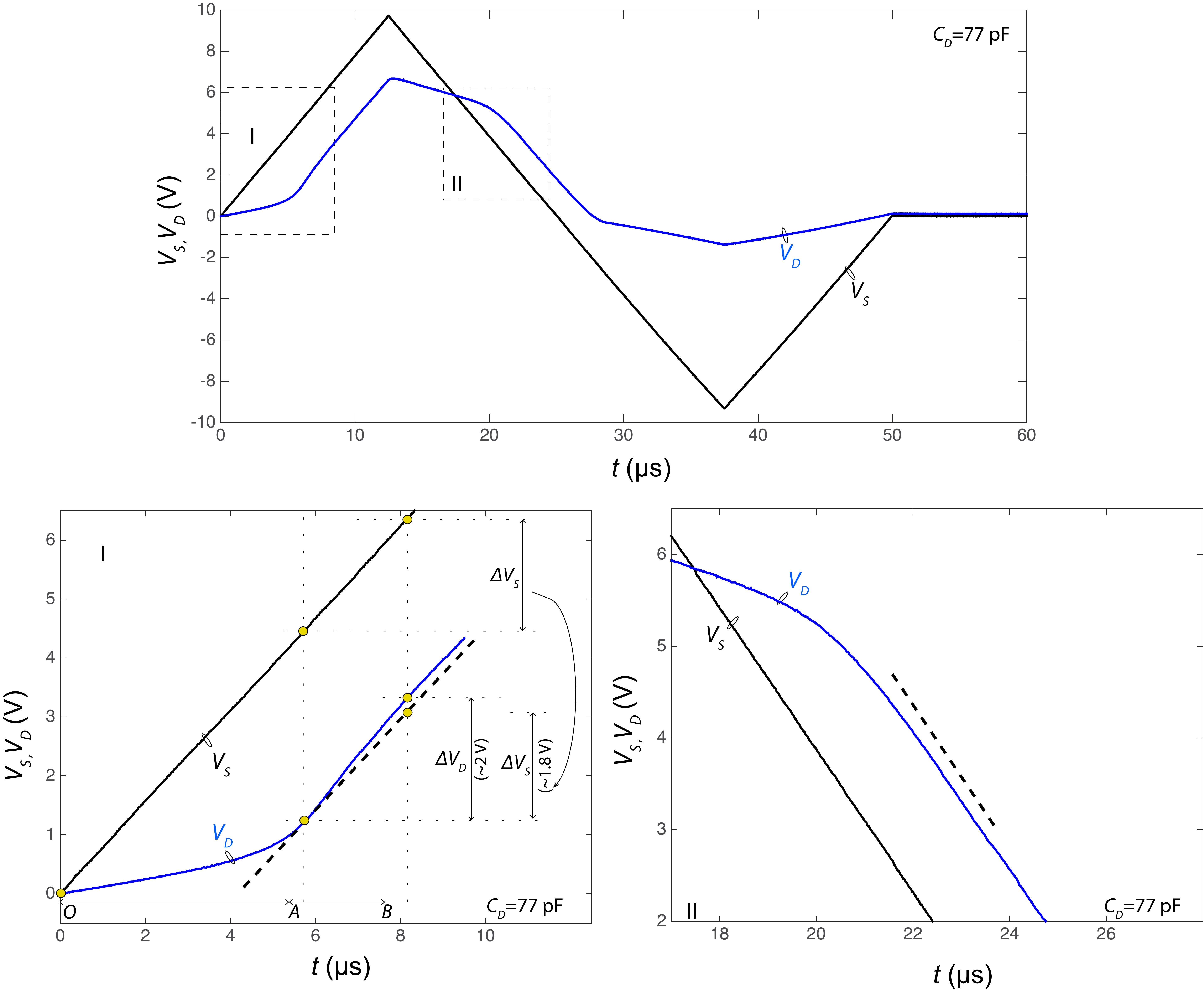}
 \end{center}
\caption{  $V_D$ and $V_S$ of the ferroelectric-dielectric series circuit in response to a bipolar triangular pulse $V_S$: 0$\rightarrow$+10 V$\rightarrow$ -10 V$\rightarrow$0 V with a period $T$=50 $\mu$s for $C_D$={77 pF}. Voltage amplification is observed in the main panel in the regions enclosed by boxes I and II. Magnified versions of the $V_D$ waveforms corresponding to boxes I and II are shown in the bottom panels separately.   In the bottom panels   the dashed  lines have the same slew rates as those of $V_S(t)$ in these durations. }
 \label{77pF}
\end{figure*}

\begin{figure*}[!h]
\begin{center}
 \includegraphics[width=6.5in]{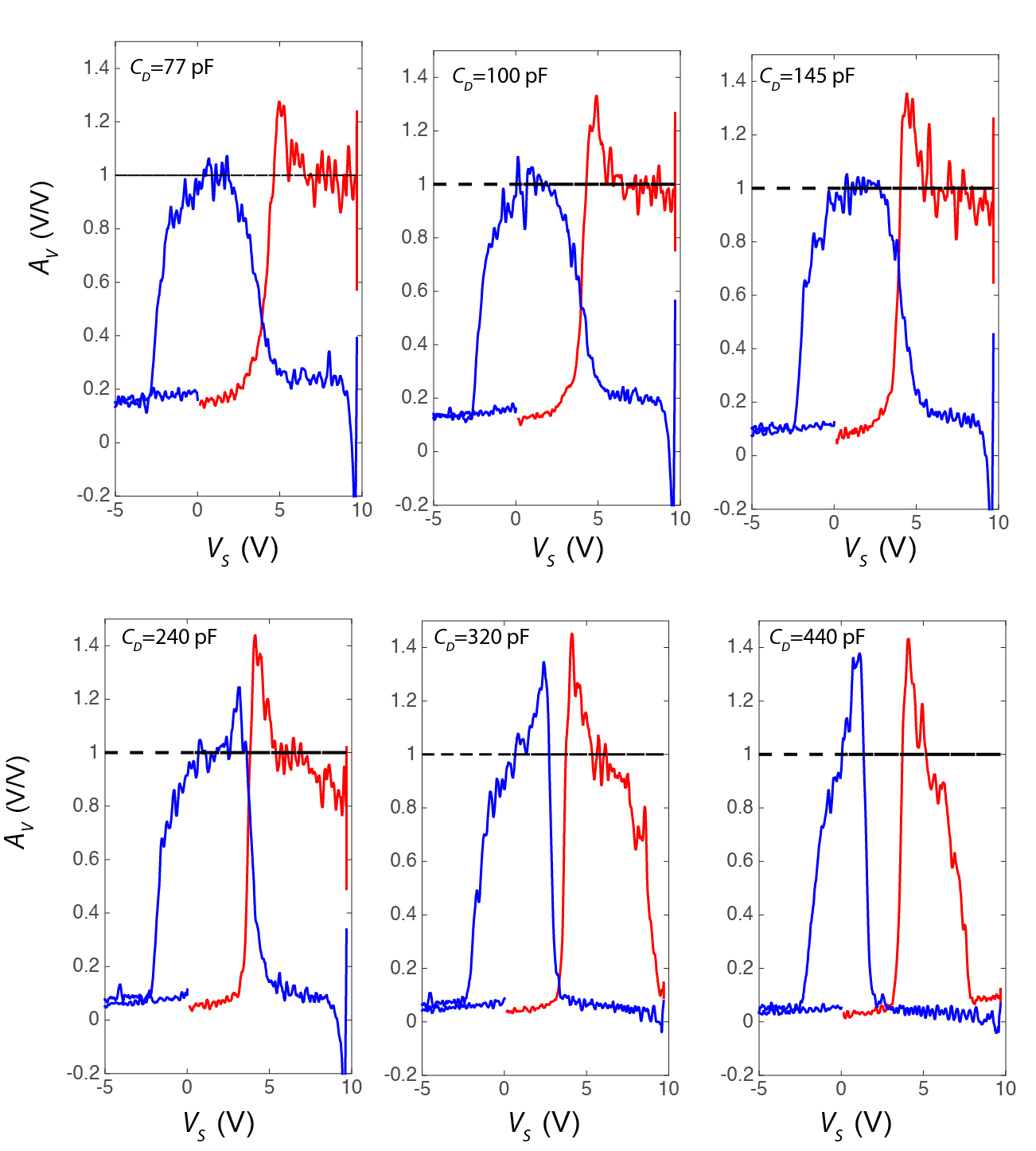}
 \end{center}
\caption{Amplification $A_V$ (=$dV_D/dV_S$) as a function $V_S$ for $C_D$={440 pF}, {320 pF},  {240 pF},  {145 pF}, {100 pF} and  {77 pF}.}
 \label{amp_all}
\end{figure*}

\begin{figure*}[!h]
\begin{center}
 \includegraphics[width=6.5in]{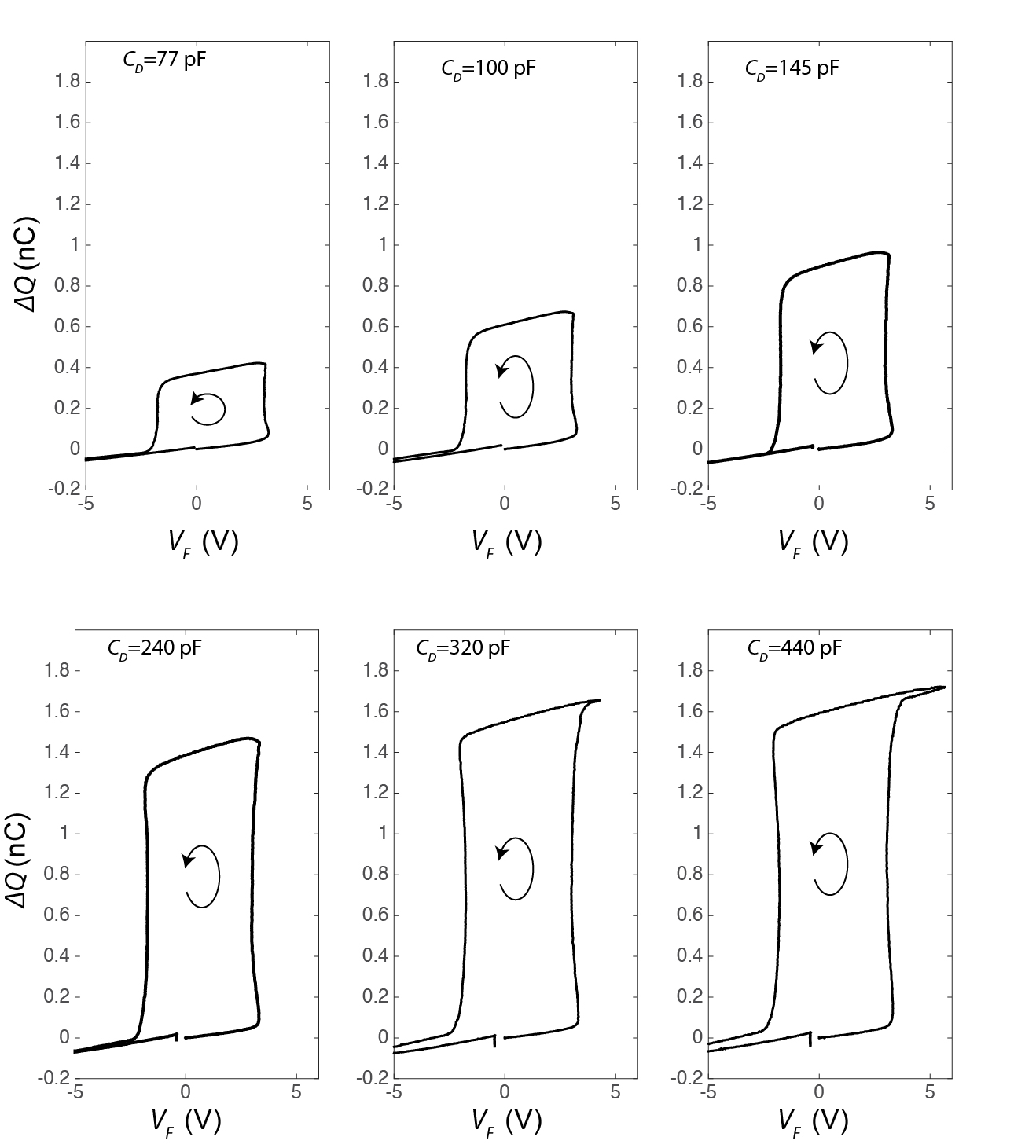}
 \end{center}
\caption{The extracted ferroelectric charge ($\Delta Q$)-voltage ($V_F$) characteristics extracted from the waveforms for $C_D$={440 pF}, {320 pF},  {240 pF},  {145 pF}, {100 pF} and  {77 pF}.}
 \label{hys_all}
\end{figure*}

Fig. \ref{setup} shows the schematic diagram for the experimental set-up. Fig. \ref{440pF}, \ref{320pF},  \ref{240pF},  \ref{145pF},  \ref{100pF} and  \ref{77pF} show the waveforms corresponding to $V_D$ and $V_S$ of the ferroelectric-dielectric series circuit in response to a bipolar triangular pulse $V_S$: 0$\rightarrow$+10 V$\rightarrow$ -10 V$\rightarrow$0 V with a period $T$=50 $\mu$s for $C_D$={440 pF}, {320 pF},  {240 pF},  {145 pF}, {100 pF} and  {77 pF}, respectively. Magnified versions of the $V_D$ waveforms in the main-panels are also shown separately.   In the $V_D$ waveform, amplification corresponds to the section $AB$  for the forward sweep direction (Fig. \ref{440pF}, \ref{320pF},  \ref{240pF},  \ref{145pF},  \ref{100pF} and  \ref{77pF}), and,  the section $CD$ for the reverse sweep direction (Fig. \ref{440pF}, \ref{320pF} and  \ref{240pF}). Amplification $A_V$ (=$dV_D/dV_S$) as a function $V_S$ is plotted in Fig. \ref{amp_all} for $C_D$={440 pF}, {320 pF},  {240 pF},  {145 pF}, {100 pF} and  {77 pF}. Fig. \ref{hys_all} shows the extracted ferroelectric charge ($\Delta Q$)-voltage ($V_F$) characteristics extracted from the waveforms for $C_D$={440 pF}, {320 pF},  {240 pF},  {145 pF}, {100 pF} and  {77 pF}.

\paragraph{Calculation of $A_V$}
The differential amplification at a given time $t$, $A_v(t)$, is calculated to be the ratio of the time rate of change of $V_D(t)$ and $V_S(t)$  ($i.e$  $A_v(t)=(dV_D/dt)_{(t)}/(dV_S/dt)_{(t)}$). A low-pass filter was applied on the waveforms for $V_S(t)$ and $V_D(t)$   to reduce the effects of noise on differentiation. 

\paragraph{Extraction of ferroelectric charge-voltage characteristics}
 By charge across the ferroelectric at a given time $t$ ($\Delta Q(t)$), we refer to the charge supplied to it by the voltage pulse source at $t$.  Since, in a series network of two capacitors, the changes in the charge across each of the capacitors at a given $t$ are the same (assuming minimal leakage), the ferroelectric charge at a given time $t$ is calculated using the relation: $\Delta Q(t)=C_D\times V_D(t)$. The voltage across the ferroelectric capacitor at $t$, $V_F(t)$, is equal to $V_S(t)-V_D(t)$.

\section{Load line analysis of the ferroelectric-dielectric series network}
\begin{figure*}[!h]
\begin{center}
 \includegraphics[width=5.in]{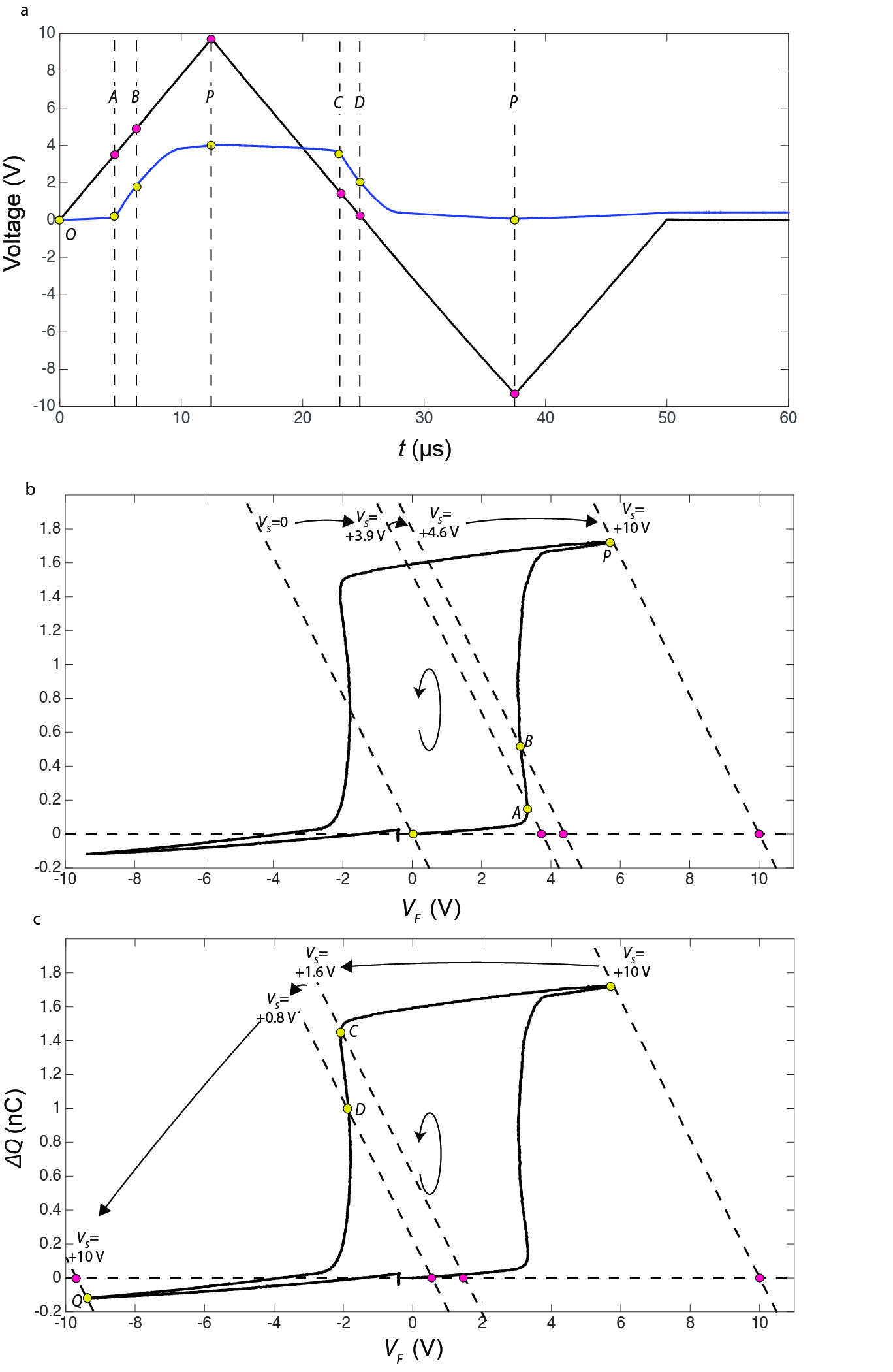}
 \end{center}
\caption{ Load lines for different $V_S$ during  0$\rightarrow$+10 V$\rightarrow$ (b) and +10 V$\rightarrow$-10 V sweeps (c) for $C_D$=440 pF. }
 \label{ll440pF}
\end{figure*}

\begin{figure*}[!h]
\begin{center}
 \includegraphics[width=5.in]{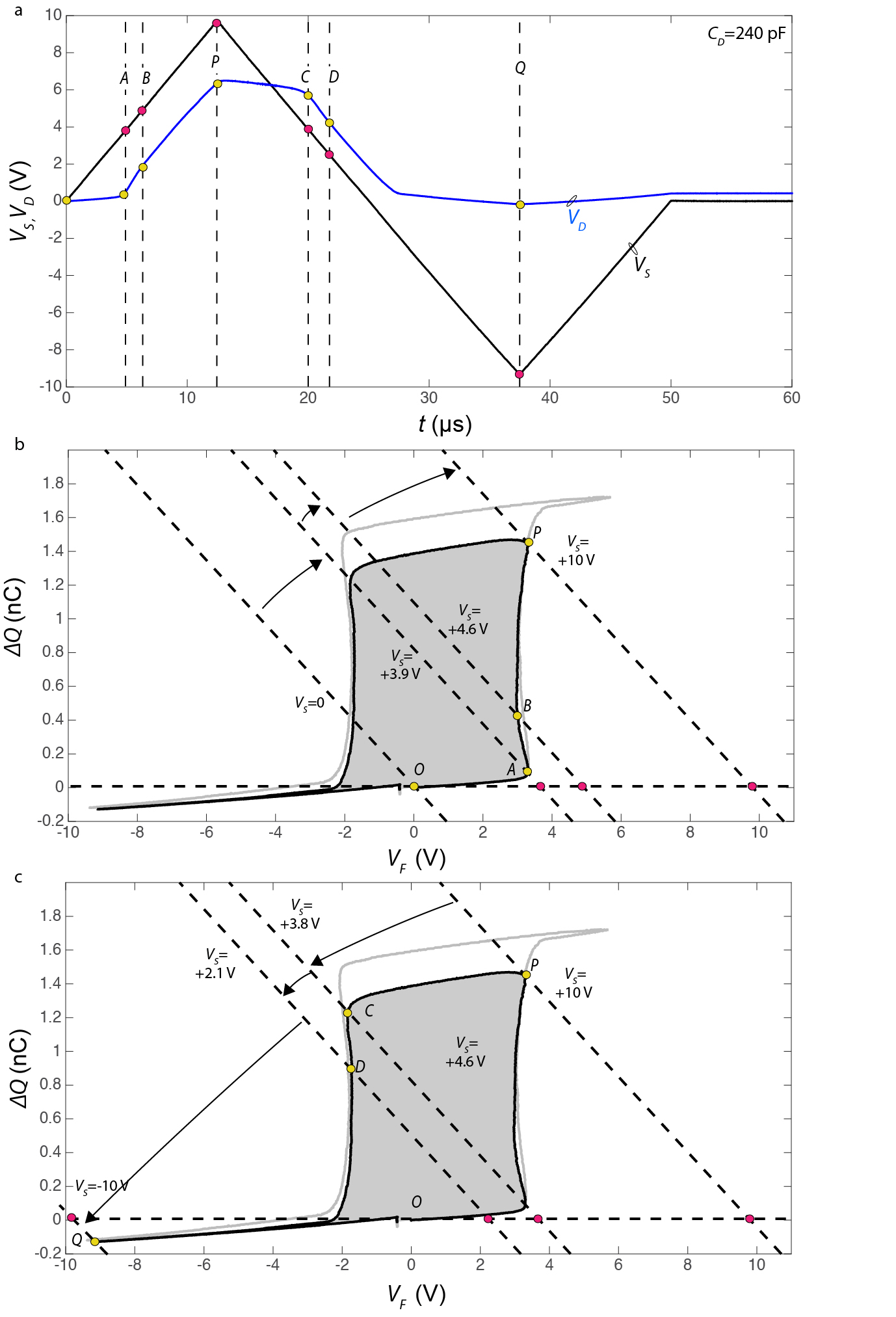}
 \end{center}
\caption{  Load lines for different $V_S$ during  0$\rightarrow$+10 V$\rightarrow$ (b) and +10 V$\rightarrow$-10 V sweeps (c) for $C_D$=240 pF. }
 \label{ll240pF}
\end{figure*}

\begin{figure*}[!h]
\begin{center}
 \includegraphics[width=5.in]{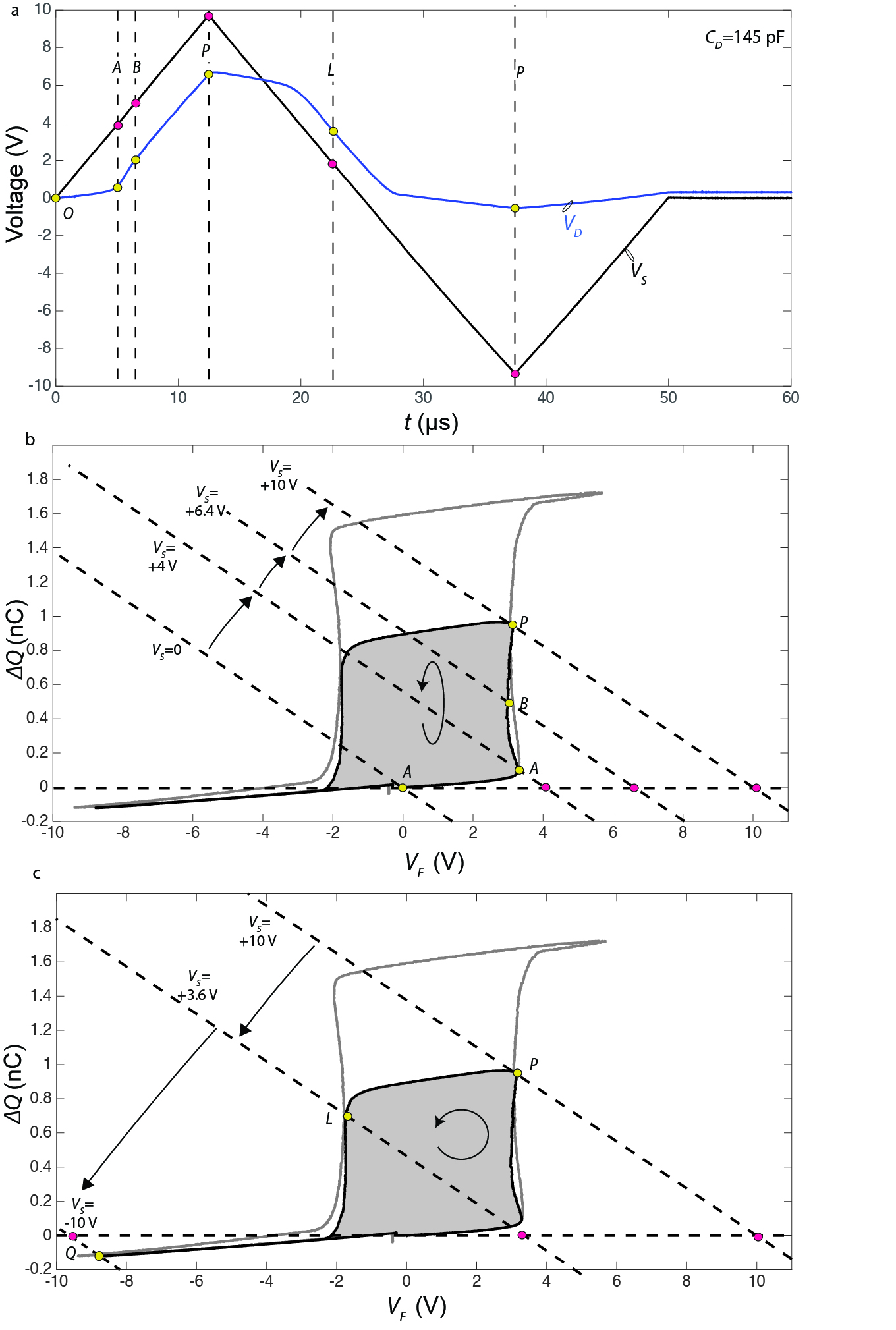}
 \end{center}
\caption{  Load lines for different $V_S$ during  0$\rightarrow$+10 V$\rightarrow$ (b) and +10 V$\rightarrow$-10 V sweeps (c) for $C_D$=145 pF.  }
 \label{ll145pF}
\end{figure*}

Fig. \ref{ll440pF}(a) shows  the waveforms corresponding to $V_D$ and $V_S$ of the ferroelectric-dielectric series circuit in response to a bipolar triangular pulse $V_S$: 0$\rightarrow$+10 V$\rightarrow$ -10 V$\rightarrow$0 V with a period $T$=50 $\mu$s for $C_D$={440 pF}.  Fig. \ref{ll440pF}(b) and \ref{ll440pF}(c) plot the extracted ferroelectric charge ($\Delta Q$)-voltage ($V_F$) characteristics extracted from the waveforms in Fig. \ref{ll440pF}(a) along with the load-lines: $\Delta Q=C_D\times (V_S-V_F)$ for different $V_S$. In Fig. \ref{ll440pF}(b) and \ref{ll440pF}(c), we show with the aid of load-lines  how the state of ferroelectric traces along its charge-voltage characteristics for $V_S$: 0$\rightarrow$+10 V$\rightarrow$ and +10 V$\rightarrow$-10 V, respectively. Similar analysis for $C_D$=240 pF and 145 pF are shown in Fig. \ref{ll240pF} and \ref{ll145pF}, respectively. 

\section{Effect of the magnitude of triangular pulse and period}
\begin{figure*}[!h]
\begin{center}
 \includegraphics[width=6.in]{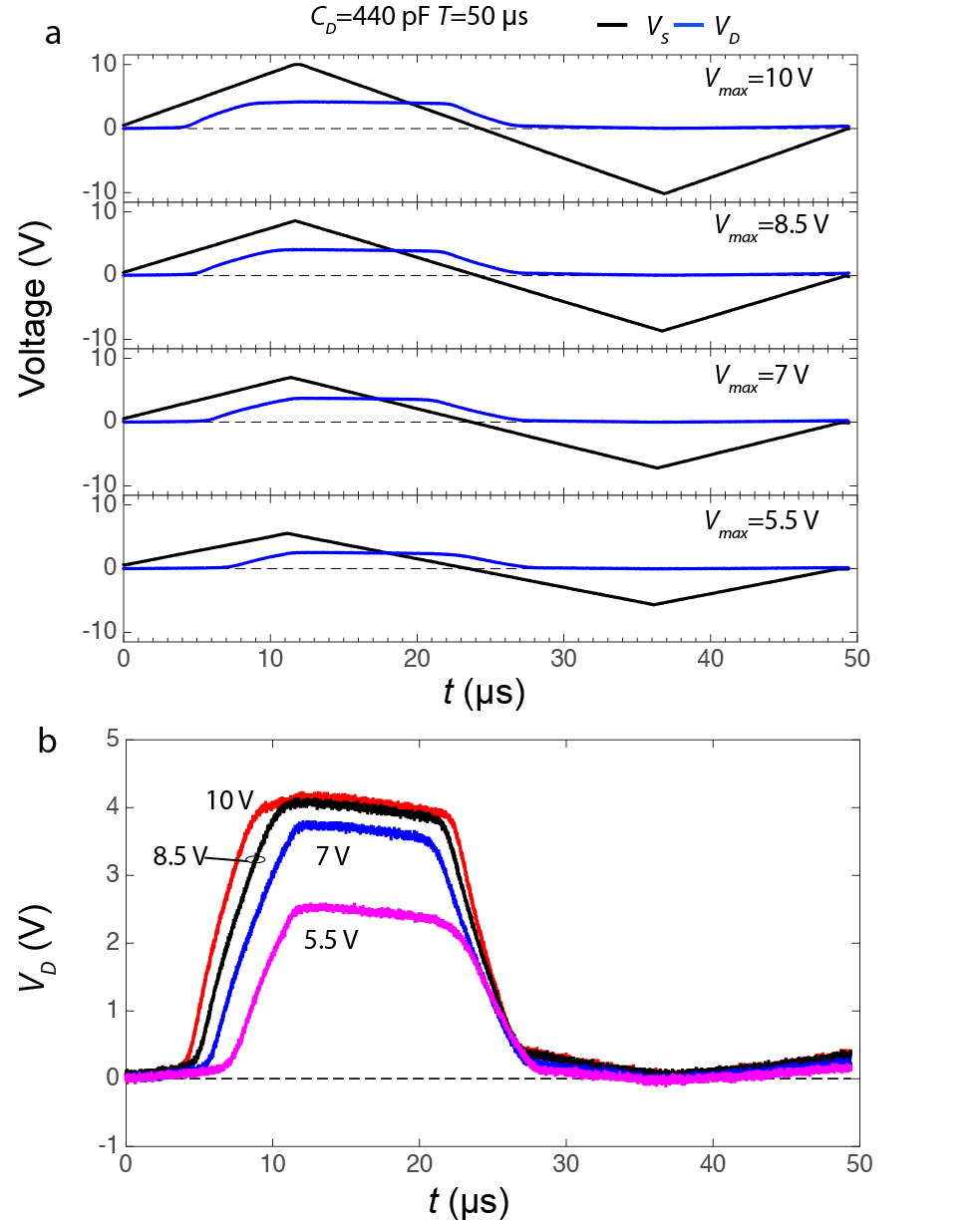}
 \end{center}
\caption{Effects of triangular pulse magnitude: $C_D$=440 pF, $T$=50 $\mu$s }
 \label{440pF_amp_50us}
\end{figure*}

\begin{figure*}[!h]
\begin{center}
 \includegraphics[width=6.in]{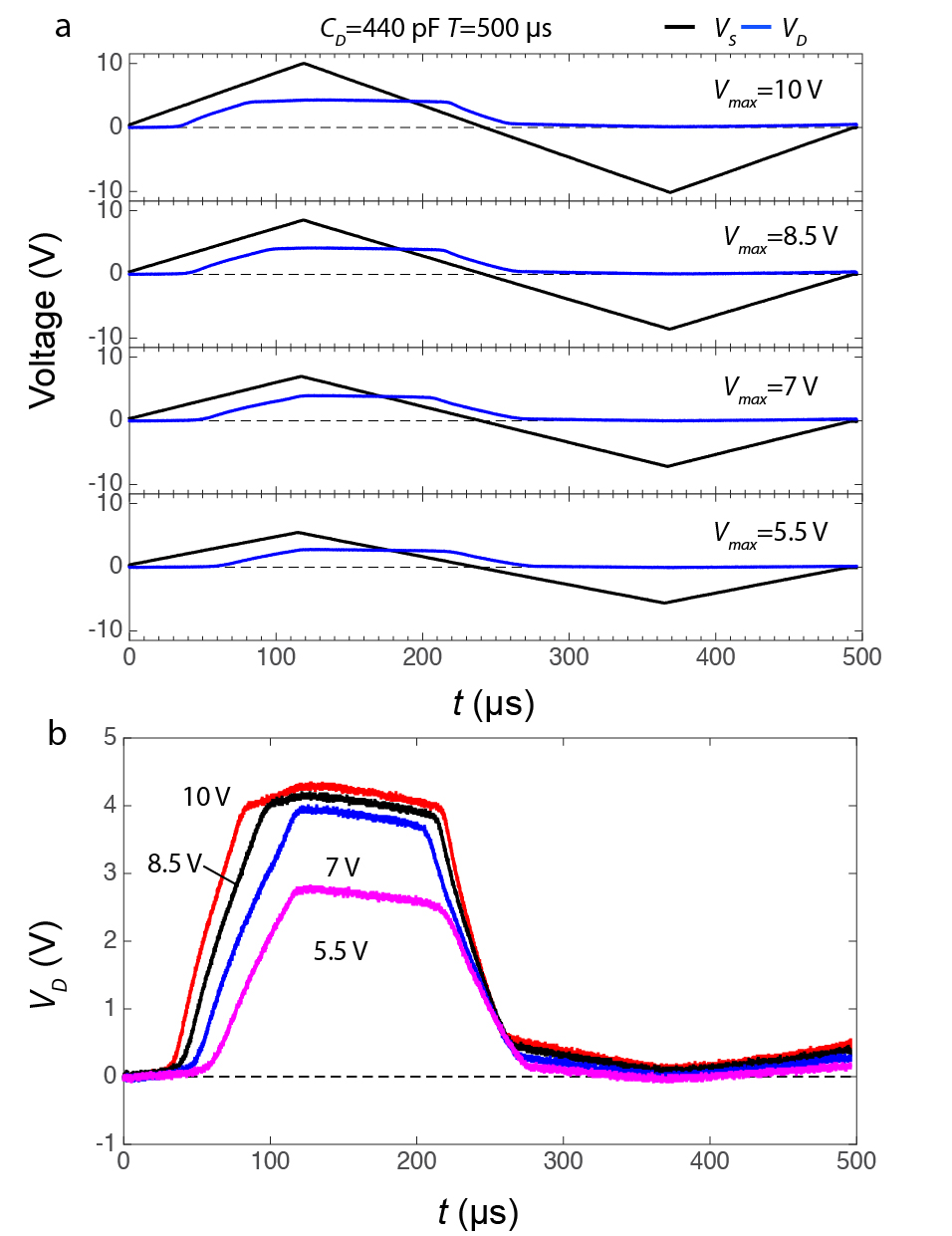}
 \end{center}
\caption{Effects of triangular pulse magnitude: $C_D$=440 pF, $T$=500 $\mu$s }
 \label{440pF_amp_500us}
\end{figure*}

\begin{figure*}[!h]
\begin{center}
 \includegraphics[width=5.4in]{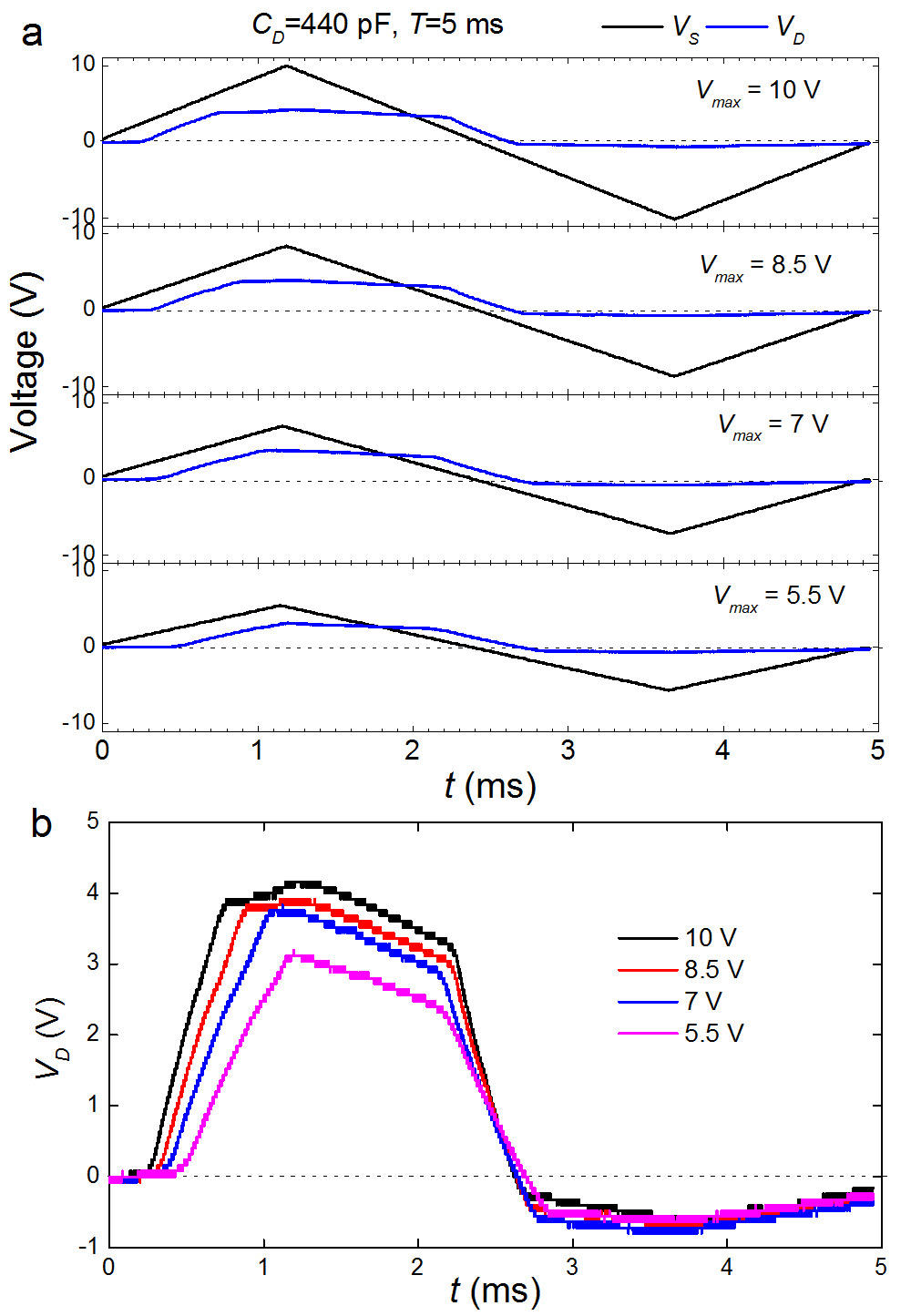}
 \end{center}
\caption{Effects of triangular pulse magnitude: $C_D$=440 pF, $T$=5 ms }
 \label{440pF_amp_5ms}
\end{figure*}

\begin{figure*}[!h]
\begin{center}
 \includegraphics[width=6.in]{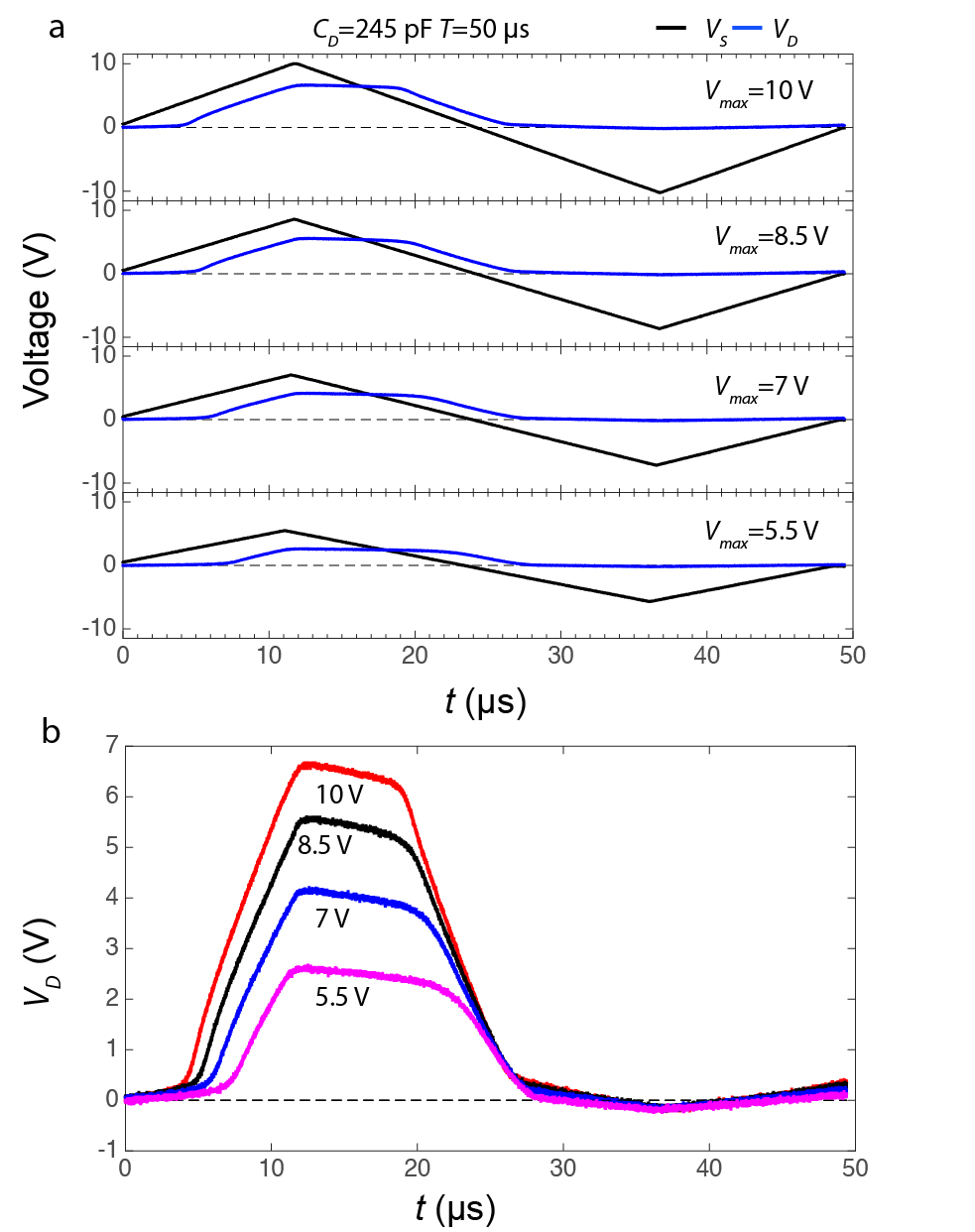}
 \end{center}
\caption{Effects of triangular pulse magnitude: $C_D$=245 pF, $T$=50 $\mu$s }
 \label{240pF_amp_50us}
\end{figure*}

\begin{figure*}[!h]
\begin{center}
 \includegraphics[width=6.in]{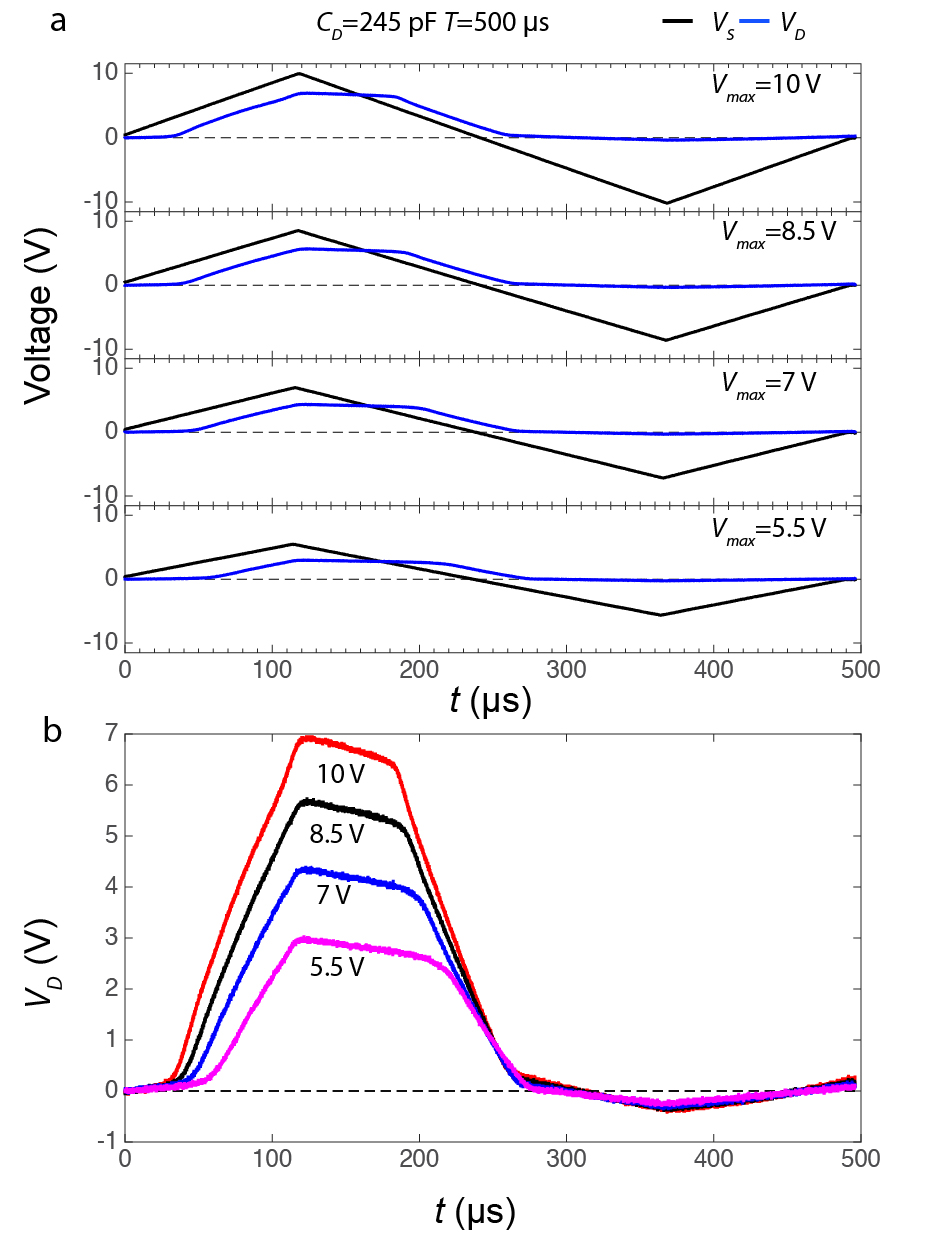}
 \end{center}
\caption{Effects of triangular pulse magnitude: $C_D$=245 pF, $T$=500 $\mu$s }
 \label{240pF_amp_500us}
\end{figure*}

\begin{figure*}[!h]
\begin{center}
 \includegraphics[width=5.4in]{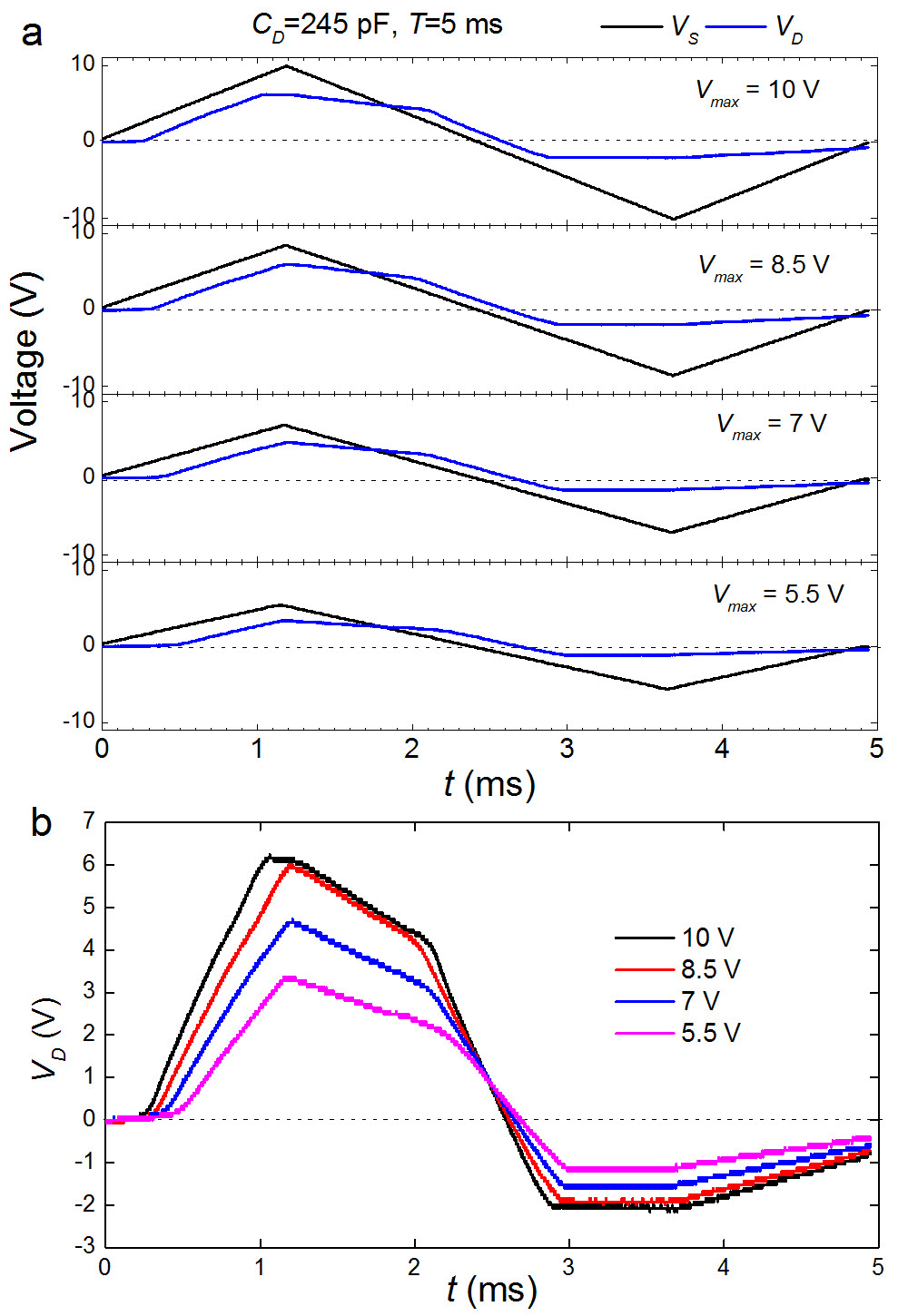}
 \end{center}
\caption{Effects of triangular pulse magnitude: $C_D$=245 pF, $T$=5 ms }
 \label{240pF_amp_5ms}
\end{figure*}

\begin{figure*}[!h]
\begin{center}
 \includegraphics[width=6.in]{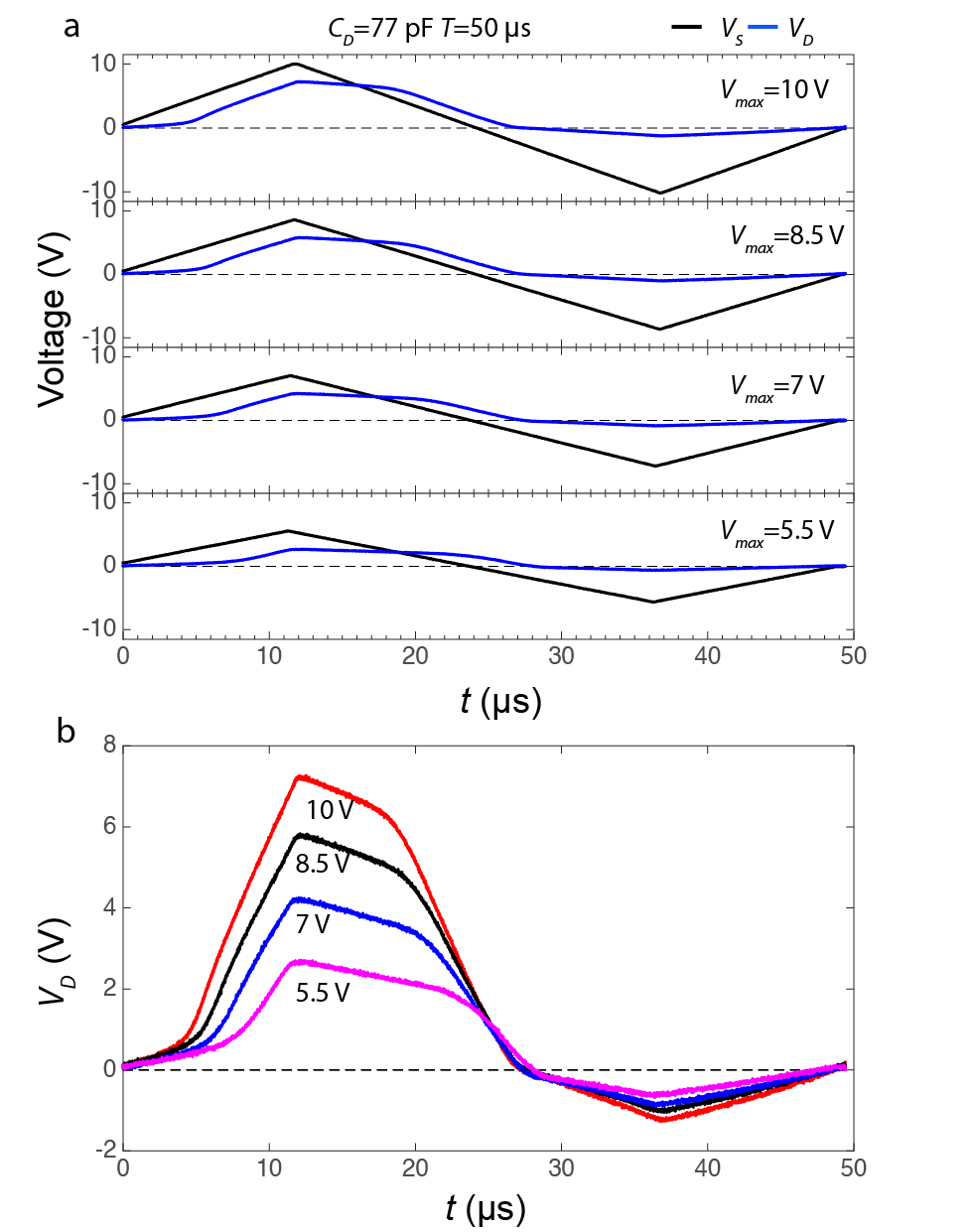}
 \end{center}
\caption{Effects of triangular pulse magnitude: $C_D$=77 pF, $T$=50 $\mu$s }
 \label{77pF_amp_50us}
\end{figure*}

\begin{figure*}[!h]
\begin{center}
 \includegraphics[width=6.in]{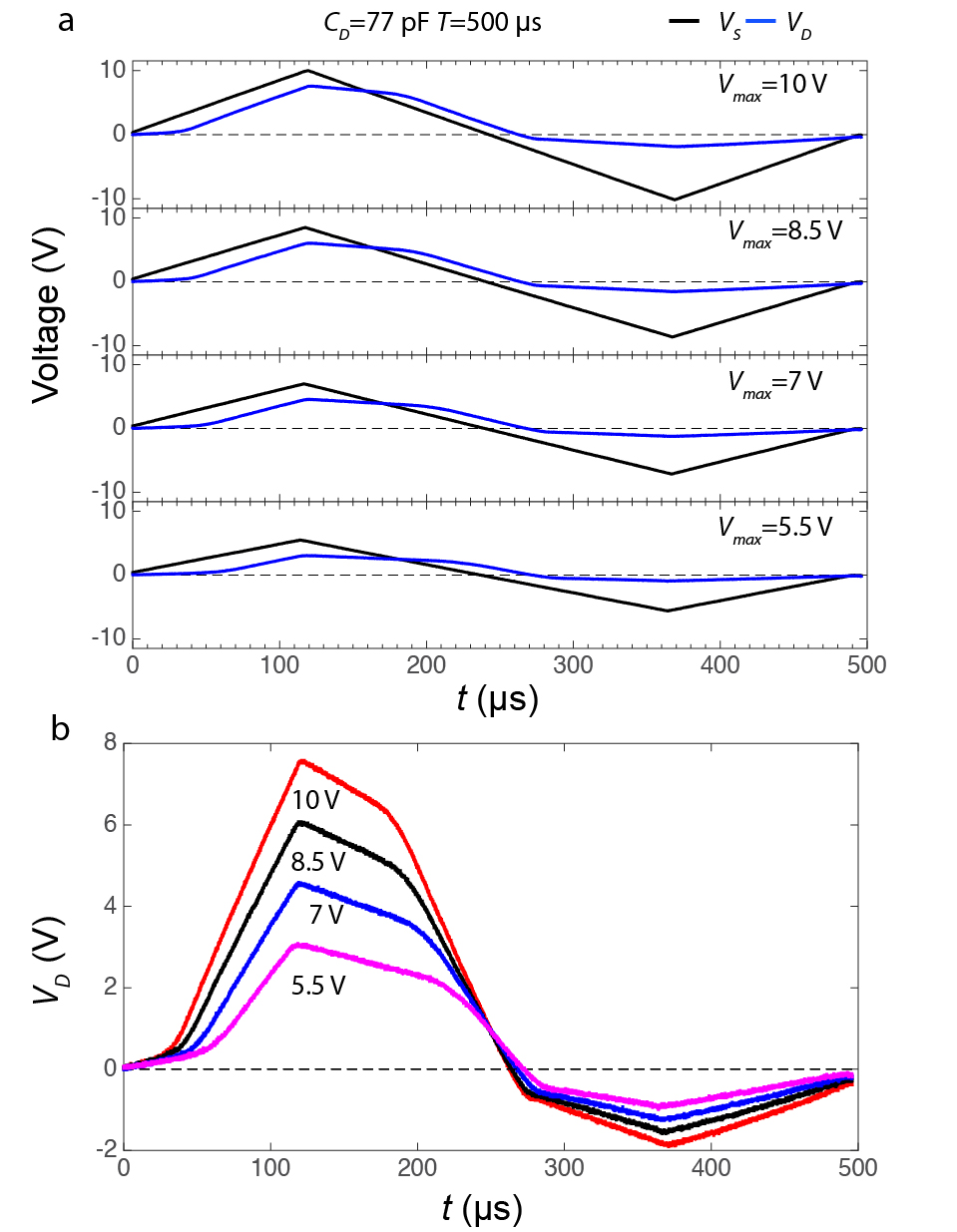}
 \end{center}
\caption{Effects of triangular pulse magnitude: $C_D$=77 pF, $T$=500 $\mu$s }
 \label{77pF_amp_500us}
\end{figure*}

\begin{figure*}[!h]
\begin{center}
 \includegraphics[width=5.4in]{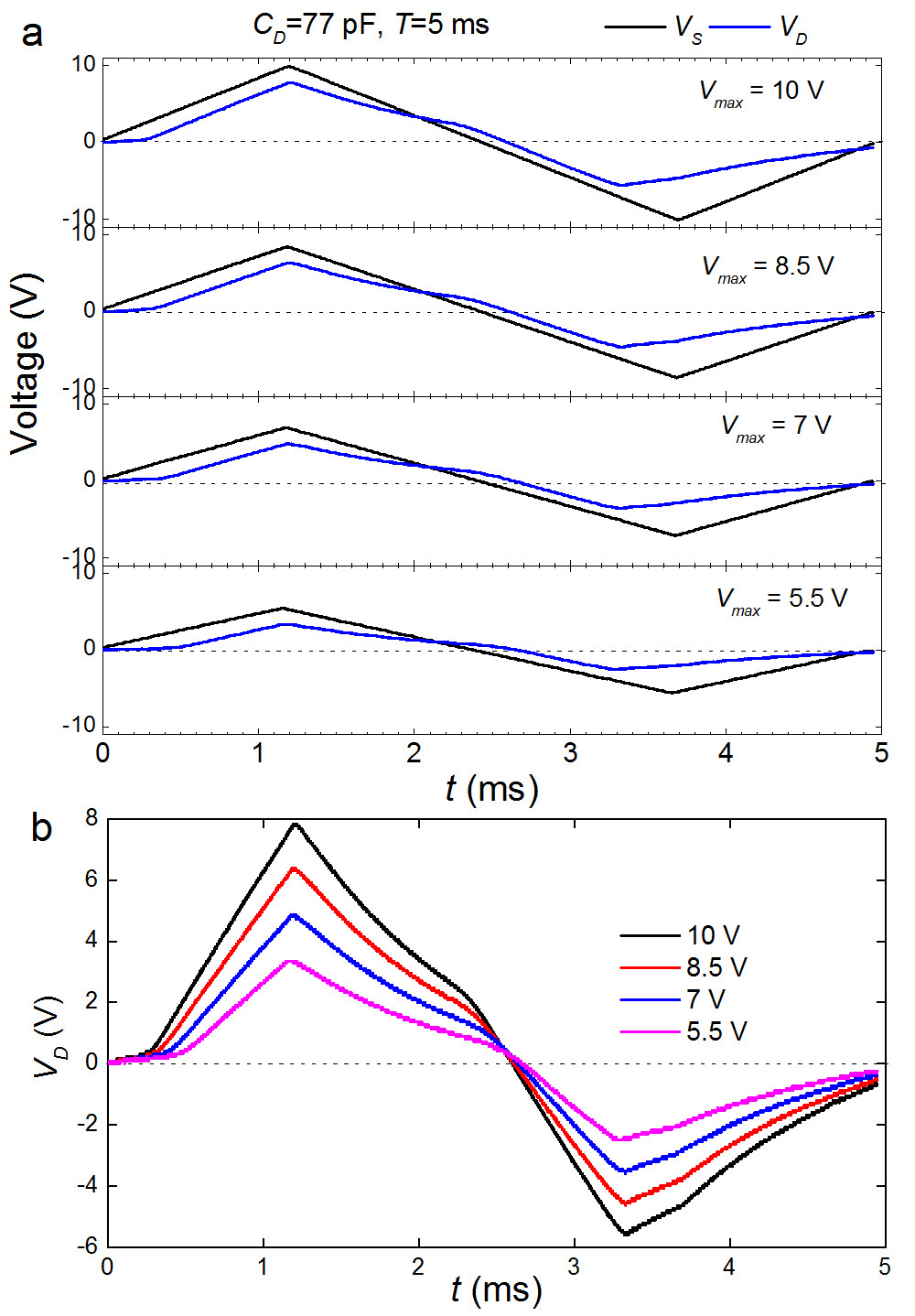}
 \end{center}
\caption{Effects of triangular pulse magnitude: $C_D$=77 pF, $T$=5 ms }
 \label{77pF_amp_5ms}
\end{figure*}

\begin{figure*}[!h]
\begin{center}
 \includegraphics[width=6.in]{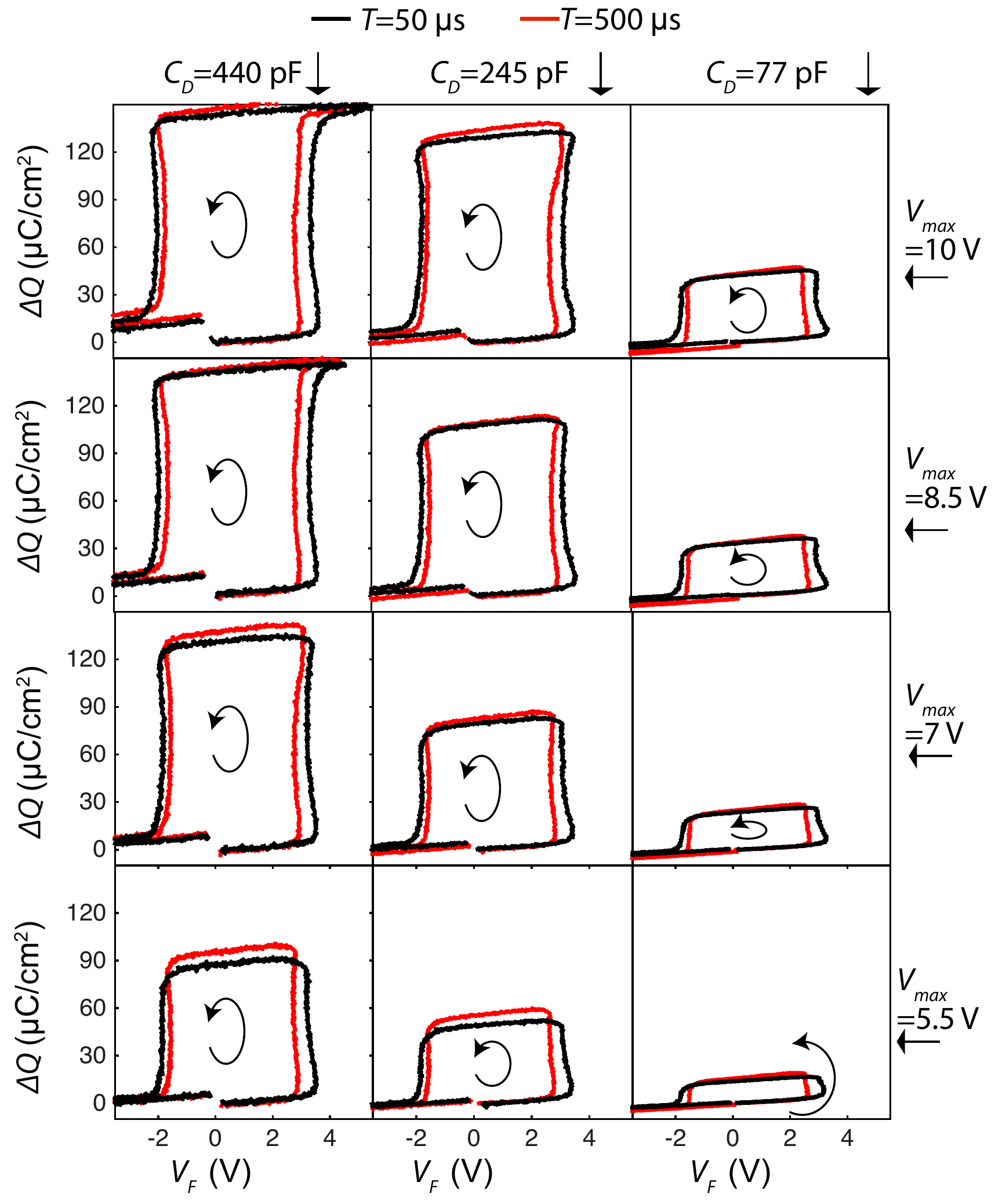}
 \end{center}
\caption{The evolution of the ferroelectric charge ($\Delta Q$)-voltage ($V_F$) with respect to $C_D$, $V_{max}$ and $T$. These are extracted from  waveforms shown in Fig. \ref{440pF_amp_50us},  \ref{440pF_amp_500us}, \ref{240pF_amp_50us}, \ref{240pF_amp_500us}, \ref{77pF_amp_50us} and \ref{77pF_amp_500us}. In fig. \ref{hys_mag_cap}, black and red curves correspond to $\Delta Q$-$V_F$ characteristics for $T$=50 $\mu$s and 500 $\mu$s, respectively. }
 \label{hys_mag_cap}
\end{figure*}

\begin{figure*}[!h]
\begin{center}
 \includegraphics[width=6.5in]{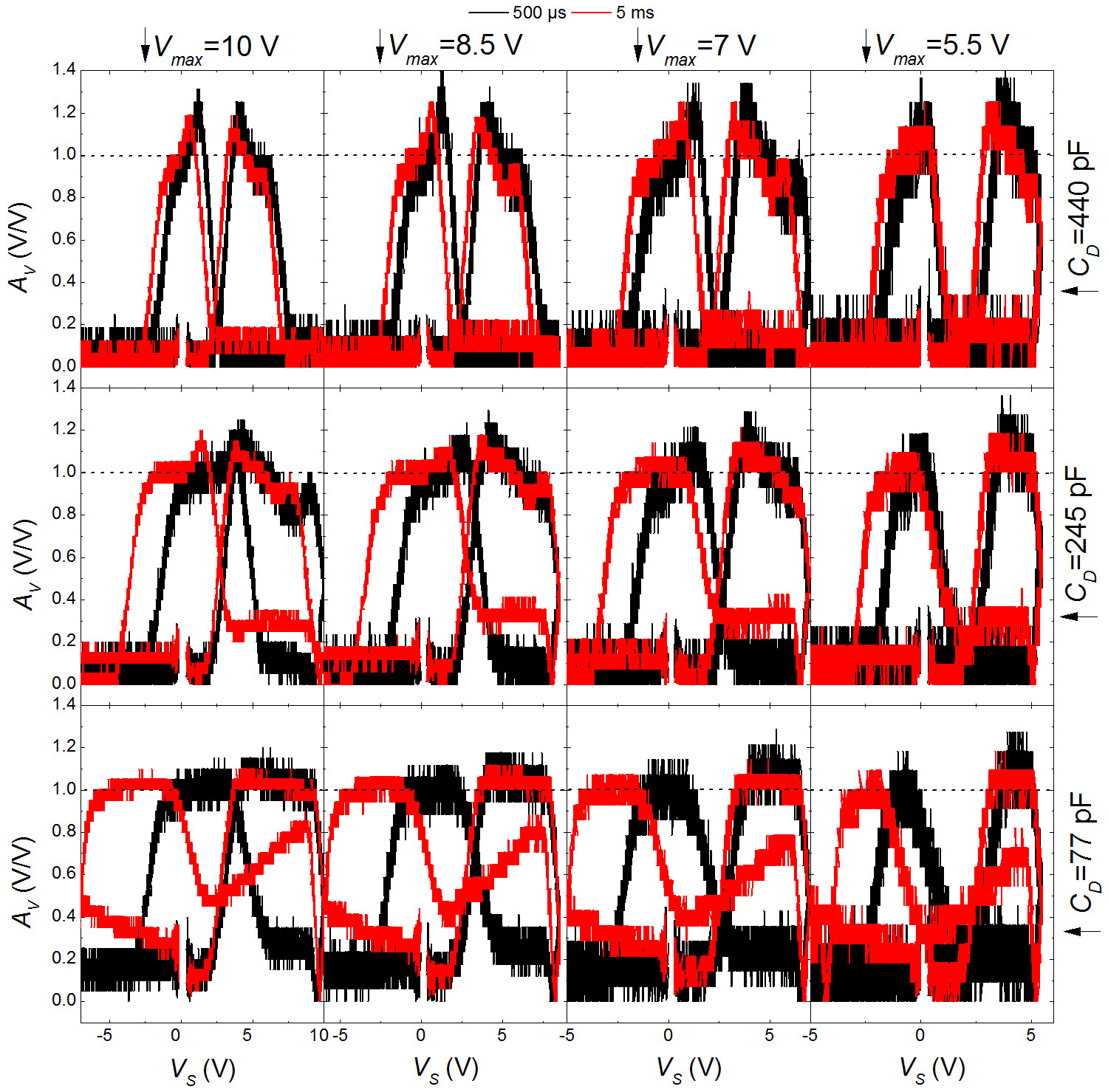}
 \end{center}
\caption{Comparison of the gain $A_V$ as a function of $V_S$ for different $C_D$, $V_{max}$ and $T$.}
 \label{gain}
\end{figure*}

Fig. \ref{440pF_amp_50us}, \ref{440pF_amp_500us} and \ref{440pF_amp_5ms} show the waveforms of the ferroelectric-dielectric series circuit for $C_D$=440 pF in response to bipolar triangular pulses of  ($V_{max}$) 10 V, 8.5 V, 7 V and 5.5 V with period $T$=50 $\mu$s, 500 $\mu$s and 5 ms, respectively. Fig. \ref{240pF_amp_50us}, \ref{240pF_amp_500us} and \ref{240pF_amp_5ms}  show the waveforms of the ferroelectric-dielectric series circuit for $C_D$=240 pF in response to bipolar triangular pulses of magnitude 10 V, 8.5 V, 7 V and 5.5 V with period $T$=50 $\mu$s, 500 $\mu$s and 5 ms, respectively. Fig.  \ref{77pF_amp_50us}, \ref{77pF_amp_500us} and \ref{77pF_amp_5ms} show the waveforms of the ferroelectric-dielectric series circuit for $C_D$=77 pF in response to bipolar triangular pulses of magnitude 10 V, 8.5 V, 7 V and 5.5 V with period $T$=50 $\mu$s, 500 $\mu$s and 5 ms, respectively. Fig. \ref{hys_mag_cap} shows the evolution of the ferroelectric charge ($\Delta Q$)-voltage ($V_F$) with respect to $C_D$, $V_{max}$ and $T$. These are extracted from  waveforms shown in Fig. \ref{440pF_amp_50us},  \ref{440pF_amp_500us}, \ref{240pF_amp_50us}, \ref{240pF_amp_500us}, \ref{77pF_amp_50us} and \ref{77pF_amp_500us}. In fig. \ref{hys_mag_cap}, black and red curves correspond to $\Delta Q$-$V_F$ characteristics for $T$=50 $\mu$s and 500 $\mu$s, respectively. $\Delta Q$-$V_F$ curves are not shown for $T = 5$ ms, because of increased leakage currents for these measurements. Fig. \ref{gain} shows the calculated gain $A_V$ as a function of $V_S$ for  $T$=50 $\mu$s, $T$=500 $\mu$s and 5 ms for different $C_D$ and $V_{max}$.

In figure S23, we observe that the $A_V-V_S$ characteristics for all three different periods (50 $\mu$s, 500 $\mu$s and 5 ms) exhibit the same functional variation with respect to $C_D$ (i.e. the amplification $A_V$ in the ramp down segment decreases with decreasing $C_D$). This confirms that the same physics of voltage amplification at play in all three different time periods. It is also important to note in figure S23 that, for a given $C_D$ and $V_{max}$, the maximum $A_V$ is always smaller for higher $T$. In fact, leakage plays an important role in determining the evolution of shape of the $A_V-V_S$ characteristics with respect of the period $T$. This phenomenon is indeed discussed in Ref. 22 (in the appendix A of the reference). To understand the effect of the leakage, we refer to the equivalent circuit of the ferroelectric-dielectric series network in figure 4a. In the limit when the period is much larger than the effective time constant of the circuit (i.e. $T>>\tau_{eff}$), the network behaves like a resistor divider, and the internal node voltage is determined by the resistors only (i.e. $V_D=R/(R+R_F)V_S$). In such a case, the amplification due to negative capacitance ceases to exist. In other words, if the time period of the measurement is fast enough, the leakage effects tend to diminish the voltage amplification due to negative capacitance, which we observe experimentally.

\section{Standard electrical characterization}
\begin{figure*}[!h]
\begin{center}
 \includegraphics[width=3.5in]{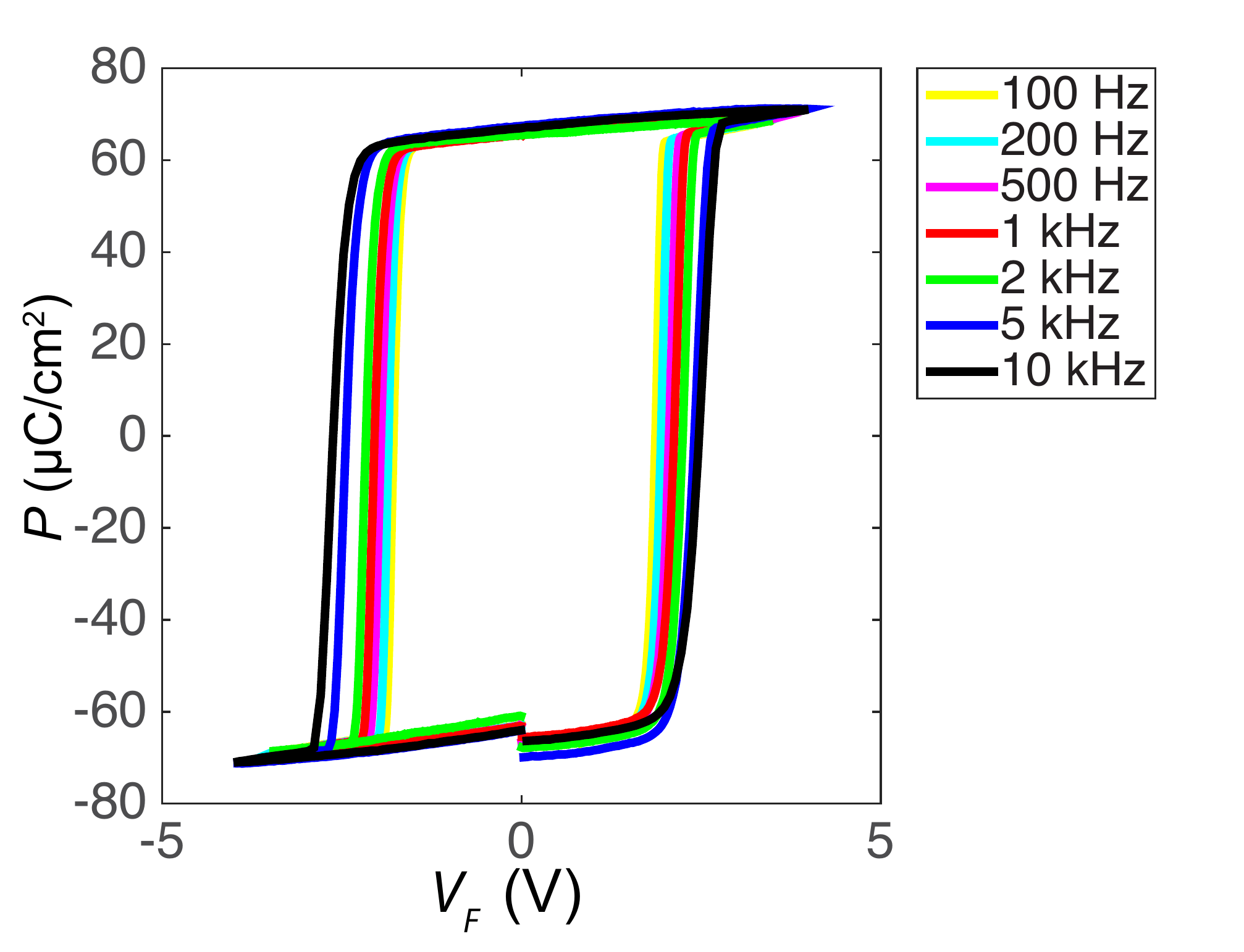}
 \end{center}
\caption{ The polarization ($P$)-voltage ($V_F$) hysteresis curve of the ferroelectric at different loop periods ($T$) measured using a Sawyer-Tower  set-up. }
 \label{PV}
\end{figure*}

\paragraph{Hysteresis loop measured using the Sawyer-Tower technique}
 Ferroelectric hysteresis loops of the ferroelectric capacitors was also measured using a Radiant Precision  Multiferroic Tester. The measurement is based on the virtual ground technique.  This technique is based on a current-to-voltage amplifier where the output is connected to the inverting input of the operational amplifier via the feedback resistance. Since the non-inverting input is connected to the ground in the setup, the inverting input is virtually on ground level that helps minimize the cable capacitance and enables the highest precision for ferroelectric measurements. Fig. \ref{PV} shows the ferroelectric hysteresis loop of the 60 nm PZT film grown on SRO buffered STO (001) substrate measured by this technique.   The remnant polarization  is measured to be $\sim$0.7 C/m$^2$.

\begin{figure*}[!t]
\begin{center}
 \includegraphics[width=6.5in]{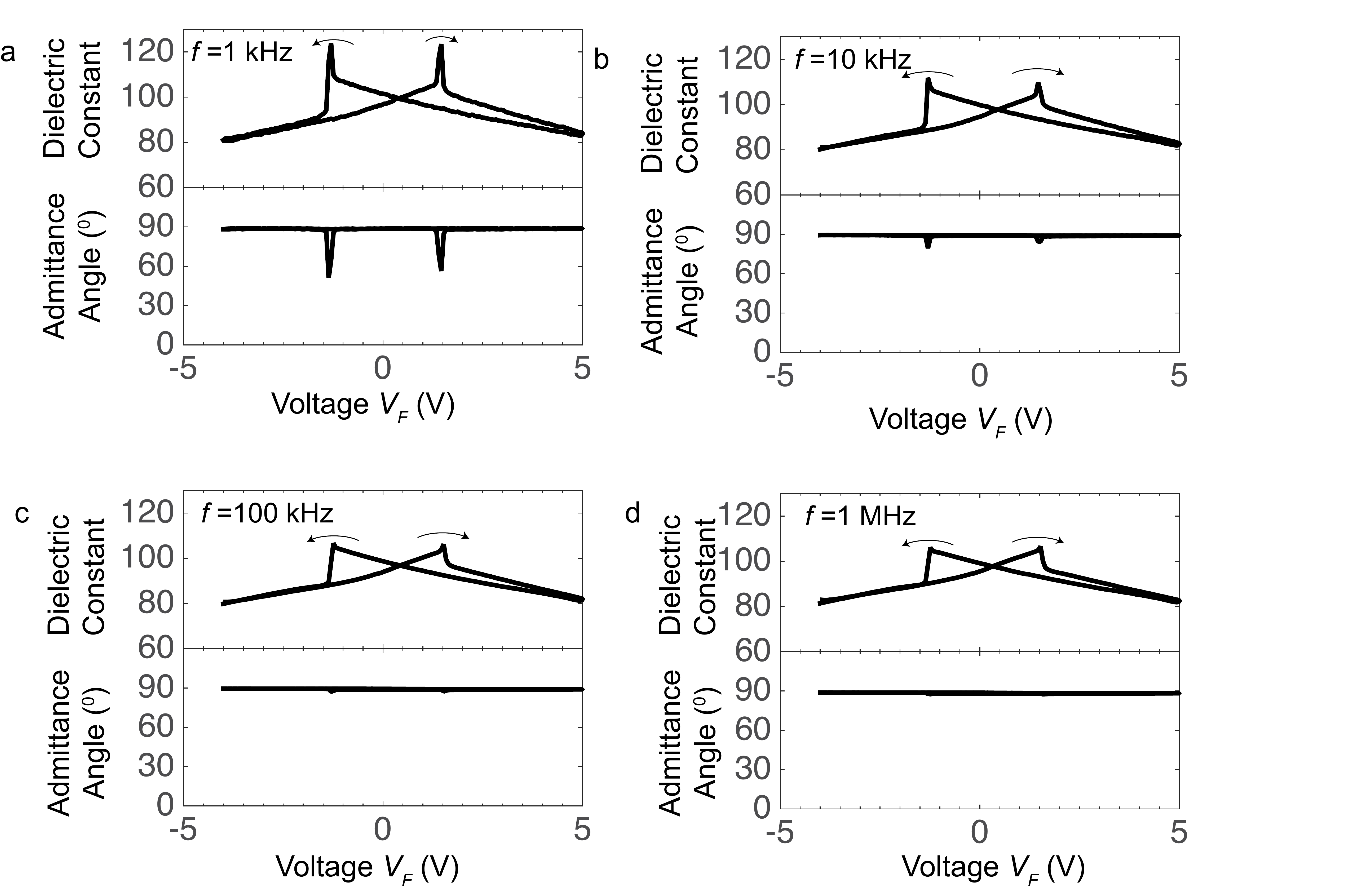}
 \end{center}
\caption{ Dielectric constant-voltage and the admittance angle-voltage characteristics
of the PZT sample at 1 kHz (a), 10 kHz (b), 100 kHz (d) and 1 MHz (d). }
 \label{CV}
\end{figure*}

\paragraph{Dielectric characterization }
Fig. \ref{CV}(a), \ref{CV}(b), \ref{CV}(c) and \ref{CV}(d) show the dielectric constant-voltage characteristics and the admittance angle-voltage of the PZT sample at 1 kHz, 10 kHz, 100 kHz and 1 MHz,  respectively.  We observe that the admittance angle is $\sim$90$^\circ$ at all frequencies. This points to negligible DC leakage through our samples. Furthermore, we   note that  the dispersion of the dielectric constant over 3 orders of magnitude variation in frequency is less than 5\%. Generally, the defect dynamics that include trapping of electrons and charging/discharging of the same in the defect states results in large frequency dispersion of the dielectric constant in ferroelectrics\cite{Taylor}. This points to the fact that there is a very small amount of defects in the PZT sample, if any. 

\section{Growth and  structural characterization}
\begin{figure*}[!h]
\begin{center}
 \includegraphics[width=6.5in]{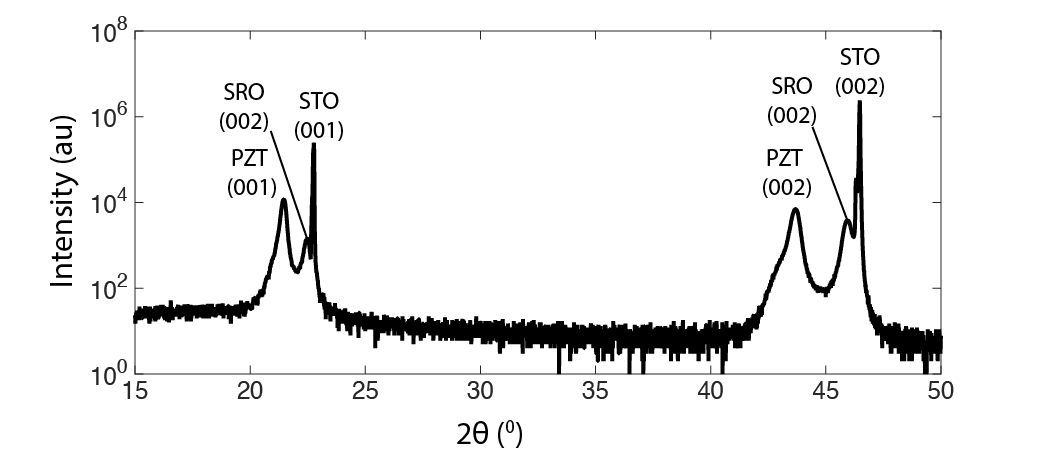}
 \end{center}
\caption{The X-ray diffraction spectrum of the PZT(100 nm)/SRO(20nm) heterostructure on the STO substrate corresponding (001) and (002) reflections. }
 \label{XRD}
\end{figure*}

A 100 nm Pb(Zr$_{0.2}$Ti$_{0.8}$)O$_3$ thin film was grown on 20 nm metallic SrRuO$_3$ (SRO) buffered SrTiO$_3$ (STO) substrate using the pulsed laser deposition technique. The epitaxial growth of the heterostructure results in a high crystalline quality and robust ferroelectricity in PZT and minimal defects. SRO and PZT films were grown at 630 $^\circ$C and 720 $^\circ$C respectively. During the growth, the oxygen partial pressure was kept at 100 mTorr and afterwards the heterostructure was slowly cooled down at 1 atm of oxygen partial pressure and a rate of -5 $^0$C/min to the room temperature.  Laser pulses of  100 mJ of energy and   $\sim$4 mm$^2$ of spot size were used to ablate the targets.  Ti/Au top electrodes were ex-situ deposited by e-beam evaporation and then patterned using standard lithographic techniques into square dots of different dimensions.  The SRO layer was used as the bottom electrode. Fig. \ref{XRD} shows the X-ray diffraction spectrum of the heterostructure around the (002) reflections.  The PZT sample is $c$-axis oriented with no contribution from any other secondary phases. 

\begin{figure*}[!h]
\begin{center}
 \includegraphics[width=1.5in]{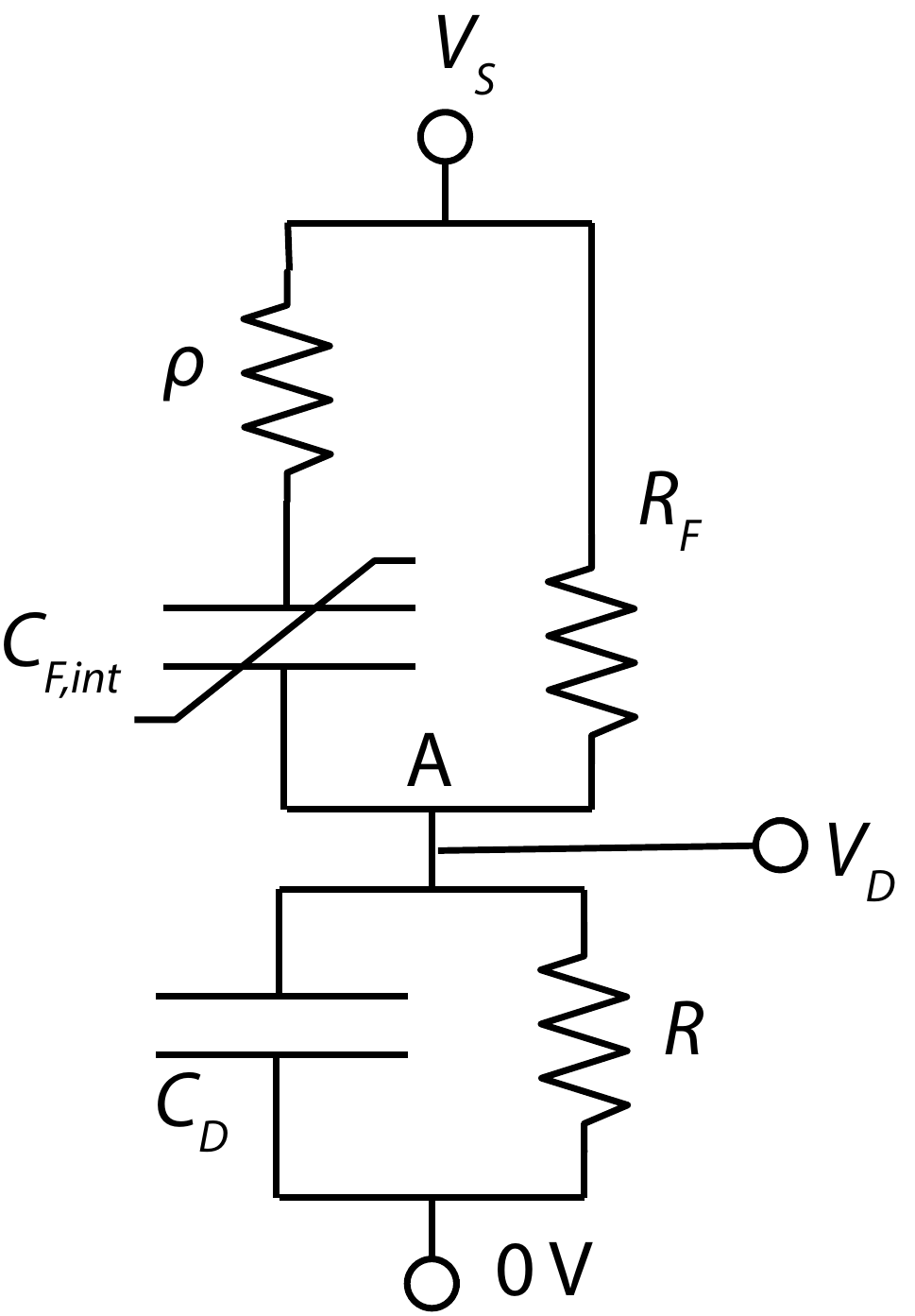}
 \end{center}
\caption{Circuit diagram for the simulation of the circuit. }
 \label{ckt}
\end{figure*}

\section{Landau-Khalatnikov model based simulation: Homogeneous switching scenario }
Following the Landau-Khalatnikov equation \cite{landau1954anomalous}, the rate of change in polarization charge $Q_F=PA$ in the ferroelectric capacitor, where $P$ is the polarization (in $C/m^2$) and $A$ is the capacitor area, can be written as:
\begin{equation}
\rho \frac{dQ_F}{dt}=-\frac{dU}{dQ_F}\label{LK}
\end{equation}
\noindent Here, $\rho>0$ signifies the frictional inertia of the system and $U$  (in eV) is the free energy of the ferroelectric material that can be written as:
\begin{equation}
U=\frac{\alpha}{2}Q_F^2+\frac{\beta}{4}Q_F^4-Q_F(V_F-\Delta V)\label{landau},
\end{equation}
\noindent where $\alpha$ and $\beta$ are anisotropy constants ($\alpha$ $<$ 0) and $V_F$ is the voltage across the ferroelectric capacitor. $\Delta V$ represents an internal bias in the ferroelectric capacitor, which can occur due work-function difference between the electrodes, defect dipoles, polarization gradients etc.  For simplicity, we have only included up to the fourth order of $Q_F$, which corresponds to a second order phase transition.  The analysis is valid for higher order ($>$4) expansions of the free energy as well.  
Combining Eq. \ref{LK} and \ref{landau} one finds
\begin{equation}
V_F=\rho \frac{dQ_F}{dt}+(\alpha Q_F+\beta Q_F^3)+\Delta V \label{eq3}
\end{equation}

Eq. \ref{eq3} can further be modified into the following form. 

\begin{equation}
V_F=\rho \frac{dQ_F}{dt}+\frac{Q_F}{C_F}+\Delta V \label{eq4}
\end{equation}

\noindent where, $C_F=(\alpha+\beta Q_F^2)^{-1}$ represents the internal capacitance of the ferroelectric capacitor. $C_F$ is highly non-linear and is negative for certain values of $Q_F$.  From Eq.  \ref{eq4}, we note that the equivalent circuit for a ferroelectric capacitor consists of an internal resistor $\rho$ and a nonlinear capacitor $C_F(Q_F)$ connected in series \cite{khan2015negative}.

Fig. \ref{ckt} shows the equivalent circuit diagram of the ferroelectric-dielectric series network. $R_F$ represents the leakage resistance of the ferroelectric.  $C_D$ and $R$ represent the  capacitance and the leakage resistance of the dielectric capacitance. Applying Kirchhoff's current law at the node $A$ in Fig. \ref{ckt}, we obtain the following equation.

\begin{eqnarray}
\frac{dQ_F}{dt}+\frac{V_F}{R_F}&=&C_D\frac{dV_D}{dt}+\frac{V_D}{R}\label{eq5}
\end{eqnarray}

The simulation results shown in main text Fig. 4b,c are obtained by solving Eq. \ref{eq4} and \ref{eq5} self-consistently as a function of time ($t$). In this calculation, the area of the ferroelectric $A$=35 $\mu$m $\times$35 $\mu$m, the ferroelectric thickness $t_{FE}$=100 nm, the internal bias $\Delta V$=1.5 V and $C_D$=440 pF. The anisotropy constants for the ferroelectric material are $\alpha_{1,FE}$=-5$\times$10$^7$ m/F and $\alpha_{11,FE}$=4.5$\times$10$^7$ m$^5$/F-C$^2$. The corresponding anisotropy constants for the ferroelectric capacitor are calculated by $\alpha=\alpha_{1,FE}\times t_{FE}/A$ and $\beta=\alpha_{11,FE}\times t_{FE}/A^3$. $R$ and $R_F$ are both equal to 1 V$/(J\times A)$, where $J$=10 $^{-2}$ A/cm$^2$.

\section{Time-Dependent Ginzburg-Landau based simulation: Inhomogeneous switching scenario }

To understand the effects of inhomogeneous switching on the amplification in the FE-DE series system, we  consider a ferroelectric  polarization  field $\vec{P}$ which can vary in the in-plane directions ($x$- and $y$-) and is homogeneous along the out-of-plane direction ($z$-).  Our PZT film is $c$-axis oriented with no contribution from the $a$-axis orientation. Due to the thickness and the compressive strain in this film,  the switching  occurs through 180$^\circ$ domain wall motion with minimal contribution from ferroelastic 90$^\circ$ domain wall motion. As such, we constrain the direction of the polarization along the $z$-direction ($i.e.$, $\vec{P}=P \hat{z}$). The evolution of $P(x,y)$ is governed by  the time-dependent Ginzburg-Landau (TDGL) formalism \cite{chen2008phase} which is as follows. 
 \vspace{-1mm}
\begin{equation}
-L\frac{\partial F}{\partial P(x,y)} = \frac{\partial P(x,y)}{\partial t}.
\label{eq6}
\end{equation}

\noindent where $L$ is a parameter related to the domain wall mobility and $F$ is the free energy of the ferroelectric thin film, which is given by the following equation. 
 \vspace{-1mm}
\begin{equation}
F=t_{FE}\int\int (F_{Landau-Devonshire}+F_{electrostatic}+F_{grad})\textrm{d}x\textrm{d}y
\label{eq7}
\end{equation}

\noindent Here, $F_{Landau-Devonshire}$, $F_{electrostatic}$, $F_{grad}$ is the Landau-Devonshire free energy per unit volume, the electrostatic energy per unit volume and the free energy per unit volume due to the polarization gradient at and around the domain walls, respectively.  These terms are defined as follows. 
 \vspace{-1mm}
\begin{eqnarray}
F_{Landau-Devonshire}&=& \alpha_1 P^2 + \alpha_{11} P^4 + \alpha_{111} P ^6\\
F_{electrostatics}&=&-EP\\
F_{grad}&=&k(\nabla_{xy}P)^2
\label{eq8}
\end{eqnarray}

\noindent Here, $\alpha_1$, $\alpha_{11}$ and $\alpha_{111}$ are the anisotropy coefficients, $E$ is the electric field across the ferroelectric thin film and $k$ is a coefficient related to the domain wall energy. To account for inhomogeneous switching,  we considered spatial distributions of the parameters $\alpha_1$, $\alpha_{11}$ and $\alpha_{111}$.

\begin{figure*}[!h]
\begin{center}
 \includegraphics[width=4in]{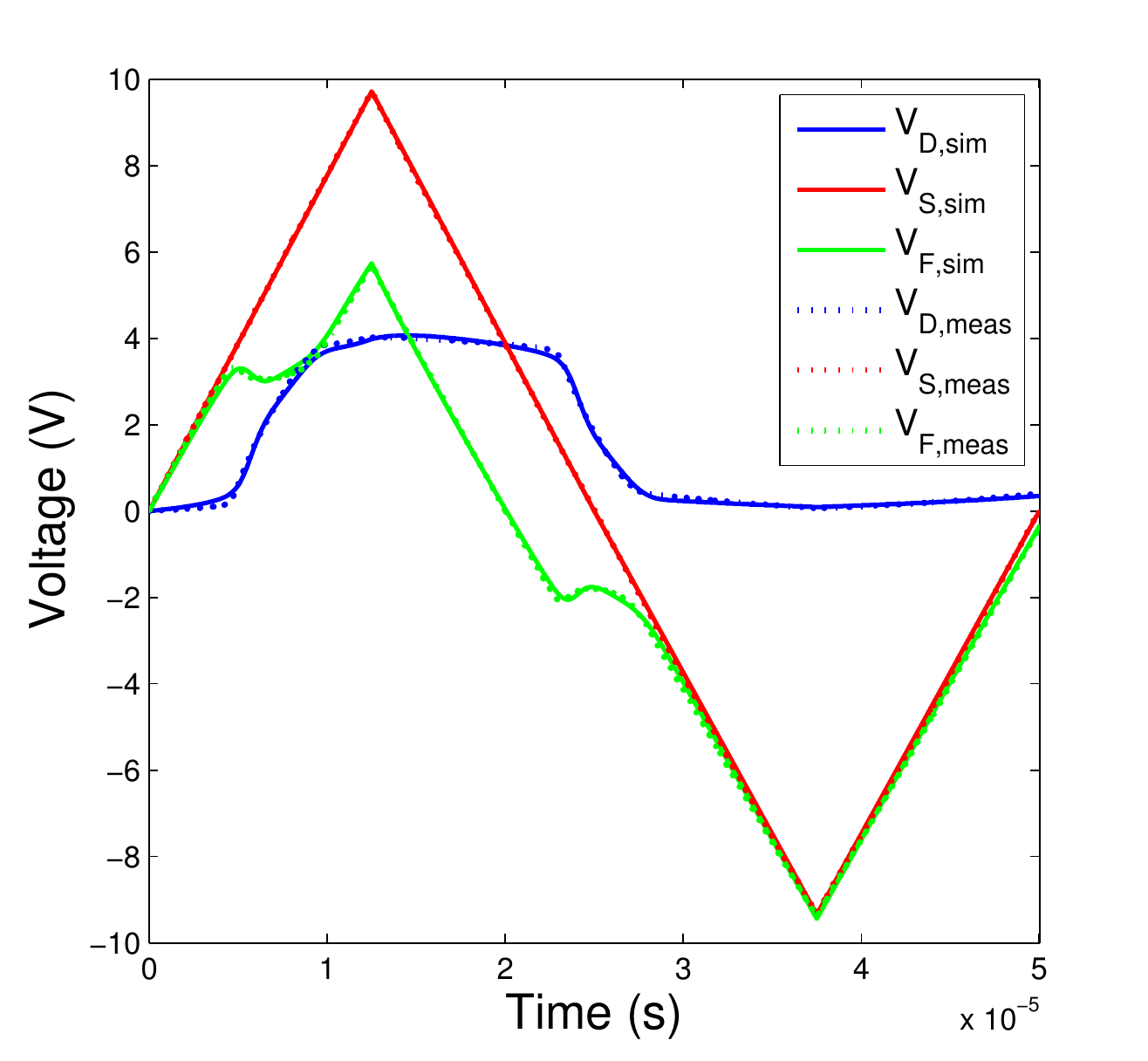}
 \end{center}
 \vspace{-5mm}
\caption{Comparison of experimental waveforms and simulated TDGL waveforms for $C_D = 440$ pF.  }
 \label{multi1}
\end{figure*} 

\begin{figure*}[!h]
\begin{center}
 \includegraphics[width=6.7in]{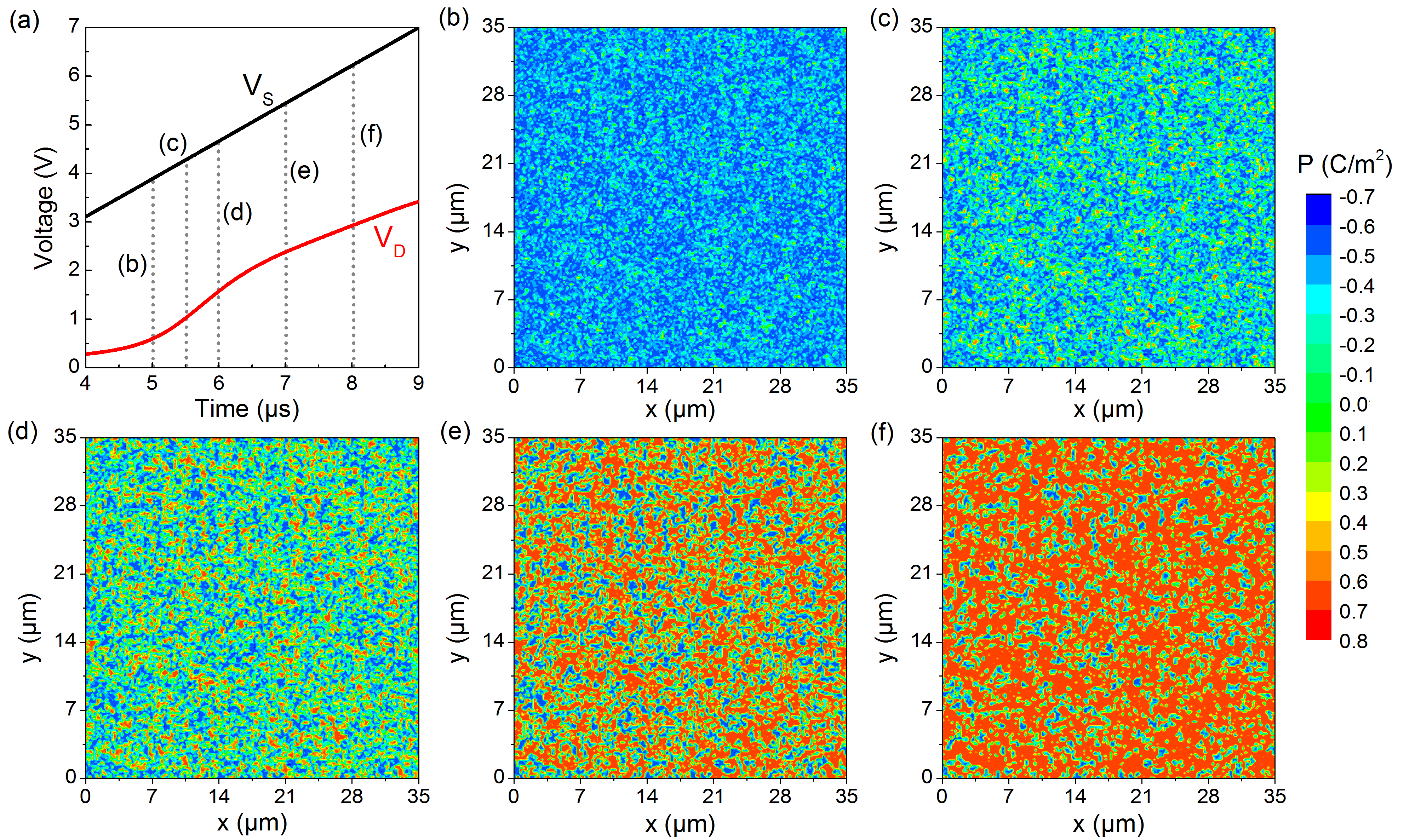}
 \end{center}
\caption{(a) Simulated voltage transients and (b)-(f) polarization distributions during different stages of the waveform. }
 \label{multi2}
\end{figure*} 

The total charge on the ferroelectric capacitor $Q_F$ is calculated as follows. 

\begin{equation}
Q_F = \epsilon_0 E A +\int P \textrm{d}x \textrm{d}y
\label{eq11}
\end{equation}

\noindent Here, $A$ is the total area of the ferroelectric capacitor and $\epsilon_0$ is the vacuum permittivity. The voltage across the ferroelectric is given by $V_F = t_{FE} E$ where $t_{FE}$ is the thickness of the ferroelectric. The voltage defined by the pulse generator $V_S$ can also be written as 

\begin{equation}
V_S =  V_F+V_D=V_F+Q_D/C_D ,
\label{eq12}
\end{equation}

where $Q_D$ is the charge of the dielectric capacitor $C_D$. When $i_R$ and $i_{RF}$ are the leakage currents flowing through the ferroelectric and dielectric capacitor, respectively, the charges of both capacitors are related by

\begin{equation}
\frac{\partial Q_F}{\partial t} + i_{RF} =  \frac{\partial Q_D}{\partial t} + i_{R},
\label{eq13}
\end{equation}

\noindent which is the same expression as in Eq. \ref{eq5}. Solving Eqs. \ref{eq6}-\ref{eq13} at each time step for the waveform $V_S$, we can simulate the transient evolution of the ferroelectric polarization during switching and the differential voltage amplification it causes. Fig. \ref{multi1} shows the comparison of the experimental data with the TDGL simulation for the case of $C_D = 440$ pF. $V_F$ is also shown to make the region of differential amplification easier to observe. An excellent agreement between data and theory is observed for $\alpha_1 = -4.39\times 10^5$ F/m, $\alpha_{11} = -1.61\times 10^8$ m$^5$/(FC$^2$), $\alpha_{111} = 2.97\times 10^8$ m$^9$/(FC$^4$), $k = 3.28 \times 10^{-8}$ m$^3$/F and $L = 3.55\times 10^{14}$ $\Omega^{-1}$m$^{-4}$. A standard deviation of 7.5 $\%$ with respect to the mean values was assumed for $\alpha_1$, $\alpha_{11}$ and $\alpha_{111}$. To confirm the inhomogeneous switching of polarization, in Fig. \ref{multi2} we plot the simulated polarization distribution corresponding to different stages in these voltage transients. One of the interesting observations based on these simulations is that  the negative capacitance effects begin at the onset of domain nucleation correspond to the knee of the hysteresis curves shown in the main-text Fig. 1c inset, Fig. 3c,d.

\begin{figure*}[!h]
\begin{center}
 \includegraphics[width=6.5in]{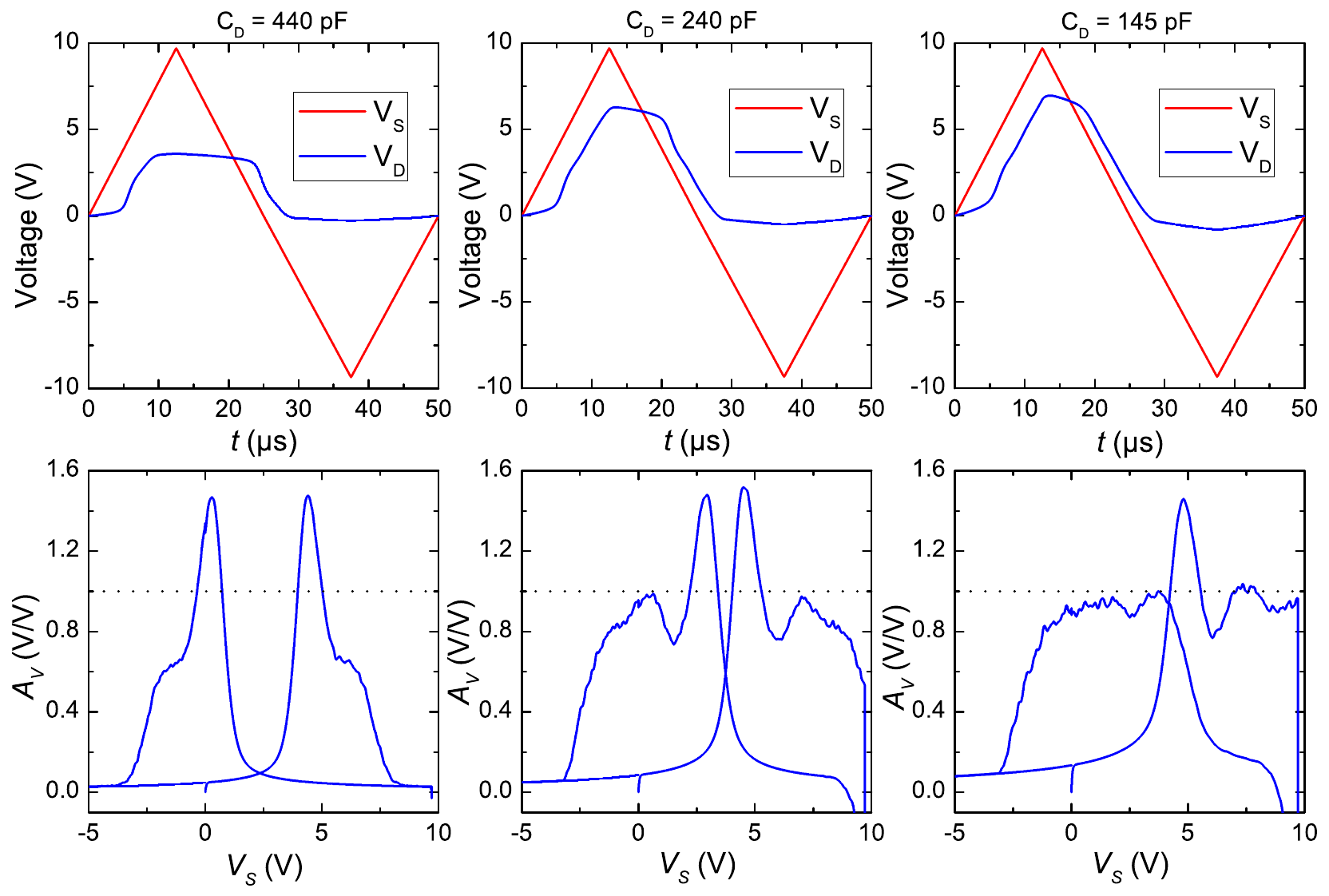}
 \end{center}
  \vspace{-5mm}
\caption{ TDGL simulation results of $V_D$ and $A_V$ for different $C_D$. }
 \label{multi4}
\end{figure*}

Fig. \ref{multi4} shows TDGL simulation results for different $C_D$ which accurately reproduce the behavior observed in the experiment, see also Fig. 2 and 3 in the main text. In the experiment, for $C_D = 145$ pF, it is observed that the amplification $A_V$ goes below 1 in the on the ramp down segment, while still being above 1 during the ramp up. The exact same behavior is observed in Fig. \ref{multi4} for $C_D = 145$ pF in the right panels, confirming the validity of the TDGL approach. 
%

\section{Negative capacitance transients with a series resistor}
\begin{figure*}[!h]
\begin{center}
 \includegraphics[width=6.5in]{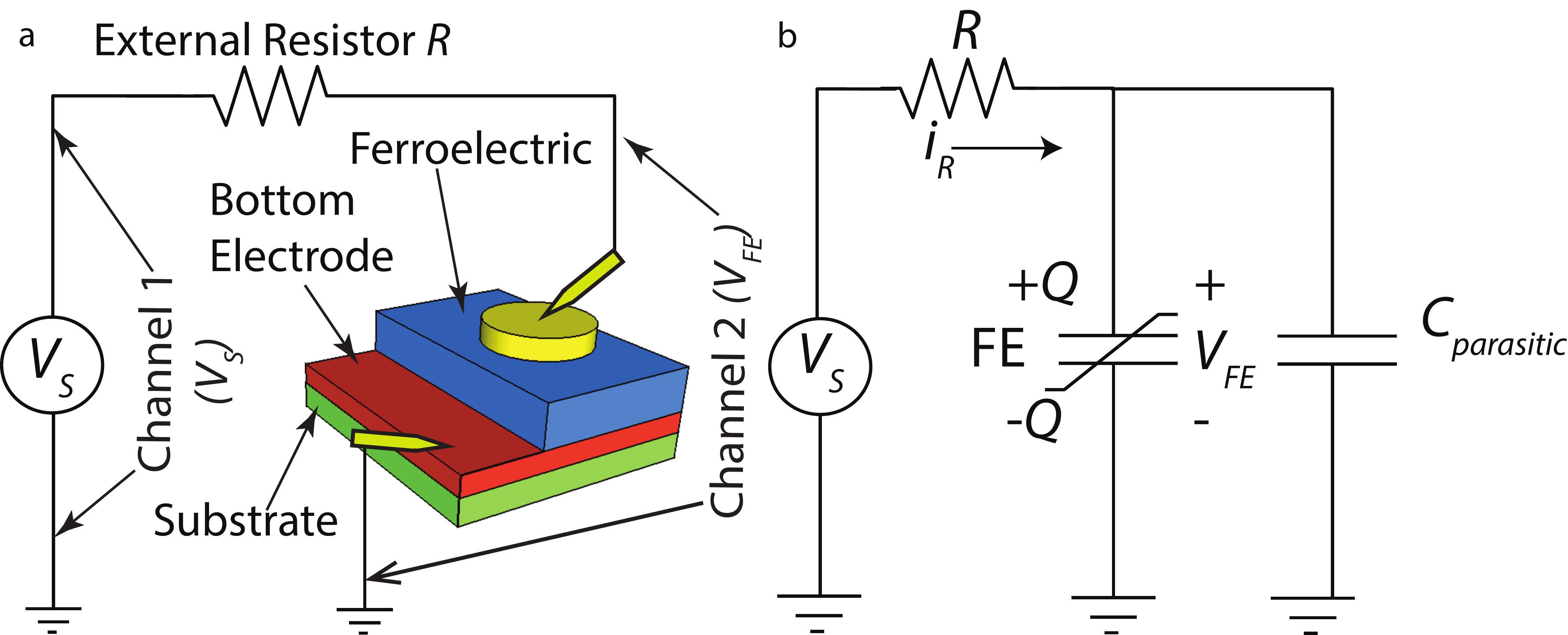}
 \end{center}
\caption{(a) Schematic diagram of the experimental set-up for direct measurement of the negative capacitance. (b) Equivalent circuit diagram of the experimental set-up. }
 \label{R_schematic}
\end{figure*}

\begin{figure*}[!h]
\begin{center}
 \includegraphics[width=6.5in]{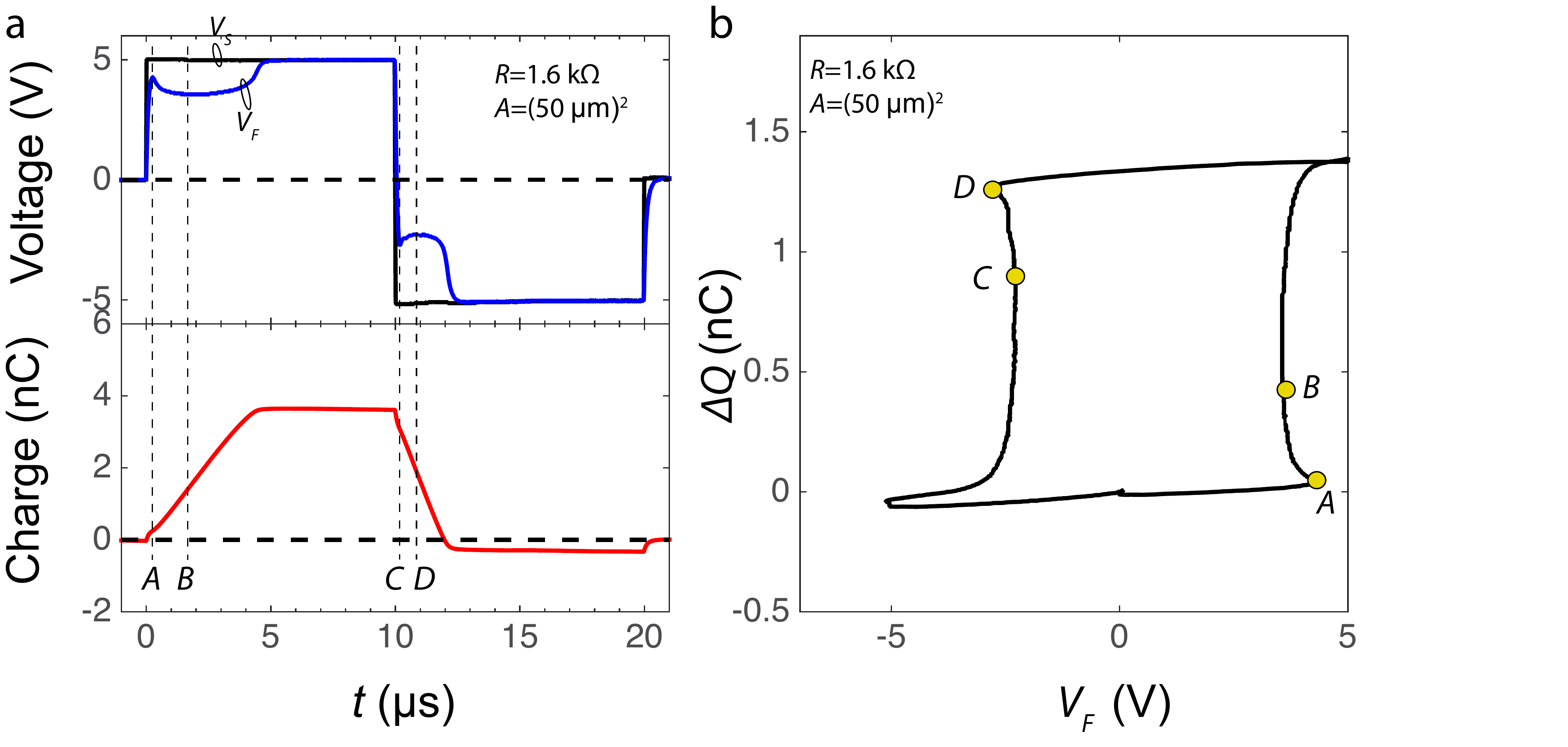}
 \end{center}
\caption{ (a) Waveforms corresponding to the source voltage $V_S$, the ferroelectric voltage $V_{F}$ and the charge $Q=\int i_R(t)dt$ in the $R-FE$ circuit ($R=$ 1.6 k$\Omega$) upon the application of a step voltage pulse $V_S$: 0 V $\rightarrow$ +5 V $\rightarrow$ -5 V$\rightarrow$0 V.  (b) The change in the charge in the circuit $\Delta Q(t)$ as a function of $V_{F}(t)$ calculated from the waveforms in Fig. a. }
 \label{PZT_1600}
\end{figure*}

Based on the methodology presented in Ref. \cite{khan2015negative}, we studied the negative capacitance response of the same Pb(Zr$_{0.2}$Ti$_{0.8}$)O$_3$ sample by constructing  a series circuit where the ferroelectric capacitor was connected to a voltage source through an in-series external resistor. Fig. \ref{R_schematic}(a) shows the experimental setup.  Fig. \ref{R_schematic}(b) shows the equivalent circuit diagram where $C_{parasitic}$ is  the parasitic
capacitance contributed by the probe station and the oscilloscope in the experimental set-up (measured to be $\sim$35 pF). Fig. \ref{PZT_1600}(a) shows the waveforms corresponding to the source voltage $V_S$, the ferroelectric voltage $V_{F}$ and the change in charge $\Delta Q=\int i_R(t)dt$ in the $R-FE$ circuit ($R=$ 1.6 k$\Omega$) upon the application of a step voltage pulse $V_S$: 0 V $\rightarrow$ +5 V $\rightarrow$ -5 V$\rightarrow$0 V. We observe  during the subsequent period $AB$, $V_{F}$ decreases while $Q$ increases. As a result, during $AB$, $C_{F}=d\Delta Q(t)/dV_{F}(t)<0$. The behavior of the ferroelectric during $AB$ in fig. \ref{PZT_1600}(a) is non-trivial; this is exactly opposite to what happens in a regular $R-C$ circuit where both $Q$ and the capacitor voltage change in the same direction in response to a step $V_s$ pulse. This points to the fact that, starting from equilibrium state with positive capacitance, the ferroelectric charge  enters into the negative capacitance state  at time $A$. After this duration, $V_{F}$ and $\Delta Q$ again change in the same direction which indicates the ferroelectric has returned to a positive capacitance state ($C_{F}>0$) . Similarly, during the time segment $CD$ in the negative voltage pulse cycle in fig. \ref{PZT_1600}(a), $C_{F}<0$ is observed.  In fig. \ref{PZT_1600}(b), we plot the change in the charge in the circuit $\Delta Q(t)$ as a function of $V_{F}(t)$. We observe in fig. \ref{PZT_1600}(b) that the slope of $\Delta Q(t)-V_{FE}(t)$ curve  is negative at the knee of hysteresis loop. In other words, negative capacitance is observed at the on-set of ferroelectric switching \cite{khan2015negative}. Note that the shape of the $\Delta Q(t)-V_{FE}(t)$ curve obtained from the ferroelectric-resistor series circuit measurement (the location of the negative slopes, to be precise) is very similar to that of the $\Delta Q(t)-V_{FE}(t)$ curve obtained from the ferroelectric-dielectric series circuit measurement (for example, the one shown in main-text Fig. 2C inset). 

\begin{figure*}[!h]
\begin{center}
 \includegraphics[width=6.5in]{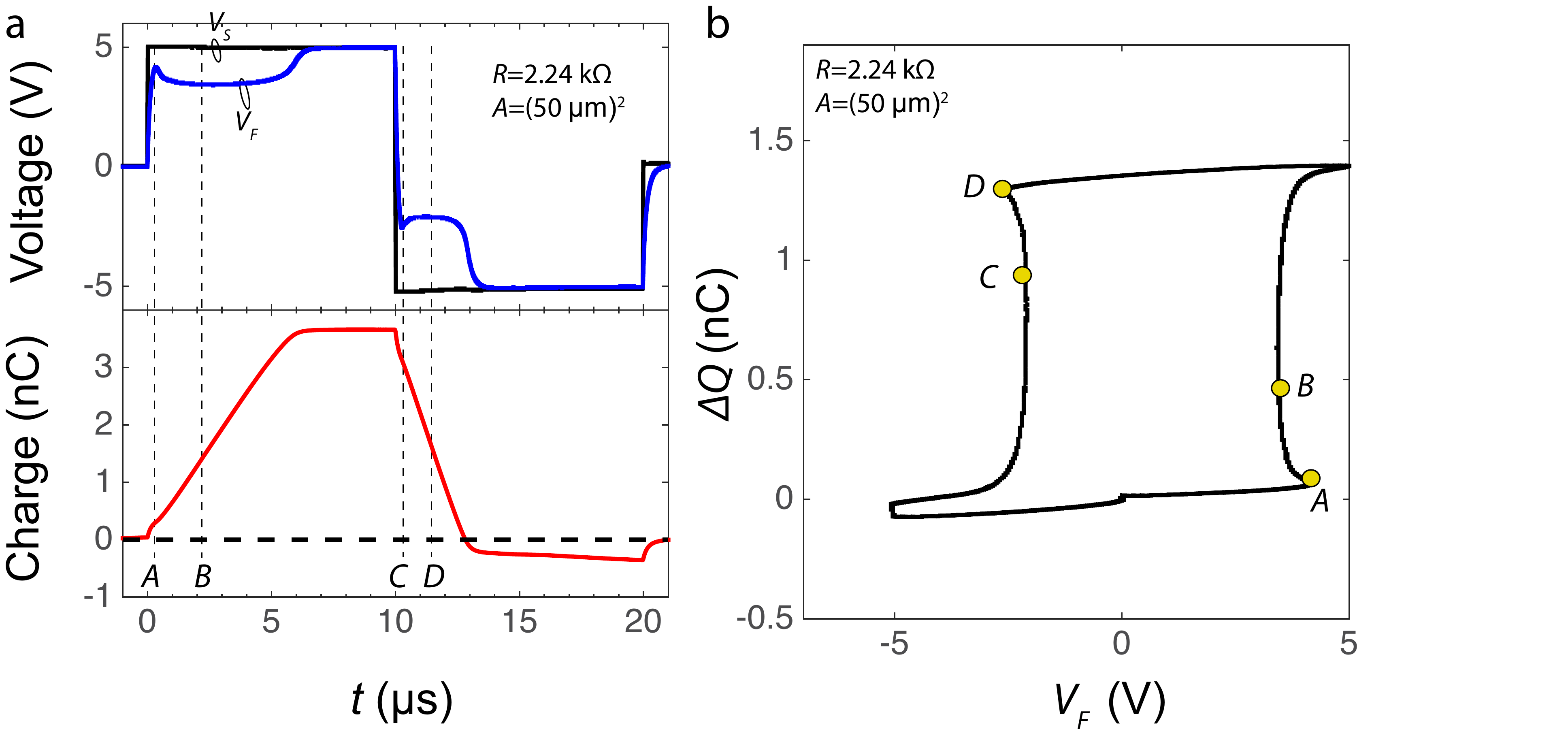}
 \end{center}
\caption{ (a) Waveforms corresponding to the source voltage $V_S$, the ferroelectric voltage $V_{F}$ and the charge $Q=\int i_R(t)dt$ in the $R-FE$ circuit ($R=$ 2.24 k$\Omega$) upon the application of a step voltage pulse $V_S$: 0 V $\rightarrow$ +5 V $\rightarrow$ -5 V$\rightarrow$0 V.  (b) The change in the charge in the circuit $\Delta Q(t)$ as a function of $V_{F}(t)$ calculated from the waveforms in Fig. a. }
 \label{PZT_2240}
\end{figure*}

\begin{figure*}[!h]
\begin{center}
 \includegraphics[width=6.5in]{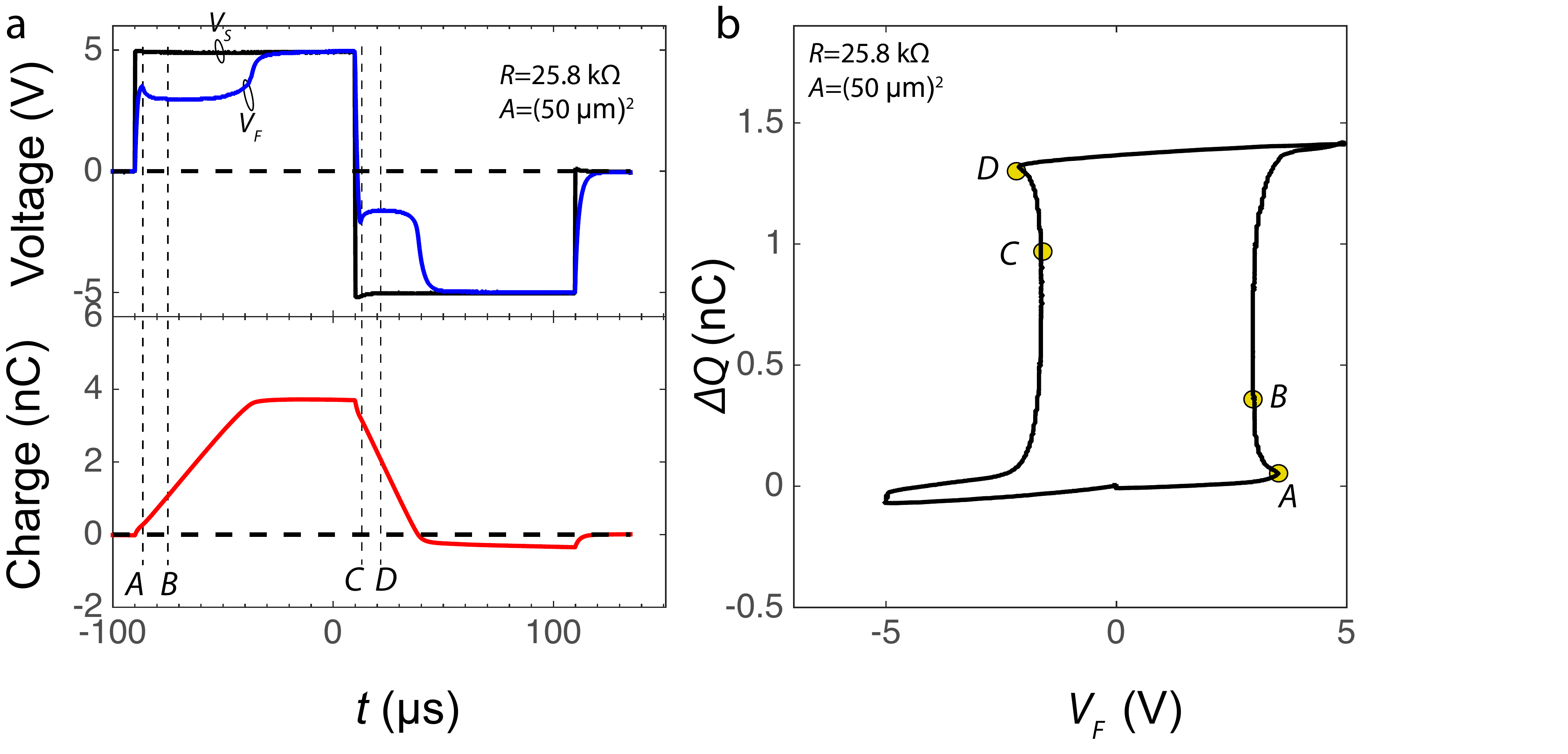}
 \end{center}
\caption{ (a) Waveforms corresponding to the source voltage $V_S$, the ferroelectric voltage $V_{F}$ and the charge $Q=\int i_R(t)dt$ in the $R-FE$ circuit ($R=$ 25.8 k$\Omega$) upon the application of a step voltage pulse $V_S$: 0 V $\rightarrow$ +5 V $\rightarrow$ -5 V$\rightarrow$0 V.  (b) The change in the charge in the circuit $\Delta Q(t)$ as a function of $V_{F}(t)$ calculated from the waveforms in Fig. a. }
 \label{PZT_25800}
\end{figure*}

\begin{figure*}[!h]
\begin{center}
 \includegraphics[width=4.5in]{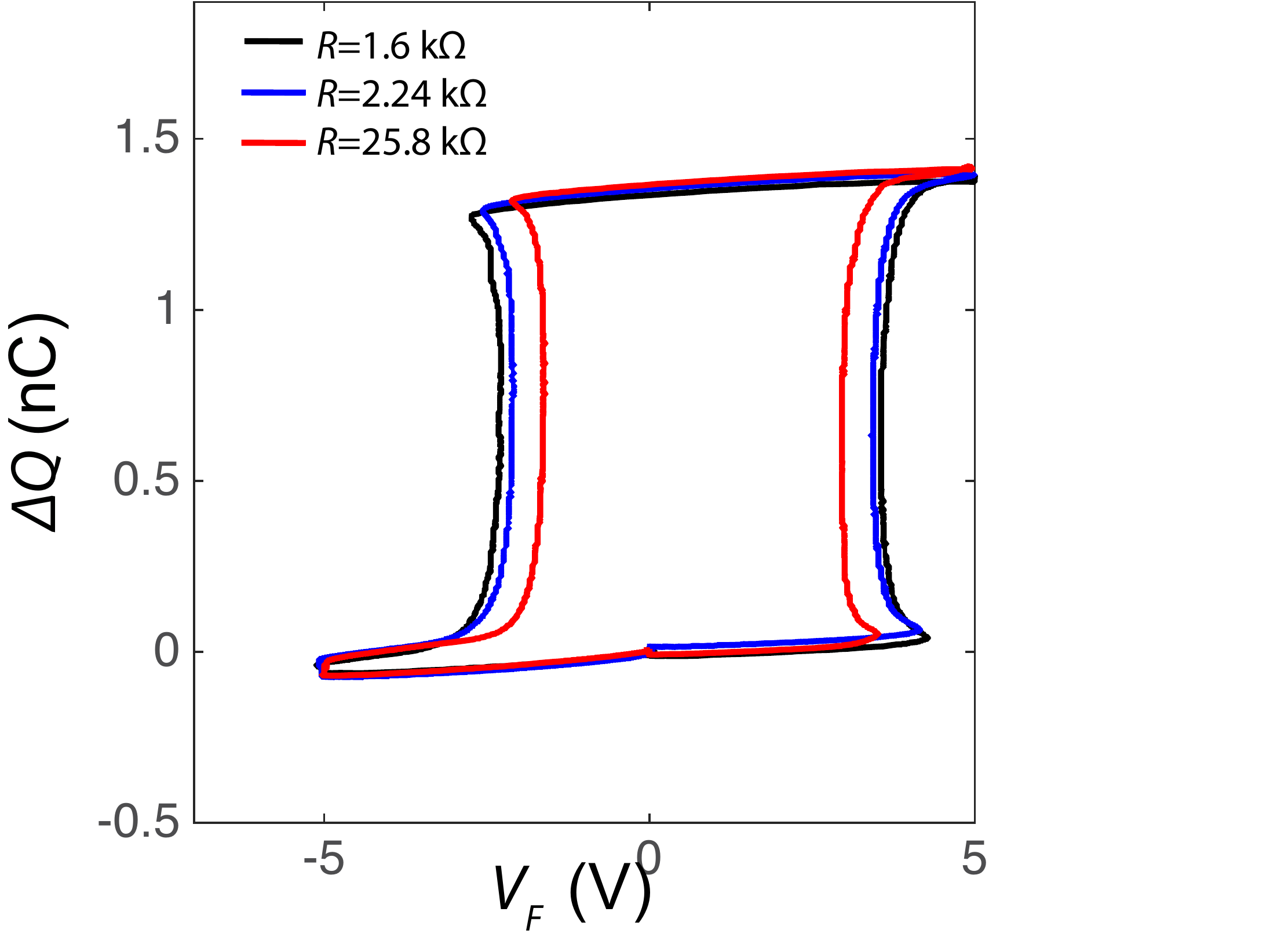}
 \end{center}
\caption{Comparison of the $\Delta Q(t)-V_{FE}(t)$ curves obtained for  $R$=1.6 k$\Omega$, 2.24 k$\Omega$ and  25.8 k$\Omega$.}
 \label{PV_all_PZT}
\end{figure*}

Fig. \ref{PZT_2240} and \ref{PZT_25800} show the response of the $R$-FE circuit with the same PZT sample and $R$= 2.24 k$\Omega$ and  25.8 k$\Omega$, respectively.  Fig. \ref{PV_all_PZT} compares the $\Delta Q(t)-V_{FE}(t)$ curves obtained for these three different values of $R$ (1.6 k$\Omega$, 2.24 k$\Omega$ and  25.8 k$\Omega$). We note in Fig. \ref{PV_all_PZT} that, with the increase of $R$, the width of the hysteresis loop decreases which is in line with the observations made in Ref. \cite{khan2015negative}.

\section{Voltage amplification in BiFeO$_3$}
\begin{figure*}[!h]
\begin{center}
 \includegraphics[width=2.5in]{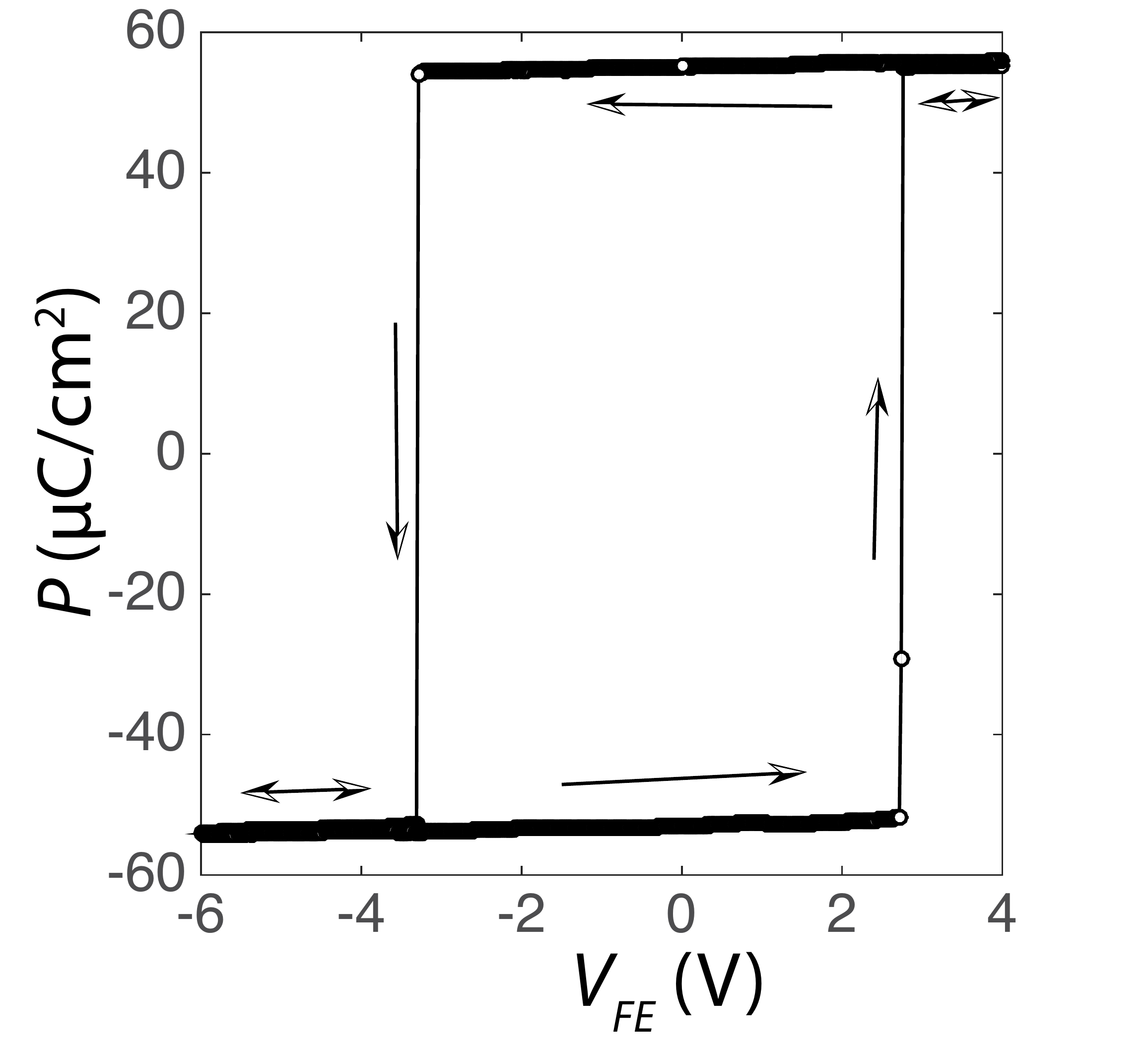}
 \end{center}
\caption{ Ferroelectric polarization-voltage hysteresis of the epitaxial 300 nm thick  BiFeO$_3$ thin film grown on metallic SrRuO$_3$ buffered SrTiO$_3$ measured using a ferroelectric tester. Measurement frequency was 10 kHz. }
 \label{BFO_hys}
\end{figure*}

\begin{figure*}[!h]
\begin{center}
 \includegraphics[width=6.5in]{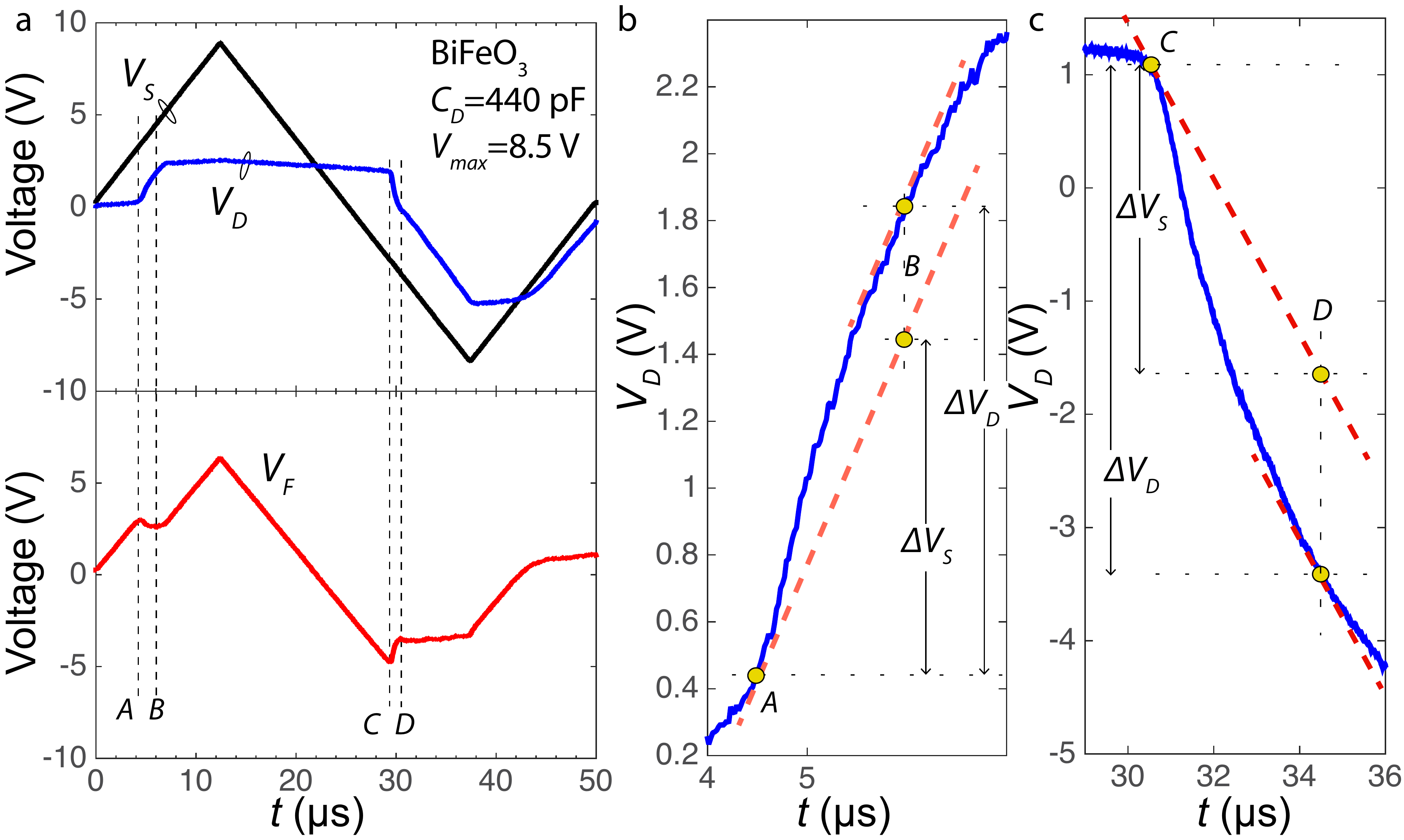}
 \end{center}
\caption{ (a) $V_D$, $V_F$ and $V_S$ of the BiFeO$_3$ (300 nm)/SrRuO$_3$/SrTiO$_3$  circuit in response to a bipolar triangular pulse $V_S$: 0$\rightarrow$+8.5 V$\rightarrow$ -8.5 V$\rightarrow$0 V with a period $T$=50 $\mu$s for $C_D$={440 pF}. (B, C) Magnified version of $V_D$ waveform during $AB$ and $CD$. The dashed red line has the same slew rate as that of $V_S(t)$ in these durations. }
 \label{BFO_440pF_1}
\end{figure*}

\begin{figure*}[!h]
\begin{center}
 \includegraphics[width=6.5in]{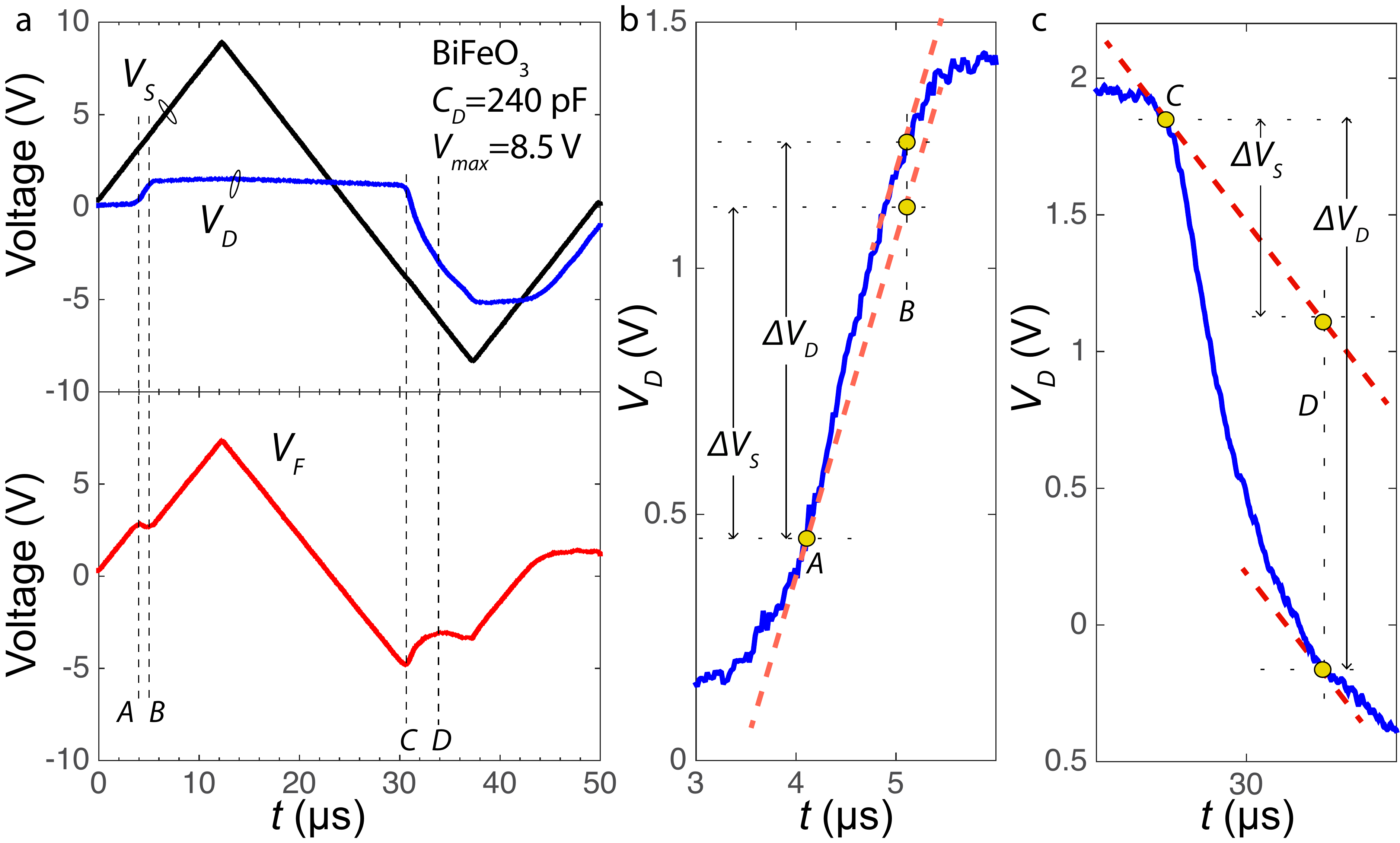}
 \end{center}
\caption{(a) $V_D$, $V_F$ and $V_S$ of the BiFeO$_3$ (300 nm)/SrRuO$_3$/SrTiO$_3$  circuit in response to a bipolar triangular pulse $V_S$: 0$\rightarrow$+8.5 V$\rightarrow$ -8.5 V$\rightarrow$0 V with a period $T$=50 $\mu$s for $C_D$={240 pF}. (B, C) Magnified version of $V_D$ waveform during $AB$ and $CD$. The dashed red line has the same slew rate as that of $V_S(t)$ in these durations.  }
 \label{BFO_240pF_1}
\end{figure*}

\begin{figure*}[!h]
\begin{center}
 \includegraphics[width=6.5in]{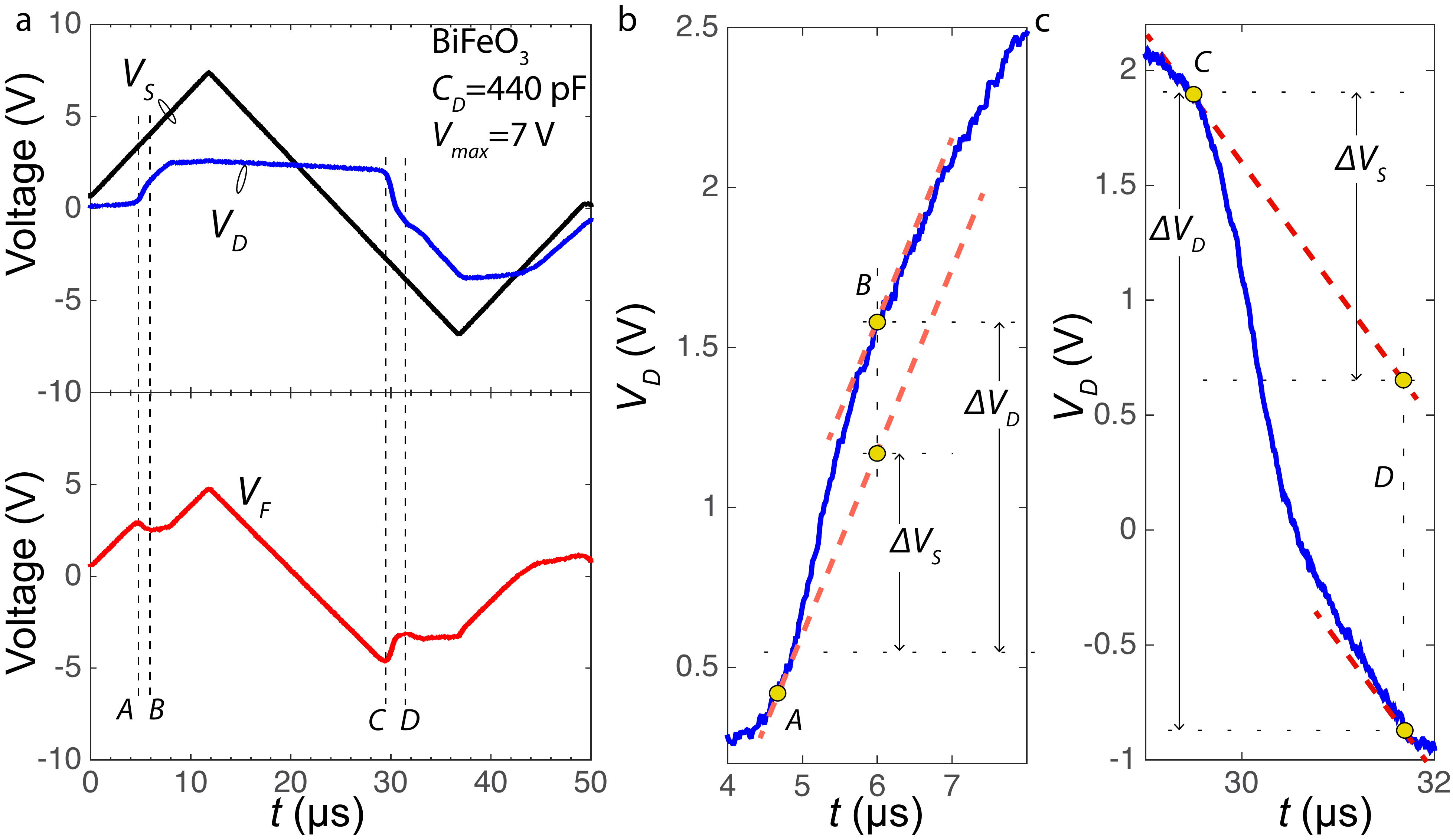}
 \end{center}
\caption{ (a) $V_D$, $V_F$ and $V_S$ of the BiFeO$_3$ (300 nm)/SrRuO$_3$/SrTiO$_3$  circuit in response to a bipolar triangular pulse $V_S$: 0$\rightarrow$+7 V$\rightarrow$ -7 V$\rightarrow$0 V with a period $T$=50 $\mu$s for $C_D$={440 pF}. (B, C) Magnified version of $V_D$ waveform during $AB$ and $CD$. The dashed red line has the same slew rate as that of $V_S(t)$ in these durations. }
 \label{BFO_440pF_2}
\end{figure*}

\begin{figure*}[!h]
\begin{center}
 \includegraphics[width=6.5in]{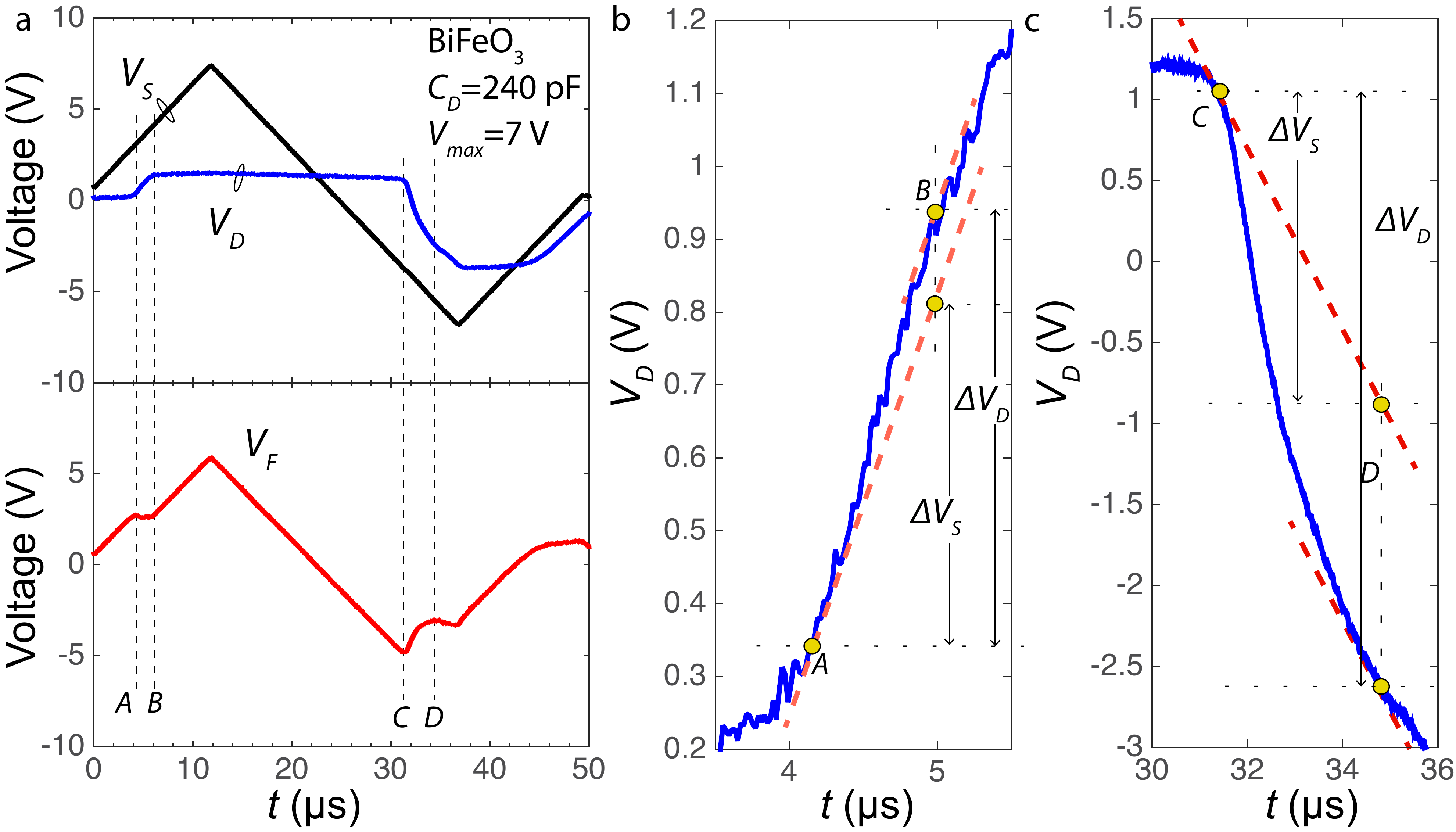}
 \end{center}
\caption{ (a) $V_D$, $V_F$ and $V_S$ of the BiFeO$_3$ (300 nm)/SrRuO$_3$/SrTiO$_3$  circuit in response to a bipolar triangular pulse $V_S$: 0$\rightarrow$+7 V$\rightarrow$ -7 V$\rightarrow$0 V with a period $T$=50 $\mu$s for $C_D$={240 pF}. (B, C) Magnified version of $V_D$ waveform during $AB$ and $CD$. The dashed red line has the same slew rate as that of $V_S(t)$ in these durations. }
 \label{BFO_240pF_2}
\end{figure*}

We also used an epitaxial 300 nm thick  BiFeO$_3$ thin film grown on metallic SrRuO$_3$ buffered SrTiO$_3$ substrate Au top electrode (area $A$=$\pi$(50 $\mu$m)$^2$) as the ferroelectric capacitor.   The sample shows extremely sharp jumps in polarization-voltage hysteresis curve as shown in Fig. \ref{BFO_hys}. The same sample was used in Ref. \cite{khan2016negative}. Fig. \ref{BFO_440pF_1} and \ref{BFO_240pF_1} show the waveforms corresponding to $V_D$, $V_F$ and $V_S$ of the ferroelectric-dielectric series circuit in response to a bipolar triangular pulse $V_S$: 0$\rightarrow$+8.5 V$\rightarrow$ -8.5 V$\rightarrow$0 V with a period $T$=50 $\mu$s for $C_D$={440 pF} and 240 pF, respectively. Fig. \ref{BFO_440pF_2} and \ref{BFO_240pF_2} show the waveforms corresponding to $V_D$ and $V_S$ of the ferroelectric-dielectric series circuit in response to a bipolar triangular pulse $V_S$: 0$\rightarrow$+7 V$\rightarrow$ -7 V$\rightarrow$0 V with a period $T$=50 $\mu$s for $C_D$={440 pF} and 240 pF, respectively. Magnified versions of the $V_D$ waveforms in the main-panels are also shown separately. In the bottom panels of Fig. \ref{BFO_440pF_1},  \ref{BFO_240pF_1}, \ref{BFO_440pF_2} and \ref{BFO_240pF_2}, we observe that the $V_F$ decreases (and increases) during the time durations $AB$ and $CD$, while $V_S$ actually increases (and decreases, respectively). As a result,  $V_D$ changes faster than $V_S$ in these durations ($i.e.$ $dV_D/dV_S\ge1$). 

\section{Domain Dynamics and Energy Considerations}
We note that as the ferroelectric switches from one state to the other, it imparts some of its stored energy onto the dielectric, leading to the amplification. This is possible due to the energy barrier between opposite polarization states corresponding to the degenerate energy minima in Figure 1a. When a field is applied opposite to the spontaneous polarization (say +$P_s$), it does not switch immediately to the opposite state (-$P_s$), even if this other state is lower in energy. Some energy is still stored due to the barrier and will only be transferred to the dielectric, when the barrier vanishes (i.e. when the voltage across the ferroelectric $V_F$ is larger than its coercive voltage). Microscopically, this means that opposite domains are nucleating and expanding throughout the ferroelectric, which can reduce the voltage across the ferroelectric as was shown before in a capacitor-resistor series circuit [3]. In a series circuit of two capacitors, if the voltage across one of them suddenly drops by -$\Delta V$, the voltage across the other one has to increase by +$\Delta V$. This then results in the differential voltage gain we report in this study.

\clearpage

\end{document}